%% file: paper.tex
\documentclass[journal]{IEEEtran}

\IEEEoverridecommandlockouts                              
\usepackage[usenames,dvipsnames]{xcolor}
\usepackage{tikz}

\usepackage{epsfig} 
\usepackage{enumerate}
\usepackage{graphics} 
\usepackage{amsmath} 
\usepackage{amssymb}  
\usepackage{cite}	
\usepackage{graphicx}
\usepackage{tabularx}
\usepackage{lipsum}
\usepackage{multirow}
\usepackage{psfrag}
\usepackage{float}
\usepackage{placeins}
\usepackage{caption}
\usepackage{soul}
\usepackage{bm}

\usepackage{subcaption}

\usepackage[]{units}
\usepackage{siunitx}
\usepackage{hyperref}
\usepackage{adjustbox}

\usetikzlibrary{shapes,arrows,fit,calc}

\usepackage{pgfplotstable}

\pgfplotsset{compat=newest} 
\pgfplotsset{plot coordinates/math parser=false} 
\newlength\figureheight 
\newlength\figurewidth

\newcommand*{\MyPath}{.}%

\newcommand{\ie}{\textit{i.e.}~}
\newcommand{\iec}{\textit{i.e.},~}

\newcommand{\eal}{\textit{~et al.}~}
\newcommand{\cfr}{cfr.~}

\newcommand{\fig}{Fig.~}

\newcommand{\tabb}{Table~}

\newcommand{\sect}{Sect.~}

\newcommand{\TL}[1]{{\color{black} #1}}

\title{\Huge 
Wearable Haptics for Remote Social Walking}

\author{Tommaso Lisini Baldi$^{1}$,
 Gianluca Paolocci$^{1,2}$,
 Davide~Barcelli$^{1}$,
 and Domenico Prattichizzo$^{1,2}$
\thanks{$^1$Tommaso Lisini Baldi, Gianluca Paolocci, Davide Barcelli, and Domenico Prattichizzo are with the Department of Information Engineering and Mathematics, University of Siena, Via Roma 56, I-53100 Siena, Italy. $\{$lisini, paolocci, barcelli, prattichizzo$\}$@diism.unisi.it}%
\thanks{%
$^2$Gianluca Paolocci and Domenico Prattichizzo are with the Department of Advanced Robotics, Istituto Italiano di Tecnologia, Genova, 16163, Italy. {\{gianluca.paolocci, domenico.prattichizzo\}@iit.it}}%
}
\date{}

\begin{document}

\maketitle

\begin{abstract}
Walking is an essential activity for a healthy life, which becomes less tiring and more enjoyable if done together. 
Common difficulties we have in performing sufficient physical exercise, for instance the lack of motivation, can be overcome by exploiting its social aspect. 
However, our lifestyle sometimes makes it very difficult to find time together with others who live far away from us to go for a walk. 
In this paper we propose a novel system enabling people to have a \lq remote social walk' by streaming the gait cadence  between two persons walking in different places, increasing the sense of mutual presence.
Vibrations provided at the users' ankles display the partner's sensation perceived during the heel-strike.
In order to achieve the aforementioned goal in a two users experiment, we envisaged a four-step incremental validation process: \textit{i)} a single walker has to adapt the cadence with a virtual reference generated by a software; \textit{ii)} a single user is tasked to follow a predefined time-varying gait cadence; \textit{iii)} a leader-follower scenario in which the haptic actuation is mono-directional; \textit{iv)} a peer-to-peer case with bi-directional haptic communication.
Careful experimental validation was conducted involving a total of 50 participants, which confirmed the efficacy of our system in perceiving the partners' gait cadence in each of the proposed scenarios.
\end{abstract}


\setcounter{secnumdepth}{2}

\section{Introduction\label{SEC:intro}}
Clapping hands in an audience, playing music in an orchestra, training in sports and dance represent a tiny fraction among the countless situations in which humans perform coordinate actions.
Coordinated motion is probably one of the most ancient and exploited human behaviors. 
For instance, religions around the world incorporate synchronous singing and gestures into their rituals. Psychologists, anthropologists, and sociologists have speculated that rituals involving synchronous moves may produce positive emotions that encourage participation. 
Wiltermuth and Heat in~\cite{wiltermuth2009synchrony} studied whether synchronous activities serve as a partial solution to the free-rider problem facing groups that need to motivate their members to contribute toward the collective good. The physical synchronization mechanism, which occurs when people move in time with one another, has been studied and discussed for decades. In \cite{hannah1977african, ehrenreich2007dancing}, and \cite{st2004connectedness} authors demonstrate that dance can also weaken the boundaries between the self and the group.

\begin{figure}
\centering
\includegraphics[width=\columnwidth]{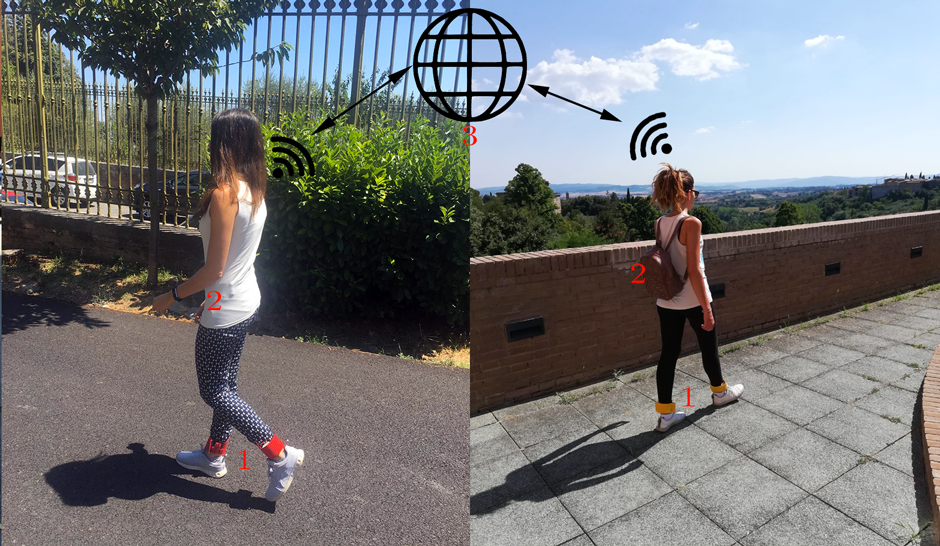}
\caption{The proposed remote social walking system: 
$i$) the anklet device (1) measures \textit{user A} gait cadence; $ii$) the smart-phone (2) receives the measurement via Bluetooth; $iii$) the information is sent to a server via the Internet (3); $iv$) the server sends the update to the \textit{user B}'s smart-phone; $v$) the smart-phone (2) commands vibrations at the user's anklets (1).}
\label{fig:systemSchema}
\end{figure}

The motivation generated by being part of a group can encourage healthy behaviors, as in the \textit{\lq Social Walking'}~\cite{hanson2015there}. Studies on its impact showed an increase in positive attitude toward physical activity, social cohesion, as well as reducing disease risks~\cite{kwak2006your}. 
\TL{In fact, it has been demonstrated that walking experiences contribute to participants'  well-being \cite{yang2018perceived}.}
Therefore, its promotion as part of the lifestyle has become one of the main aims of governments, as it is expected to significantly reduce health care costs and increase the quality of life~\cite{hardy2011ability}. 

Not many years ago, walking was the primary mean of transport people used to get to work. Nowadays, the great majority of industrialized countries population use cars or public transports to move within the city for daily activities~\cite{mckenzie2014modes}. 
It follows that walking for a sufficient distance on a weekly basis must be incorporated in the spare time. A recurrent cause of physical inactivity is the lack of time, or often, an appropriate motivation. 
%
Although participating in group activities proved a powerful mean to fight physical inactivity, finding the right walking partner may be difficult. In fact, to have a social walk the group needs to 
share the same time slots and, obvious but essential, 
the common location where the tour takes place, \textit{e.g.} the city park. 
Connecting people far in space may solve the problem. With this work, we aim at creating a system for \lq\lq feeling'' a partner remotely while walking, by exploiting the internet connection to exchange vibrotactile cues, as sketched in \fig\ref{fig:systemSchema}.
Towards the same goal, Muller\eal  presented a headset to transmit gait cadence information through audio cues~\cite{mueller2007jogging}. 
\TL{Audio signals are often exploited to suggest body postures. For instance, in~\cite{menzer2010feeling} the authors investigated agency for the entire body by testing auditory action effects related to gait, while Murata\eal proposed a system that synchronizes each step with the music being listened to 
and creates a feeling of generating music through walking~\cite{murata2017walk}.}
Those systems, however, occupy the hearing, one of the most used input channels.

In our experience the audio channel 
is a primary source of information about the surroundings, thus cadence cues should be conveyed using different modalities. Haptic communication is generally the preferred option, due to its intuitiveness, efficacy, and intimate nature. Following these motivations, we decided to display cadence information through the use of touch. Tactile feedback has been demonstrated to be an effective way to  significantly improve the perceived virtual and social presence of a remote companion. In  \cite{sallnas2010haptic}, authors show that haptic feedback reinforces the impression that one is actually there in a mediated environment.

Tactile perceptions can be rendered using haptic interfaces, which apply different kinds of stimuli to the human body that are easily associated with realistic sensations. 
Haptic interfaces can either rely on kinaesthetic or cutaneous interaction~\cite{lederman2009haptic}. The former consists in the proprioception of ligaments and muscles tension, necessary for the awareness of limb posture and to estimate external forces.
The cutaneous counterpart relies on skin receptors to perceive details like texture, temperature, and shape. Kinaesthetic actuators are not suitable for the proposed work since they are mostly grounded or bulky, while cutaneous stimuli can be displayed by means of small and wearable devices, making them the most appropriate choice for the purpose.

The main categories of wearable cutaneous devices are: skin indentation, skin stretch, temperature, and vibrations~\cite{pacchierotti2017wearable}. 
Skin indentation and stretch devices usually can exert forces (normal or tangential, respectively), while vibrations can be modulated to display textures or can be used to render events. 

Over the years, cutaneous stimuli have been found an effective, yet non-intrusive, way for suggesting directions and pace cues to users. A vibrotactile waist belt composed of eight tactors was used for waypoint navigation in outdoor scenarios \cite{van2005waypoint}. The waist belt displayed both the direction and distance to the next
waypoint. A similar device was used to provide vibrotactile
cues for improving situational awareness of soldiers in a
simulated building-clearing exercise \cite{lindeman2005effectiveness}. In \cite{cosgun2014guidance}, a vibrotactile 
belt was used for human guidance in indoor environments.
Continuous stimuli were used to display directional and rotational motions to the blindfolded users. Vibrotactile armbands were used to navigate subjects along fixed paths using three different stimuli \cite{AgScPr-iros15}.
Similar devices and strategies were used to guide blindfolded users in dynamic environments autonomously \cite{LiScAgPr-RAL2017} or assisted by a mobile robot~\cite{scheggi2014cooperative}.

For what concerns suggesting the step cadence, exploratory research in this direction revealed the potentiality of using haptics for suggesting rhythm. For instance, in \cite{paolocci2018human} and \cite{koutamerd} authors exploited vibrations as a metronome for suggesting tempo in walking/running activities. 

With the objective of removing the spatial constraint and establishing a remote presence, we propose a system that measures the gait cadence of each participant and transmits it to the partner, allowing each walker to `feel' the other. The system, is composed of vibrotactile devices worn at the ankles which provide timing cues 
displaying the partner's walking pace. The gait cadence is measured using a pressure sensor immersed into a silicon heel insole and connected with one of the haptic interfaces. The detected steps are then sent to the user's smart-phone that communicates with a dedicated server. Finally, each social walker's smart-phone receives the gait cadence update from the server and adjusts the vibration pattern consequently.
To the best of our knowledge, this work represents one of the first attempts to allow user dyads to walk together, despite their distance.

This paper is organized as follows. \sect\ref{SECT:setup} provides a detailed description of the proposed system from an engineering perspective, including hardware (haptic anklets) and software (firmware, application and server) details. The third section (\sect\ref{SECT:gait}) presents a \textit{divide et impera} approach to the problem. We identified four objectives of incremental difficulty for remote social walking to succeed. \sect\ref{SECT:Experimental} describes the experiments performed to verify the achievement of these objectives and reports \textit{a-posteriori} discussions, enriched with correlational analysis of participants' baseline walking parameters and performance in \sect\ref{SECT:CORR}. In \sect\ref{SECT:qualitative}, qualitative results and users' feedback are reported. Conclusions are drawn in \sect\ref{SECT:conclsion}, along with a brief discussion on the range of possible reach directions that the developed system may enable.

\section{System overview\label{SECT:setup}}
The proposed system is composed of three elements: 
$i$) an anklets pair: two wearable fabric-made bands equipped with an electronic board; $ii$) an application running on a smart-phone; $iii$) a remote server for broadcasting and logging data. During the remote social walk each user has an anklets pair and a smart-phone connected to the server.

\subsection{Description of the anklet device}\label{SECT:descriptionBracelet}

The purpose of the wearable devices is twofold: providing the user with vibrotactile stimuli and extracting the gait cadence. The two devices, worn on the two ankles, equip an electronic board with a Bluetooth module and two vibro-motors. The devices in a pair differ by the presence of a pressure sensor for cadence sampling. From here on, the sensing anklet will be referred to as master.

Tactile vibratory sensitivity is influenced by the physical location of tactile receptors on the body, their distance with respect to the actuators, the stimulation frequency, and the age of the user. Studies demonstrated that vibrations are better sensed on hairy skin due to its thickness and nerve depth, and that vibrotactile stimuli are best detected in bony areas~\cite{gemperle03}. In particular, wrists and spine are generally preferred for detecting vibrations, with arms and ankles next in line~\cite{Karuei-2011}. 
A previous study showed that during walking activities participants preferred being informed with haptic cues delivered at the ankles~\cite{baldi2018human}.
Due to the aforementioned considerations and since our aim is to design an intuitive and non-obtrusive device which could be easily worn, we concentrated on the development of vibrotactile anklets. 
Starting from the results presented by Scheggi \eal in \cite{ScAgPr-THMS16}, we decided to use the bilateral configuration, that required two vibro-tactile interfaces, one per ankle. 

From a technical point of view, the vibrotactile anklets are composed by cylindrical vibro-motors, independently controlled via Bluetooth with a custom communication protocol (see \fig\ref{fig:anklets}). 
The communication between the haptic interface and the smart-phone is realized with an RN-42 \mbox{Bluetooth} antenna connected to the serial port of a \unit[3.3]{V} Arduino pro-mini. The wireless connection baud rate is \unit[57600]{bps}.  The microcontroller installed on the  board is used to pilot the motors activation and to receive data from an external source. Note that each vibro-motor is voltage controlled, which implies a strict coupling between frequency and amplitude that cannot be varied independently.
As the user's maximal sensitivity is achieved around \unit[200-300]{Hz} \cite{riener2010} (the human \TL{sensitive range} is between \unit[20]{Hz} and \unit[400]{Hz}), two Precision Microdrives Pico Vibe vibration motors are placed into two fabric pockets (the width of the fabric band is about \unit[60]{mm}), with vertically aligned shafts. The motors have a vibration frequency range of \unit[100-300]{Hz}, lag time of about \unit[20]{ms}, rise and stop time of \unit[35]{ms}. The bracelet guarantees about $4$ hours of battery life with one motor permanently activated. Each bracelet weights about \unit[90]{g}.

The micro-controller permanently checks for incoming data on the serial port and, upon reception, activates the motors for \unit[150]{ms}. This design choice allows for a finer control on the vibrational cues timing at the smart-phone application level, and proved to be the most effective strategy for a prompter adaptation to new gait cadences during the system development.
In addition to the previously described hardware, the master anklet is equipped with a flexible force sensor (FSR~400, manufactured by Interlink Electronics, Inc.) connected to the controller. 
During experimental trials, the force sensor is placed inside the right shoe (under the heel) to record the force pattern due to the contact between foot and ground. 
The force sensing resistor measures the force applied through the deformation of the active surface, which produces a resistance variation.
The force value is converted into a 10 bit digital signal. 
The stride extraction procedure exploits a single-threshold value, defined as the double of the standard deviation of the  data, measured during a calibration phase.
The sensor records the pressure under the heel at \unit[100]{Hz}. Thus, we are able to extract the stride temporal sequence from the obtained data. 
The stride-detection procedure consists of three phases. In the first step, raw force data are acquired by the system, normalized, and transformed into a two-levels signal using the computed threshold.
A square wave is generated as follows: the signal assumes logical value $1$ whenever the foot is in contact with the ground, and $0$ otherwise. Before further processing, a debounce software mechanism is adopted to rejects variations that are physically improbable. Then, the algorithm extracts positive edges representing the heel contact with the floor, identifying the current stride duration as the time interval between two consecutive edges. 
When a cadence variation is detected, the new value is sent over the serial port so that the Bluetooth module can deliver it to the paired smart-phone.
Moreover, upon establishing a solid proof of concept and for a wider experimental validation, it would be possible to move to a more robust, small and easy-to-use device to estimate the user's gait cadence. In that respect, we investigated the usage of one of the most famous sport gadgets: the \textit{Nike+iPod Sport Kit} (Apple Computer, Inc.) \cite{NikeSensor}. This solution removes the wear and tear of the pressure sensor and increases the battery lifetime. Details, interfacing, and recording strategies are reported at the end of this manuscript. 
However, we opted to use the pressure sensor for the present work, because the \textit{Nike+iPod Sport Kit} guarantees an inferior temporal resolution and introduces additional delay. Such inconveniences would have a negligible impact on a larger scale session as they would be canceled out in the long run.

\input{figures/setup}

\subsection{Smart phone app and remote server}

The experimental evaluations presented in this paper leveraged on different pieces of software whose purpose was to regulate the haptic stimulation in real time and to record the gait data during trials, even in an unstructured environment.
The overall software architecture employed in this work is organized as follows.

A Java TCP server, hosted in our university facility, permanently accepts incoming connections on a dedicated pair of public IP address and port, while the smart-phones use their cellular network to connect to it over the Internet. The server handles each connection in a dedicated thread, so as to ensure scalability. When the desired number of walkers have established a connection (without loss of generality, in this paper we study the case of two) the server notifies all actors and the information stream is started: whenever a client sends an update the server broadcasts it to all the other clients. In correspondence to any such event, the server also logs the entire state in a text file, for later post-processing.
On the smart-phone side, an application is organized with background services to handle both connections, \textit{i.e.} TCP toward the remote server and Bluetooth toward the Arduino boards. Two main cases are envisaged: $i$) a gait update is received from an anklet, which implies an immediate transmission of that information to the server and, in turns, to the other smart-phone; $ii$) an update is received from the network and the local vibration frequency needs to be adjusted accordingly.
Vibrations are managed by the application, depicted in \fig\ref{fig:AndroidApp}, which employs timers to send the start vibration command to the Arduino boards. Clearly, the anklets vibrations should exhibit a 180 degrees phase, thus the app sends the vibration signal alternately to one side or the other, then waits half a period before the next vibration.
This design allows for a quick gait cadence adaptation as only half period has to be waited to adjust the vibration timings, while also retaining operation smoothness.
The correctness, effectiveness and reliability aspects of the system were analyzed carefully, however we omit the description of the validation process due to space requirements. It is still meaningful though to mention a few considerations in this regard. The overall time needed to observe a vibration frequency update commanded by the other smart-phone is negligible with respect to the system time scales; in fact the 4G internet connection on the smart-phones introduced almost no delay and the dedicated faculty server was endowed with more than enough bandwidth and tiny ping timing. 
With regard to data protection, threats must be evaluated since user's data flow over untrustworthy networks. In our case, no sensible information was contained in the packets, which, in addition to the fact that the server port was not previously used and the server was turned on only during experiments, reduced the probability of data corruption and/or stealing.

\begin{figure}
\centering
\includegraphics[width=0.9\columnwidth]{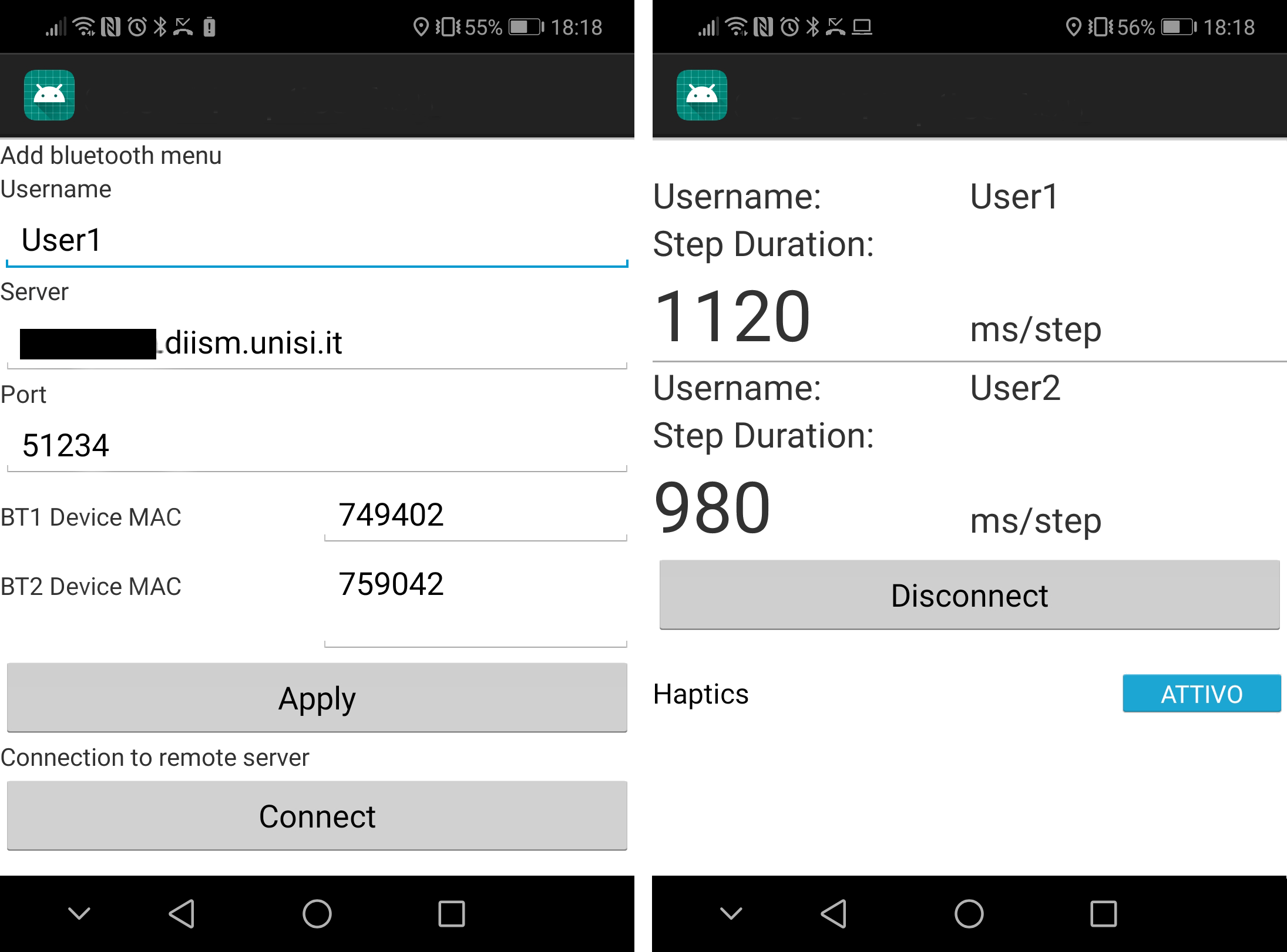}
\label{fig:AndroidApp}
\caption{The smart-phone application used to connect to the remote server via internet and to the anklet via Bluetooth. In the left panel is the initial view of the app. The user can select the username and pair the two haptic anklets using the MAC address. The right panel shows the app screen during a trial. User's and companion's stride duration are visualized by the app. Moreover, a button allows the user to enable or disable the haptic cues manually.}
\label{fig:AndroidApp}
\end{figure}

\section{Human walking pace regulation \label{SECT:gait}}
In this work, we rely on the gait cycle schema proposed by Philippson in~\cite{philippson1905autonomie}. A stride is completed after a stance and swing phase of a given foot, from heel strike to heel strike. Walking kinematics are characterized by pseudo-periodic patterns, where the stride represents the single period. The temporal duration of a stride is called gait cycle or stride duration, and the distance covered by two consecutive heel strikes of the same foot is called stride length. 
The application framework of our study is the remote connection between two or more users, thus we cannot provide intuitive indications to regulate the stride length. Instead, we investigate the temporal properties of human gait, \ie human gait cadence, without considering stride length and walking speed.
Notice that we will talk about walking cadence and stride duration referring to the same concept. 

The temporal aspects of gait are regulated by (internal) human timekeeping mechanisms and environmental parameters. 
The synchronization of brain oscillations (and thus motor coordination) with external stimuli relies on a feature of human sensory-motor system called sensory-motor entrainment~\cite{delcomyn1980neural, miyake1999internal}. If a subject is given a periodic stimulus of constant frequency and sufficient intensity to excite the thalamus, the brain has the tendency to align its dominant EEG frequency to the frequency of the external stimulus. Entrainment applies for visual, audio and haptic stimuli~\cite{wang2013preliminary}.
In our method, participants are periodically provided with vibro-tactile stimuli to suggest the appropriate gait cadence.
We positioned the cueing system following results presented in~ \cite{baldi2018human}.
Generally, the human locomotor system incorporates inputs from both the central nervous system, peripheral inputs, and sensory feedbacks. As reported in \cite{dingwell2000nonlinear}, these kinds of inputs are suggested to be possible reasons for the presence of the long range correlation in normal human walking. 
Thus, a gait cadence synchronization method based on body tactile stimulation is suitable for comfortably and effectively suggesting walking cadence. 

To test the effectiveness of our system and study the mutual effect of tactile stimuli on users' gait cadence, we introduced a set of \lq incremental' experiments, each tailored to assess a specific goal. The rest of this section is devoted to the introduction of the hypotheses motivating each experiment, to facilitate the reading and comprehension of the experimental procedure.

\TL{
\input{figures/representativeExperiments}
}

\subsection{Preparatory experiments}
As a preparatory phase, we conducted an experimental campaign to assess whether humans are able to	synchronize gait events, \textit{e.g.} the heel strike or lift off, with vibrations displaying a constant rhythm close to the participant's walking cadence. 

Two hypotheses were formulated and tested: $i$) humans are capable of aligning their stride sequence to an external rhythm displayed using vibrations; $ii$) humans can maintain the suggested walking cadence in presence of a simple secondary task. 
While the first hypothesis lays the basis for future experiments, the second is tested to understand if it is possible to perform actions which do not require heavy cognitive load while using our system.
\TL{
\subsection{Artificial constant reference \label{SEC:VirtualReference}}
As a second phase, we conducted the analysis of human synchronization with an external haptic stimulation by investigating the dependence of human alignment performance from the sign of the cadence variation. In this experiment, participants were tasked to match their stride frequency with paces both slower and faster than the baseline.
}

\subsection{Artificial leader \label{SEC:ExpArtificialLeader}}
As a further step towards the goal of creating a remote walking system, we had to verify whether humans promptly respond to frequency variations of the haptic cues. 

We designed the simplest situation for measuring to which extent a human is capable of adjusting his gait cadence to match a time-varying reference (\fig\ref{fig:exp1}). Each trial involved only one user who was instructed to follow as close as possible an external rhythm. The stimulus frequency was updated every $30$ seconds while remaining in a specific interval. 
The purpose of this experiment is to determine whether a properly instructed human can adapt with ease to a time-varying cadence. Therefore we measured the discrepancies between the reference and the human cadence.

\subsection{Human leader \label{SEC:ExpHumanLeader}}

The next step toward the remote social walking was replacing the computer program adopted in the \lq Artificial Leader' experiment with an actual human. The purpose of this scientific question was to determine whether a human can follow the gait cadence of another human using the proposed system (\fig\ref{fig:exp2}). This experiment differs from the previous one since the human walking pace is not as regular as an artificially generated rhythm, and can have small unpredictable variations.
It is essential to notice that in this scenario the communication was restricted to one direction only: gait cadence values measured on the leader side were sent over to the follower, while no action was taken upon follower cadence variations. This asymmetry was designed to nullify any synchronization dynamic. 

The human follower was instructed to adapt to the perceived gait cadence, whereas the human leader was allowed to walk at his own pace. The leader received no notification regarding the presence of the follower, thus he/she was fully unaware of whether the follower was feeling his/her steps.
We measured discrepancies between leader's and follower's gait cadences to study the time needed for the follower to adjust his/her pace and the stability level he/she was capable of maintaining.

\subsection{Peer-to-peer \label{SEC:ExpP2P}}

In an in-person social walk, \textit{i.e.} when a group of at least two humans walk together, it is an implicit rule that of adjusting the speed so that the group can clump and stick together. Therefore, in the general case there are no specific roles to be given, such as leader and follower, and each participant would contribute to reach an agreed advancement pace. 

This is the scenario in which the full capability of our system emerges: both participants were sending their respective gait cadence (sensed by their anklet devices) to the server which broadcasted them in real time to the partner's smart-phone that, in turns, applied the vibration frequency (\fig\ref{fig:exp3}). It follows that if the two walkers advanced with different cadences they would have both experienced a misalignment between their steps and the anklets vibrations.

The experimental guidelines in this case play a paramount role as the feedback loop is established and several consensus dynamics are possible: the instructions determine which of those effectively takes place. We decided to instruct the participants to pursue two competitive goals simultaneously and with the same priority: try to adapt to the other's gait cadence but also try to maintain a comfortable walking, as close as possible to one's own natural cadence.

The scientific questions relevant to this experiment are more articulated than the previous cases: while it is still worth studying the transient before stabilizing to a common cadence, it is also interesting to determine the absolute and relative discrepancies between the common and personal cadences. Moreover, comparing the agreed cadence with the average of the two personal ones would be indicative in terms of the consensus dynamics between the participants. 

\input{tablePercentage}

\TL{
\begin{figure}[t]
\centering
\resizebox{0.92\columnwidth}{!}
{
\input{figures/boxplot.tex}
}
\caption{
\TL{Boxplots represent the distributions of alignment percentages for each condition. Labels \textbf{H} and \textbf{HS} refer to data acquired during trials with haptic stimulation and haptic stimulation and secondary task, respectively. The footnote describes the tolerance band used to discriminate aligned strides from misaligned. The percentage is referred to each subject's baseline stride duration.
}}
\label{FIG:boxplot}
\end{figure}
}

\section{Experimental validation \label{SECT:Experimental}}

As previously mentioned in the introduction (\sect\ref{SEC:intro}) we conducted a step-wise validation.
In this Section, we retrace the progression of the experimental process  describing experimental protocol, setup, and results per each step.

All the experiments have been held at the open-air athletics track in Siena. Participants were provided with written informed consent and suggested to wear sport equipment. The experimental campaign was held in 5 non-consecutive days, one per experiment, and subjects could discontinue participation at any time. Some participants took part in multiple experimental sessions. All were healthy and none had neuromuscolar disorders or recent injuries at the time of study. 
It is important to point out that in all the trials involving two participants, they were instructed to walk along different paths, avoiding any visual and audio interaction. The only way of communicating was through haptic stimuli.

\begin{figure*}
\centering
\resizebox{1.9\columnwidth}{!}
{
    \begin{subfigure}{\columnwidth}
		\small
	    \centering
        \input{figures/VR1.tex}
		\caption{}
		\label{FIG:VR1}
    \end{subfigure} 
    \begin{subfigure}{\columnwidth}
		\small
    		\centering
        \input{figures/VR2.tex}
		\caption{}
		\label{FIG:VR1}
    \end{subfigure} %
    }
    \caption{\TL{
Representative epochs for faster and slower cadence  suggestion. In (a) the provided haptic rhythm has shorter period than the participant's baseline walking pace, and the user has to walk faster to match the external frequency. The green line represents the instant in which the haptic stimulation is activated, while the blue band highlights the tolerance band used to assess the alignment. Figure (b) is the symmetrical condition for slower cadence (in fact the stride duration reference increases wrt the baseline).}}
    \label{FIG:trialVirtualReference}
\end{figure*}

\subsection{Preparatory experiments \label{SEC:prepExp}}

In this experiment we evaluated the human capability in adapting the gait cadence to an external constant rhythm. Moreover, we tested whether the addition of a secondary task did affect the cadence synchronization performance.

The representative sample consisted of twenty subjects (age $31.7 \pm 10.6$) with these characteristics: 10 females and 10 males; 6 had previous experiences with haptic interfaces; 2 played music instruments at high level (drums and piano); 12 played sport, two of them in a professional league with regular training sessions. 

Participants were provided with an Android phone and two haptic interfaces. The pressure sensor, connected to the master anklet, was positioned under the right heel to sample pressure data and extract stride durations, which were then transmitted via Bluetooth to the smart-phone and logged by the server. 
In all the trials, participants walked while wearing headphones reproducing white noise to avoid entrainment due to the sound of vibrations and external stimuli.

The first experiment was performed to test the first hypothesis: \textit{\lq \lq can humans synchronize their gait cadence with the rhythm suggested by the anklets?"} For each participant the experimental session started with a preliminary acquisition of self-paced gait along a \SI{200}{\m} path, to record the user's comfortable cadence. In the second trial the haptic interfaces were activated at a frequency $10\%$ faster 
than the previously estimated baseline stride duration. Subjects were instructed to align their step sequence to the vibrations during the \unit[200]{m} walk.


The second hypothesis was tested by adding a secondary task to the experimental protocol described in the first experiment. The purpose of the secondary task was to determine whether the presence of additional mental efforts affected the users' ability to follow the rhythm dictated by the haptics.
The secondary task was selected according to the requirement of low mental efforts, described in \cite{baldi2018human}. 
Subjects were asked to answer simple mathematical questions (double-digit sums and differences) on a smart-phone app and to walk at the same time, giving the same priority to the two tasks.
We adopted the same procedure of the previous experiment: the first trial was meant to estimate the participants' baseline cadence, whereas the second trial was conducted with haptic stimulation enabled and the secondary task. Walking distance was $200$ m in both cases and gait parameters investigated in the data analysis were the same as previous experiment.


\subsubsection{\textbf{Results}}
The analysis of comfortable cadence trials yielded baseline information about the gait parameters of our sample. In average, participants' stride duration was \unit[1138 $\pm$ 36]{ms}, corresponding to a cadence of 52.72 ($\pm$2.19) strides/min. The inter-subject cadence variability, expressed in percentage, represents the 4.2\% of the average value. This result is in line with the results of \cite{menz2003acceleration}. In that work, accelerometer signals recorded during the comfortable cadence walking of 60 subjects were analyzed. The authors reported a mean walking cadence of 53.54 ($\pm$3.87) strides/min, which corresponds to an average stride duration of 1120 $\pm$64 ms. The inter-subject cadence variability was 7.2\%. Anyway, those data were recorded while walking on regular and irregular surface, which would motivate the higher variability.
We then calculated separately the stride duration variability (standard deviation) for each user, which mean value was 1.93\%, to assess the degree of physiological variability of human walking pace. 

In order to discriminate changes in walking parameters due to haptic stimulation, we defined three tolerance bands (2\%, 4\%, and 6\%, corresponding to $\sigma$, $2\sigma$ and $3\sigma$) around the reference stride duration subjects were asked to keep. In fact, we assume that cadence variations in the band reference \lq stride duration $\pm 6\%$' (three times the standard deviation) should be related to physiological variability, and higher misalignment may be due to the user being unable to follow the haptic rhythm.
The tolerance bands were then used to investigate the amount of time, expressed in percentage of the trial duration since the beginning the synchronization with the haptic stimulation, in which subjects were able to follow the external pace given the acceptable error. This quantity is referred to as \lq \lq alignment percentage".

\begin{figure*}
\centering
\begin{subfigure}{0.45\textwidth}
\centering
\null \hfill
\resizebox{0.8\columnwidth}{!}
{
\input{figures/boxplot_time_2_review.tex}
}
\caption{\label{FIG:boxplot_time2}}
\end{subfigure}
\hfill
\begin{subfigure}{0.45\textwidth}
\centering
\resizebox{0.8\columnwidth}{!}
{
\input{figures/boxplot_percent_review.tex}
}
\caption{\label{FIG:boxplot_percent}}
\end{subfigure}
\hfill \null
\caption{\TL{The left boxplot (a) represents the distribution of times required by participants to align to the suggested pace, for $10\%$ faster and slower cadence wrt to the baseline, respectively. The right boxplot (b) reports the distribution of trial time percentage (after the synchronization) during which participants' stride duration differed less than $4\%$ from the suggested pace. In both cases, performance data show no significant difference due to increasing or decreasing stride duration.}}
\end{figure*}

\TL{
In Table \ref{TABLE:synchronization_intervals} we report the synchronization rate for each subject, calculated in the three tolerance bands. The median synchronization rates in the first experiment were 78.84\%, 99.28\%, and 100.00\% for the 2\%, 4\% and 6\% tolerances, respectively. The introduction of the secondary task lowered the overall synchronization rate: 47.50\%, 85.92\%, and 99.80\% were the median values extracted.
We assessed through the  Shapiro-Wilk's test that the data were not normally distributed, so we visually depicted data by means of box-plots in \fig\ref{FIG:boxplot}, and numerically using quartiles (reported in \tabb\ref{TABLE:synchronization_intervals}). 
}

To assess if the effect of vibrations was relevant, we compared data obtained during comfortable gait and haptic stimulation trials. A paired-samples t-test revealed a statistically significant mean difference in the stride durations recorded in the two conditions ($p=0.015$). 
No outlier was detected. For both conditions, the assumption of normality was not violated, as assessed by Shapiro-Wilk's test ($p=0.195$). 

The same procedure was applied for the analysis of gait data recorded during the second experiment. Shapiro-Wilk's test confirmed the normal distribution of mean differences in stride duration per each subjects ($p=0.583$), and the paired-sample t-test assessed that participants modified their stride duration also in presence of additional cognitive load ($p=0.04$).

\input{table_constant_rhythm}
\subsubsection{\textbf{Discussion}}

Experimental results confirmed that humans can synchronize their step sequence to an external, constant rhythm provided through vibrotactile cues, with an error comparable to the natural cadence variability. 
Thus we can assume that it is possible to influence the participants' average cadence by asking them to voluntarily align to the provided rhythm.

The increase of cognitive load due to the secondary task did not have a relevant effect on the synchronization, which was achieved for most of the time by all the users, although the variability increased. Only one user could not successfully adapt to the suggested rhythm. 

This experiment paved the way and defined some evaluation criteria for the other trials. To the best of our knowledge, literature lacks a clear and unanimous way of evaluating the human cadence synchronization with an external stimulus, therefore a straightforward choice was to use results of this experiment as a metric. The users' average stride duration variability during comfortable walking, in a regular surface without disturbance, is about 2\%.  
Thus, fluctuation around the mean value in the interval of $\pm 4\%$ ($2\sigma$) could be considered an appropriate interval for including the majority of the strides walked in a correct tempo. In the following experiments, 4\% was used as reference to discriminate strides aligned and non aligned with the reference stride duration.

\TL{
\subsection{Artificial constant reference}\label{SEC:virtual_Ref}
\
Since the preparatory experiment (\sect\ref{SEC:prepExp}) only investigated the faster cadence condition, we enrolled 10 new participants (age $28.3 \pm 4.3$, 7 males) to collect data on symmetrical pace variation. 
We replicated the setup of the previous case: participants were provided with the hardware, then their baseline cadence was acquired in a \SI{100}{\m} self-paced walk. 
Each subject was asked to perform two trials, during which they had to voluntarily synchronize their strides to a reference rhythm, $10\%$ slower and faster than the baseline (the order was randomized). 
Trials were divided in 100 m of self-pace walking and 200 m of haptic-assisted walking. Stride times measured by the master anklet were logged by the system, and then compared with the reference to extract time to reach synchronization and alignment percentage (\cfr\fig\ref{FIG:trialVirtualReference}).
In this experimental campaign we did not take into consideration the disturbance due to the secondary task, since we were interested only in the effects of the slower and faster external cadence.
\subsubsection{\textbf{Results}}
Stride duration data were processed to evaluate the time necessary to achieve synchronization with the haptic stimuli frequency, and to determine the deviation from the reference after the initial synchronization. The latter parameter was represented as the percentage of trial time during which the user's cadence drifted from the suggested cadence less than $4\%$.

Shapiro-Wilk's test assessed that the distribution of times required by subjects to synchronize with the external cadence was normally distributed both for the fast ($p = 0.09$) and slow cadence ($p = 0.78$) conditions, while alignment percentages were not normally distributed in both cases ($p = 0.01$ for fast cadence, $p = 0.01$ for slow cadence).

Mean time required to match the external stride duration were $2.44 \pm 1.63 s$ for fast condition and $2.31 \pm 1.05 s$ for slow condition, while median alignment percentages were $98.8\%$ and $99.0\%$ respectively. Boxplots in Figs. \ref{FIG:boxplot_time2} and \ref{FIG:boxplot_percent} visually describe data, that are also listed in \tabb\ref{tab:constant}.

Paired t-test conducted on time to alignment values revealed no statistical difference between the two distributions ($t(9) = 0.14$, $p = 0.89$). No statistical test was conducted on alignment percentages because visual representation showed very small difference between the two distributions.

\input{table_botFollower}

\subsubsection{\textbf{Discussions}}
This experiment was aimed to assess performance asymmetries during faster and slower pace conditions. Experimental results suggest that participants managed to tune their walking pace to the external rhythm for a large portion of the trial duration, regardless of the sign of the cadence variation. The statistical analysis of time to reach synchronization also did not evidence significant differences between conditions.
For these reasons, we expect that participants abilities in synchronizing is not asymmetrically biased by the sign of the cadence variation.
Forthcoming experimental procedures investigate in detail human acceptance of fast varying gait rhythms, thus broadening the study on the participants' proficiency in aligning to faster and slower paces.
}

\begin{figure}[t]
\centering
\small
\resizebox{0.95\columnwidth}{!}
{
\input{figures/botFollower2.tex}
}
\caption{Artificial Leader representative trial. The participant was tasked to align his cadence with the one proposed via the wearable haptic devices. The user started the trial walking at his most comfortable gait. After $200$ meters (about $130$ seconds) the haptics were automatically turned on and the user was able to feel the vibrotactile stimulation. The stimulation continued for $400$ meters, than the interfaces were turned off. The participant walked for additional $200$ meters at his most comfortable cadence. Green lines identify the time-points in which haptics were turned on and off. The user's cadence is depicted with a red line, whereas the blue line and the surrounding violet area represent the suggested rhythm and the $\pm 4\%$ interval, respectively.
}
\label{FIG:trialArtificialRef}
\end{figure}

\subsection{Artificial leader}\label{SEC:Artificial}
In this experiment we examined the human capability in entraining to a time-varying rhythm generated by an algorithm. This is a common \textit{modus operandi} in training and rehabilitation bouts, sportspeople and patients have to follow an external pace with time-varying frequency in order to improve (or recover) physical abilities. 
We named this methodology {leader-follower}, borrowing the idea from robotics, because the follower is asked to align his step sequence to the haptic rhythm displaying the leader's walking cadence (in this case simulated).
Twenty participants (age 29.8 $\pm$ 5.3, 14 males) have been recruited for this phase.  

The experimental setup was composed by two anklets, one of which equipped with the force sensor for recording the stride sequence, headphones reproducing white noise, and a smart-phone with the \textit{ad-hoc} application. 
Each subject performed a single trial composed of three phases. In the first phase the participant was instructed to walk at his/hers comfortable pace for \SI{200}{\m}, to record baseline cadence. In the second phase the user was asked to align to the pace provided through haptic stimuli, for \SI{400}{\m}. In the last phase, vibrations were turned off and the subject continued walking for \unit[200]{m} at his comfortable pace. A dedicated piece of software simulated the leader's cadence updates that, though the server, instructed the app to vary the vibration frequency. 
A new reference stride duration was randomly selected every $30$ seconds in the interval \unit[900-1100]{ms} (average stride duration is \unit[1]{s}). 

\input{table_leaderFollower}

We selected a \unit[30]{s} update time to analyze the stability of gait cadence after each variation and the number of strides necessary to adapt to the new stride frequency. The second phase of the experiment in average lasted 4 minutes, resulting in at least 7 cadence variations. 
%
A representative trial is reported in \fig\ref{FIG:trialArtificialRef}.

From each trial we examined: \textit{i)} follower's comfortable cadence before vibrational cueing, \textit{ii)} strides needed by the follower to align his gait with the proposed cadence 
(considering a $4$\% tolerance), \textit{iii)} percentage of time the follower is aligned with the suggested gait.

\subsubsection{\textbf{Results}}

The primary aim of this experiment was assessing whether humans could align their cadence to a time-varying frequency. We calculated for each subject the percentage of time in which stride duration was in the range {reference stride duration} $\pm 4\%$ during the phase with haptic cues.
All followers were able to align to the leaders' rhythm for more than $94\%$ of the trial time. 
Data of the trials are reported in \tabb\ref{tab:botFollower}.

The average number of strides necessary to adapt to the new cadence is $2.2 \pm 1.2$. In particular, an asymmetry was observed between increasing and decreasing stride duration: the number of strides necessary to achieve a misalignment lesser than $4\%$ was $1.1 \pm 0.7$ and $3.1\pm 1.9$ strides for slower and higher frequency variations, respectively.
For variations of the reference stride duration below 4\% in most cases there was no transient in aligning to the new cadence.

\subsubsection{\textbf{Discussion}}
Outcomes of the test revealed that participants could  easily adapt to cadence variations, especially if the difference was small. In fact, considering the human temporal resolution and physiologic variability of gait, the user may not even notice small variations (in the order of \unit[20]{ms}).
These results allow to study the synchronization of human cadence with external rhythms which vary fast, but with limited oscillations, as in the case of human gait.
For variations greater than $4$\%, results show that users react quickly to cadence increase (\iec smaller stride duration), probably by making smaller steps to restore the synchronization with the external rhythm, whereas it seems more difficult to rapidly reduce the pace (\iec increase the stride duration).
The last phase of recording without haptics is not studied quantitatively; we plot it to demonstrate the effect of the haptic stimulation. In fact, after the vibrations were turned off, the self-selected stride duration was restored to the baseline value.

\begin{figure}
\centering
\small
\resizebox{0.95\columnwidth}{!}
{
\input{figures/leaderFollowe.tex}
}
\caption{Human Leader representative trial. The follower was tasked to align the walking cadence with the leader's one, transmitted via the wearable haptic devices. The users started the trial walking at their  comfortable pace. After $200$ meters (about $120$ seconds) the follower's haptics were automatically turned on to enable the vibrotactile stimulation, while the leader continued walking at self-selected pace. After $200$ meters, the follower's anklets were turned off and the participants walked for additional 200 meters without haptic suggestions. Green lines identify the time follower-side haptics were turned on and off. The user cadence is depicted with a red line, whereas the blue line and the surrounding violet area represent the suggested rhythm and the $\pm 4\%$ area, respectively.
}
\label{FIG:trialHumanHman}
\end{figure}

\subsection{Human leader}

The results obtained in the previous experiment encouraged the assumption that humans can adapt with ease their walking cadence to time-varying rhythms if the variability is limited (assuming the human cadence physiological variability as boundaries). 
In this experimental session the follower is provided with haptic stimuli replicating the human leader's cadence. We stress that in this experiment the leader could not feel the follower by any means.

Twenty subjects (age $27.9 \pm 6.1$, $12$ males) took part in this phase. The experimental setup for each participant was composed by two haptic interfaces, one of which equipped with the force sensor for recording the stride sequence, headphones reproducing white noise, a smart-phone with the \textit{ad-hoc} app. 
The $20$ participants, randomly labeled from U1 to U20 for convenience, were arranged in couples.
Each couple performed one trial, the role of leader and follower was selected randomly at the beginning of the trial. In the first phase of the trial, both participants were asked to walk at their comfortable cadence for \unit[200]{m}, to record gate parameters in the baseline condition. In the second phase the follower received haptic stimuli replicating the leader's gate cadence, to which he has been instructed to adapt. The leader was not notified about the beginning of the second phase, and continued walking at his own pace. The anklets were used by the leader exclusively to record the strides duration, while vibro-motors never activated.  After \unit[200]{m}, haptics were turned off and the last phase began, during which subjects were instructed to walk for \unit[200]{m} at their comfortable cadence.
A representative trial is reported in \fig\ref{FIG:trialHumanHman}.

From each trial we estimated: $i)$ comfortable cadences before haptic cueing, $ii)$ time needed by the follower to align to the leader's cadence (calculated from the initial activation of the haptic devices to the reaching of the desired cadence, considering the $4\%$ error bound), $iii)$ percentage of time follower is following the leader tempo (defined as the follower's cadence $\pm$ 4\%).

\subsubsection{\textbf{Results}}
Experimental results (detailed in \tabb\ref{tab:LeaderFollower}) show that all the followers succeeded in aligning their walking cadence to the leader's for more than $90\%$ of the time, assuming an acceptable oscillation of $4\%$ around the reference pace. 
The average time required to align with the leader was \mbox{3.31 $\pm$ 0.69} seconds.

\subsubsection{\textbf{Discussion}}
In this experimental session we evaluated the human capability in adapting the walking cadence to a fast-varying pace displayed through vibrations. Experimental results show that cadence oscillations due to natural variability do not impede the entraining with haptic cues. 

These results open a wide range of applications in which a leader guides one or more followers, as in training and rehabilitation. A more comprehensive discussion on possible future research directions is reported in \ref{subsec:future}.
Moreover, these results provide the last prerequisites for hypothesizing and testing the mutual cadence alignment, \ie remote social walking, referred in the following as \lq peer-to-peer'.

\input{table_masterMaster}

\subsection{Peer-to-peer \label{SECT:peerTopeer}}

This experimental session represents the last piece of the remote social walking step-wise validation. Once the capability in following an external  
rhythm was assessed, we tested bilateral transmission of cadence through vibrotactile interfaces to connect two people walking far from each other.
Our aim is testing if the system we developed can be successfully used to achieve the cadence synchronization between two users without direct interaction. 
An assumption we had to make was asking participants to voluntarily align to the partner's cadence, but still keeping a step frequency close to their comfortable one. In fact, the group walking (or social walking) condition is replicated if the participants agree on a common pace comfortable for everyone. As a consequence, in this experiment the users did not receive strict guidelines, they had to `negotiate' with the partner.
Although the psychological aspect plays a relevant role in the achievement of the consensus, it will be studied in a future work. In fact, before studying how people agree on a common rhythm, we need to validate the proposed system and assess if and how the cadence alignment takes place.
Thus, in this work we study temporal gait parameters to investigate the system features and capabilities.

Twenty participants were enrolled for the experimental session (age $28.1 \pm 5.4$, 8 males), randomly labeled from U1 to U20, and arranged in couples. All the users took part in a previous experiment, at least.
The experimental setup for each subject was composed by two haptic interfaces, one of which equipped with the force sensor for recording the stride sequence, headphones reproducing white noise, a smart-phone with the \textit{ad-hoc} app.

Trials were composed by three phases: users were asked to start from predefined positions and walk at their comfortable pace along different paths for \unit[200]{m}. In the second phase haptic stimuli representing the partner's cadence were delivered to each participant, who was instructed to adapt to the received rhythm and, simultaneously, try to pull the partner toward his own pace. After both participants walked \unit[200]{m}, the haptic stimulation was turned off and the users walked at their own cadence for 200 meters.
A representative trial is reported in \fig\ref{FIG:peerToPeer}.

In order to give a quantitative evaluation of the effectiveness of our system, here we defined the concept of {cadence alignment}: a user's stride is aligned with the partner's if the duration of the current stride is in the interval {partner's last stride duration} $\pm 4\%$. 
In this last experiment we analyzed: $i)$ comfortable cadences, $ii)$ variation of the average cadence for each user during the haptic stimulation phase (with respect to the comfortable cadence), $iii)$ time needed for reaching the alignment, $iv)$ percentage of time users strides were aligned.

\subsubsection{\textbf{Results}}
All users, with exception of two, varied their average walking cadence during the phase with haptic stimulation, as visible in \tabb\ref{tab:peerToPeer}. The average stride duration variation with respect to the comfortable pace was $4.0\%$.
In all but two cases the participants agreed on a common cadence which was intermediate between the two comfortable cadences. 
The average time to reach the cadence alignment was $4.92 \pm 2.91$ seconds.  
After the beginning of the alignment, on average, the participants maintained a similar gait frequency (in the limits of $4\%$) for the $94.5 \pm 4.1\%$ of the time.

\begin{figure}[t]
\centering
\small
\resizebox{0.95\columnwidth}{!}
{
\input{figures/peerToPeer.tex}
}
\caption{{Peer-to-peer} representative trial. The participants were tasked to tune their own gait cadence with the partner's rhythm, displayed by the anklets. The users started the trial waking at their comfortable cadence. After $200$ meters (about $120$ seconds) the haptics were activated and both users were able to feel the partner's cadence for $200$ meters. 
Then the interfaces were turned off again and the participants walked for additional $200$ meters at their comfortable cadence. Green lines identify the time-points in which haptics were turned on and off. The users' cadences are depicted respectively with red and blue lines, the surrounding violet area represents the $\pm 4\%$ area of the mean computed at each timestamp.
}
\label{FIG:peerToPeer}
\end{figure}

\subsubsection{\textbf{Discussion}}
As shown in \tabb\ref{tab:peerToPeer}, the effect of haptic stimulation is evident and results demonstrate the effectiveness of the system. All the participants changed their walking pace according to the partners' stride duration after the stimulation was activated, and immediately moved back to their comfortable cadence after the haptic phase.
It is worth pointing out that the time for aligning the cadence to the partner's is higher than the ones observed in the Leader-Follower case. This is probably a consequence of the fact the users try to follow the partners' cadence, resulting in a transient during which the users' stride durations oscillate.
In addition, for the majority of the trial, we observed a greater oscillation at the beginning of the haptic cueing, followed by a constant reduction. This is characteristic for a system with an inertia following a reference. 
Although the study of psychological aspects is not in the focus of this work, we can make two considerations:
\begin{itemize}
	\item [$i)$] in most of the trials the participants aligned their gait cadence on a common rhythm which was close to the mean value of the comfortable cadence of the two users;
	\item [$ii)$] in two trials participants achieved the consensus, but they aligned on a cadence which was close to the comfortable cadence of one of the participants. A possible explanation is that one user may not have fully understood the experimental protocol.
\end{itemize}

\TL{
\section{Human behaviors and  performance correlation analysis}\label{SECT:CORR}

The presented experimental validation and the following results discussion can be enriched by analyzing the correlation between the participants' self-pace at baseline and their ability to align the walking cadence to the reference pace.
In particular, we searched for possible relationships between subjects comfortable cadence, suggested rhythm, and success in synchronizing. 
We started by evaluating scenarios with artificial rhythms, both in case of constant (\sect\ref{SEC:virtual_Ref}) and variable reference cadence (\sect\ref{SEC:Artificial}).
In the former scenario, we tested the presence of a relationship between the suggested pace and alignment performance. 
The difference between stride duration suggested and participants' baseline cadence was correlated with the time required to synchronize with the external stimuli, and with the percentage of task time during which the participants' stride duration was comparable with the haptic stimulation period. Since synchronization percentages values were not normally distributed, we resorted to Spearman's correlation tests, while Pearson's test was used for time to alignment. The tests revealed no significant relationship of baseline both with time ($p = 0.22$) and alignment percentage ($p = 0.86$).
For what concerns the \lq Artificial Leader' data, the considered values were not normally distributed so Spearman's rank-order correlation tests were run to assess the relationship between baseline and performance. 
Results of the tests show that there was no statistically significant correlation between comfortable stride duration and percentage ($p = 0.117$). 
Similarly, there was no statistically significant correlation between baseline and time for alignment ($p = 0.794$). 

This result is not surprising, because the displayed stride duration was not constant and was updated every \SI{30}{\second} in the range going from $900$ to $1100$ \si{\milli\second}/stride. In fact, this experiment was aimed to assess the behaviour of participants when facing cadences varying in a wide range. The fact that all participants managed to align to the external rhythm with no dependence on the baseline gait parameters may prove that, as long as the suggested cadence is selected inside a feasible range, the human can successfully adapt his own walking pace. The time to achieve the alignment instead was calculated as the sum of the synchronization time after each cadence variation, thus it depends also from the randomness factor.

Then we took into consideration the social aspect and the users' response in following a partner. Outcomes from the experiments described in \sect\ref{SEC:ExpHumanLeader} and \sect\ref{SEC:ExpP2P} were analyzed.
For what concern the \lq Human leader' scenario, we evaluated the relationship between the time to reach the synchronization, the percentage of the trial in which the leader was aligned with the master, and the difference between leader's and follower's stride duration. 
All the considered  variables were normally distributed, as assessed by Shapiro-Wilk's test (difference $p = 0.391$, alignment percentage $ p =0.164$, and time to alignment $p = 0.965$).
The Pearson's product-moment correlation revealed no significant correlation between walking pace difference and alignment percentage ($p = 0.128$). 
On the contrary, Pearson's product-moment correlation between initial stride duration difference and time to alignment was statistically significant ($r = 0.714$, $p = 0.02$).

As in the \lq Artificial Leader' experiment, the lack of relationship between baseline cadence and alignment performance may imply that the self-selected pace does not affect the synchronization percentage. On the other hand, the significant correlation of baseline pace with the alignment time may be due to the fact that accommodating to a farther rhythm takes longer. This aspect is interesting on the perspective of defining effective strategies to facilitate the alignment between two or more participants: instead of providing the raw partner's cadence, it may be smoothed to avoid oscillations during transient.

Similar results were collected for data in the \lq peer-to-peer' experiment (\sect\ref{SEC:ExpP2P}).
The same metrics were exploited to evaluate the correlation between users' pace and performance. 
Shapiro-wilk's test assessed normality distribution for stride duration difference ($ p = 0.255$), alignment percentage ($ p = 0.546 $), and time to achieve alignment ($ p = 0.249 $).
Pearson's product-moment correlation revealed no statistically significant relationship between the difference in initial gait cadence and alignment percentage ($ p = 0.081$), neither between cadence deviance and time to reach the common stride duration ($p = 0.263$).
While the former result is in line with the one obtained from \lq Human Leader' data, the latter is in contrast. Further experiments are required to address this matter, but we hypothesize that the two participants' efforts in aligning their cadence may generate non-linear dynamics.
}

\section{Qualitative results and users' feedback\label{SECT:qualitative}}
Similarly to \cite{sallnas2004effect}, at the end of the \lq peer-to-peer' experiment participants were asked to fill a questionnaire comprising four multiple-choice and one open question about personal impressions and suggestions.
The aim of the questionnaire was investigating the effectiveness of the system with a qualitative approach.

It is worth pointing out that all the subjects involved in the survey participated in at least two experiments. 
The first question was about the spontaneity in aligning to the external rhythm provided by the haptics. 
With the second question we evaluated the ease of use of the vibro-tactile anklets.
The following topic under investigation was the social side of the proposed work: we asked subjects opinion on the system transparency, \iec whether the stimulation resembles a human walking cadence. Finally, we evaluated the impressions of walking with a remote companion.
 
The list of questions is reported in the following.

\vspace{1mm}

\noindent 
\begin{tabular}{cm{.41\textwidth}}
\hline
\textbf{\textit{Q1:}} & \textit{Did aligning to the vibrations come naturally to you?} \\
\textbf{\textit{Q2:}} & \textit{Could you align with ease to the rhythm received?}\\
\textbf{\textit{Q3:}} & \textit{Do you think that the vibrations you received could be associated to a human walking cadence?}\\ 
\textbf{\textit{Q4:}} & \textit{Did you perceive your partner's telepresence?}\\
\hline
\end{tabular}
\vspace{1mm}

Answers were entered on a Likert scale with range 1-7, where 1 represented \lq Strongly Disagree', and 7 \lq Fully Agree'. 

\subsubsection{\textbf{Results}}
\begin{figure}
\small
\resizebox{0.95\columnwidth}{!}
{
\includegraphics[clip,width=\columnwidth, trim={0 0.5cm 0 0}]{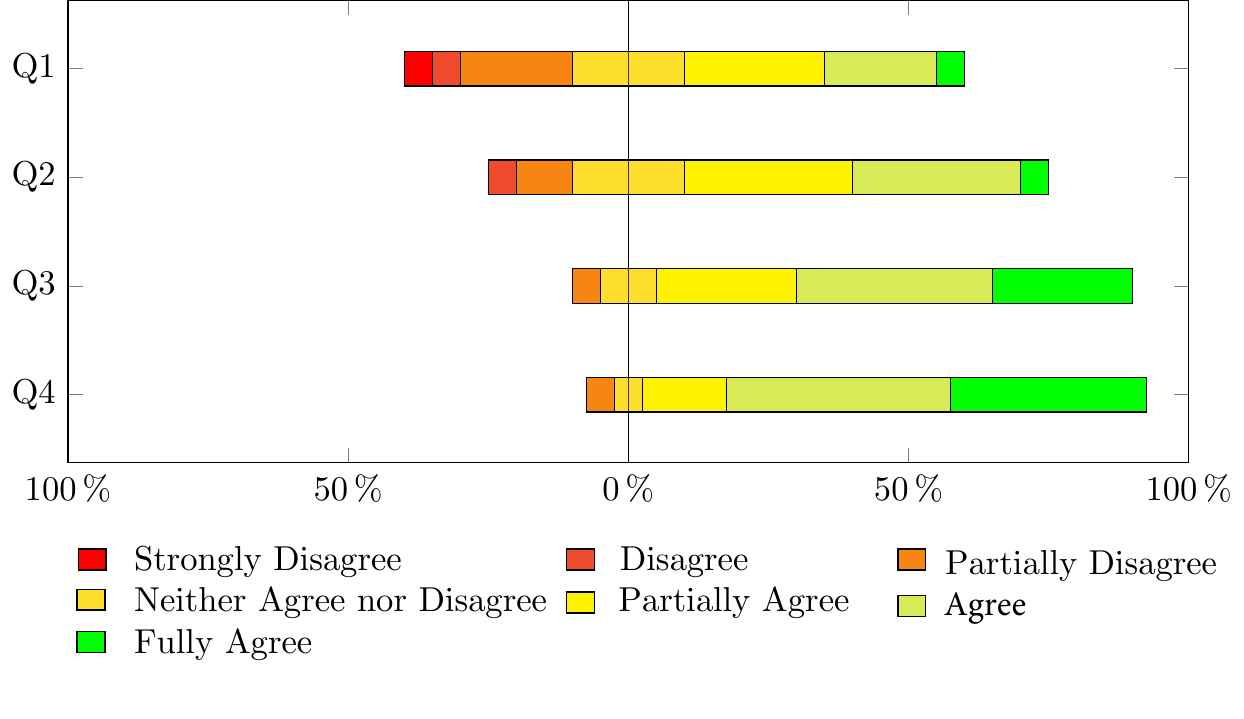}
}
\caption{Likert scale data for the proposed questionnaire. For each question the percentage of answers is reported. \label{fig:questionnarieResult}}
\end{figure}

Answers to the questionnaires are reported in what follows in terms of mean $\pm$ standard deviation.
The subjects' average ratings were 4.3 $\pm$ 1.5, 4.8 $\pm$ 1.3, 5.7$\pm$ 1.1, and 5.9 $\pm$ 1.1, respectively for questions Q1, Q2, Q3, and Q4.
A graphical representation of the users' answers is reported in \fig\ref{fig:questionnarieResult}. 

\subsubsection{\textbf{Discussion}}
The analysis of the multiple choice questions confirms that, for almost all participants, our system is an effective mean to transmit walking cadence.
Vibrational cues are generally perceived as an easy and intuitive way to \lq feel' the presence of the remote companion.
Although most of the users felt the system mechanism to convey the gait cadence as natural, some did not agree on its intuitiveness.
The cadence alignment takes place and the great majority of the users are motivated to adapt to it, even though roughly one out of four found it hard to achieve synchronization.

Moreover, the answers to the open questions revealed that not only the emotive aspect incentives the alignment, but also that synchronizing to the vibrations is satisfying. 

\TL{As a conclusive assessment we evaluated the possible correlation between the users' performance and the correspondent questionnaire responses.
A global score was calculated for each participant as the sum of the four questionnaire ratings.  For what concerns the variation of the walking rhythm during the \lq peer-to-peer' experiment (see \sect\ref{SECT:peerTopeer} and \tabb\ref{tab:peerToPeer}), we did not select the users' cadence variation from the baseline value, because it does not consider the partners' behaviour during the experiments.
Instead, we used the difference between user's and partner's cadence variation during the task. 
For instance,  the couple U7-U8 (\tabb\ref{tab:peerToPeer}) has an average variation of 4\%, but U7 modified his cadence by 7\%, while the U8's change was only 1\%. On the contrary, U5 and U6  average cadence modification was 4.5\%, with a slight difference among them (1\%). In those cases, cadence variations were $+6\%$ and $-6\%$ respectively for U7 and U8 (obtained as $variation_{U7} - variation_{U8}$ and $variation_{U8} - variation_{U7}$), and $-1\%$ and $+1\%$ for U5 and U6.

Firstly, we assessed through the Shapiro-Wilk's test the normality of data.
While the users' ratings and the percentage variations of the user' gait were normally distributed ($p = 0.358$ and $p = 0.977$), alignment percentage failed Shapiro-Wilk's test ($p = 0.037$).
Pearson's product-moment correlation was run to assess the relationship between questionnaire rates and user's cadence variation. There was a statistically significant positive correlation between percentage variation in modifying the walking cadence and answers in the survey ($r = 0.701$, $ p= 0.01 $). 
The Spearman's test between trial aligment percentage and questionnaire ratings revealed no statistically significant correlation ($p = 0.204$). 

Users' rating are not linked with the task performance (\iec walking and reaching a common rhythm), as already suggested by correlational tests in the previous section.
On the other hand, the correlation coefficient expresses a strong relationship ($r^2 = 0.49$) between questionnaire ratings and participants' relative cadence variations after the synchronization with partners. Although correlation does not imply causality, we hypothesize that participants who did accomodate to the partners' rhythm succesfully had rated their experience as positive, while users who felt uncomfortable with the haptic stimulation mainly expressed low scores.
This assumption lays the basis for the next projects, were participants' behaviour will be investigated as a factor to achieve cadence alignment with multiple partners. Moreover, we need to test whether training affects the users' acceptance of our system. 
}

\section{Conclusions \label{SECT:conclsion}}


\subsection{Summary}
In this paper a system for social remote walking was presented and incrementally tested in each of the aspects comprising its global functioning. After designing the technological parts (hardware and software), and performing engineering testing, a first experimental session confirmed that humans can follow a time varying artificial rhythm perceived via anklet vibrations. 
We then assessed that the tracking performances are retained when the virtual reference is replaced with a human gait cadence with a dedicated set of experiments. Finally, we obtained experimental evidence that two humans, walking simultaneously but not in each other proximity, can synchronize their gait cadence when perceiving the companion's walking rhythm using our system.

\subsection{Future research directions overview} \label{subsec:future}
The presented results pave the way for numerous interesting research directions that will be the subject for future works. We briefly list the most attracting.
 
The paper mainly focused on presenting the haptic system and testing its effectiveness in allowing mutual gait cadence influence in humans. Following that confirmation, we are ready to extend results to more various population including older adults.

Even if this work focused on its social aspect, the presented system may also be used by a single walker to have a personal stimulus and track a cadence profile. Such profile may come by a previous personal run, or by a friend's one; additionally it can also be prepared by a personal trainer. In a similar fashion, rehabilitation scenarios can be designed so that patients can exercise under supervision.

In presence of relevant differences in height or training condition, the synchronization may be difficult to achieve. In this case, it would be wise to investigate whether a scale factor would help to agree a common, even if different, pace cadence while retaining the feeling of `walking together'. Note that gait using a scale factor is not feasible while walking side by side.

Our study can be extended to a group of more than two humans. Game Theory provides numerous models that could potentially be suitable for the interpretation of the occurring group dynamics. Among the relevant indexes the synchronization and consensus of gait cadence are the most attracting.

One may also investigate different strategies to display information though vibrations, or new algorithms to facilitate synchronization (for two or more users) tailored on scenarios.

Finally, we believe that our results on remote social walking can be extended to jogging and running.

\input{appendixNike}

\bibliographystyle{IEEEtran}
\bibliography{conference,IEEEabrv,biblioAll}

\end{document}

%% file: figures/setup.tex
\begin{figure}[t]
  \centering
\begin{subfigure}{.6\columnwidth}
  \centering
\includegraphics[width=0.9\columnwidth]{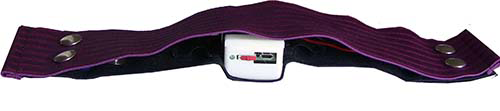}\\
\includegraphics[width=0.9\columnwidth]{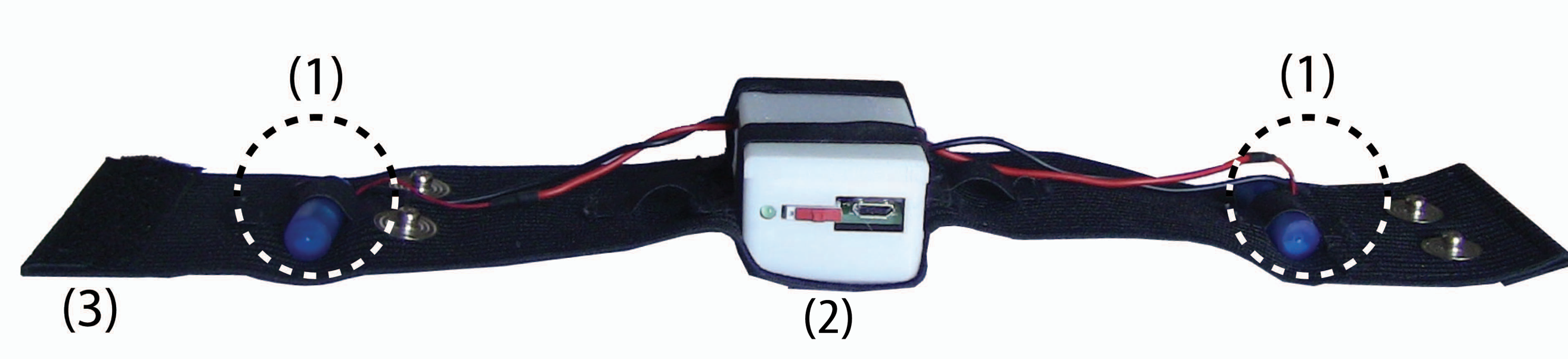}
\end{subfigure}%
\begin{subfigure}{.4\columnwidth}
  \centering
\includegraphics[width=0.9\columnwidth]{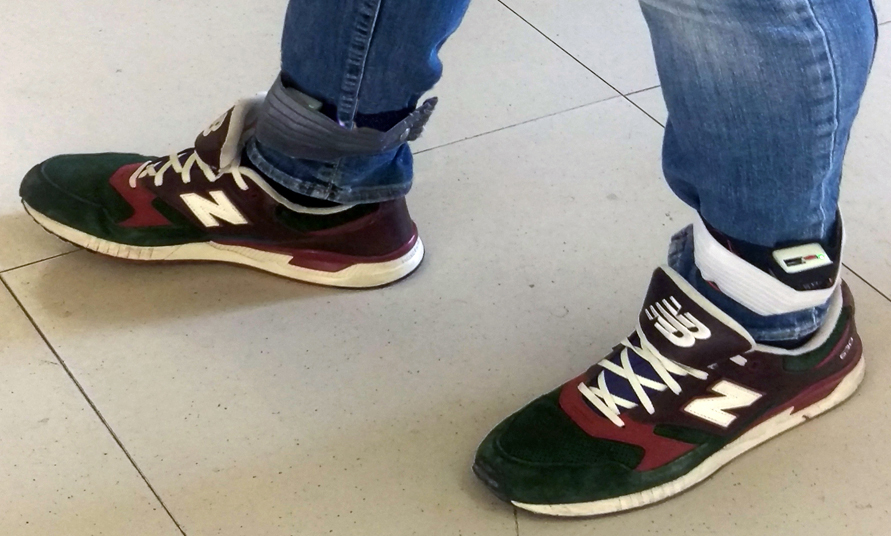}
\end{subfigure}
\caption{Haptic cues are provided to the users via two vibrotactile interfaces placed on the ankles. The interfaces are composed of two vibrating motors (1) attached to an elastic band (3). A Li-Ion battery and an Arduino board are in~(2). \label{FIG:bracelet2Motors}
}
\label{fig:anklets}
\end{figure}

%% file: figures/representativeExperiments.tex
\begin{figure*}
    \centering
    
    \begin{subfigure}{0.32\textwidth}
    \centering
        \includegraphics[width=0.95\textwidth]{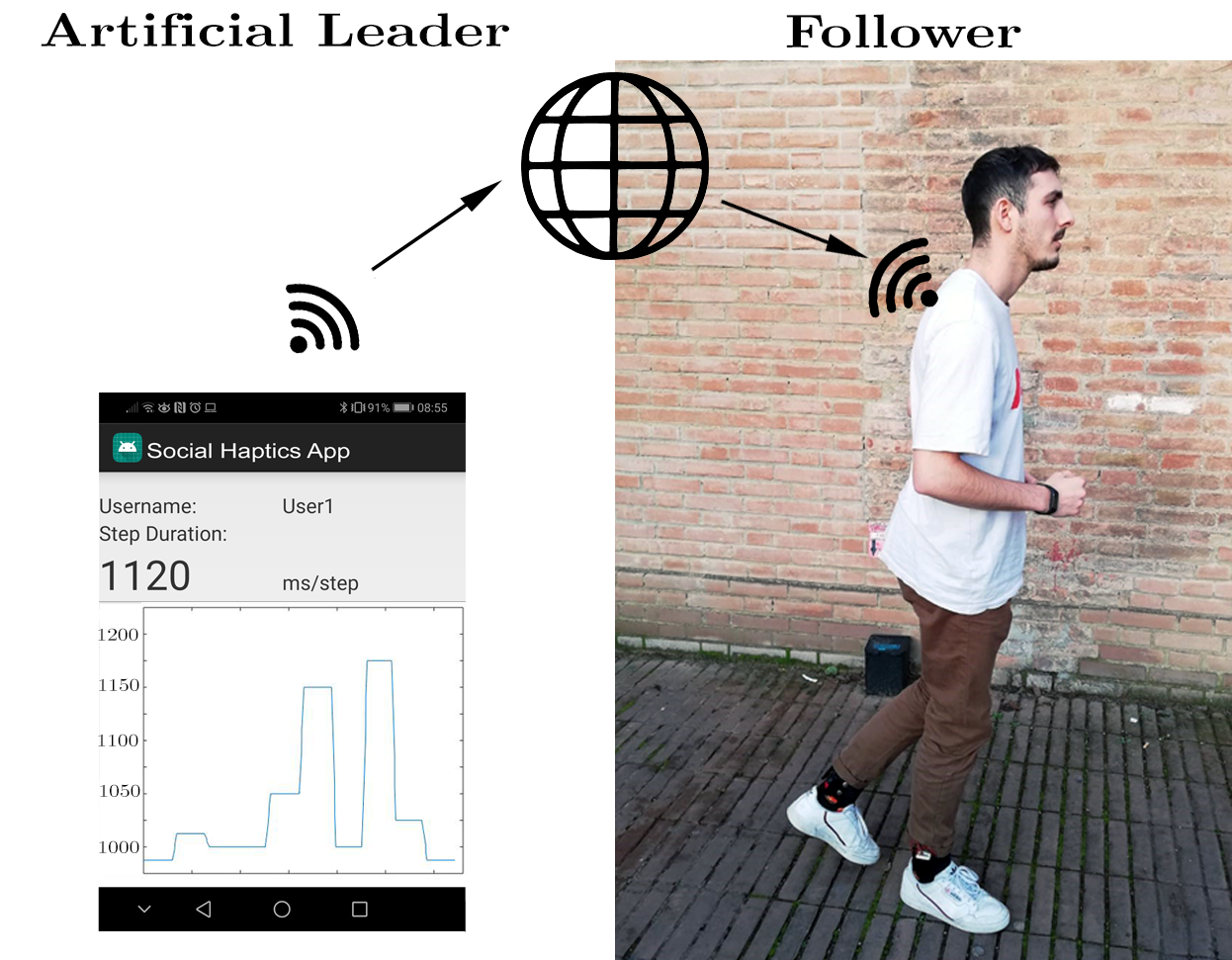} 
        \caption{Artificial Leader}
        \label{fig:exp1}
    \end{subfigure} %
    \begin{subfigure}{0.32\textwidth}
    \centering
        \includegraphics[width=0.95\textwidth]{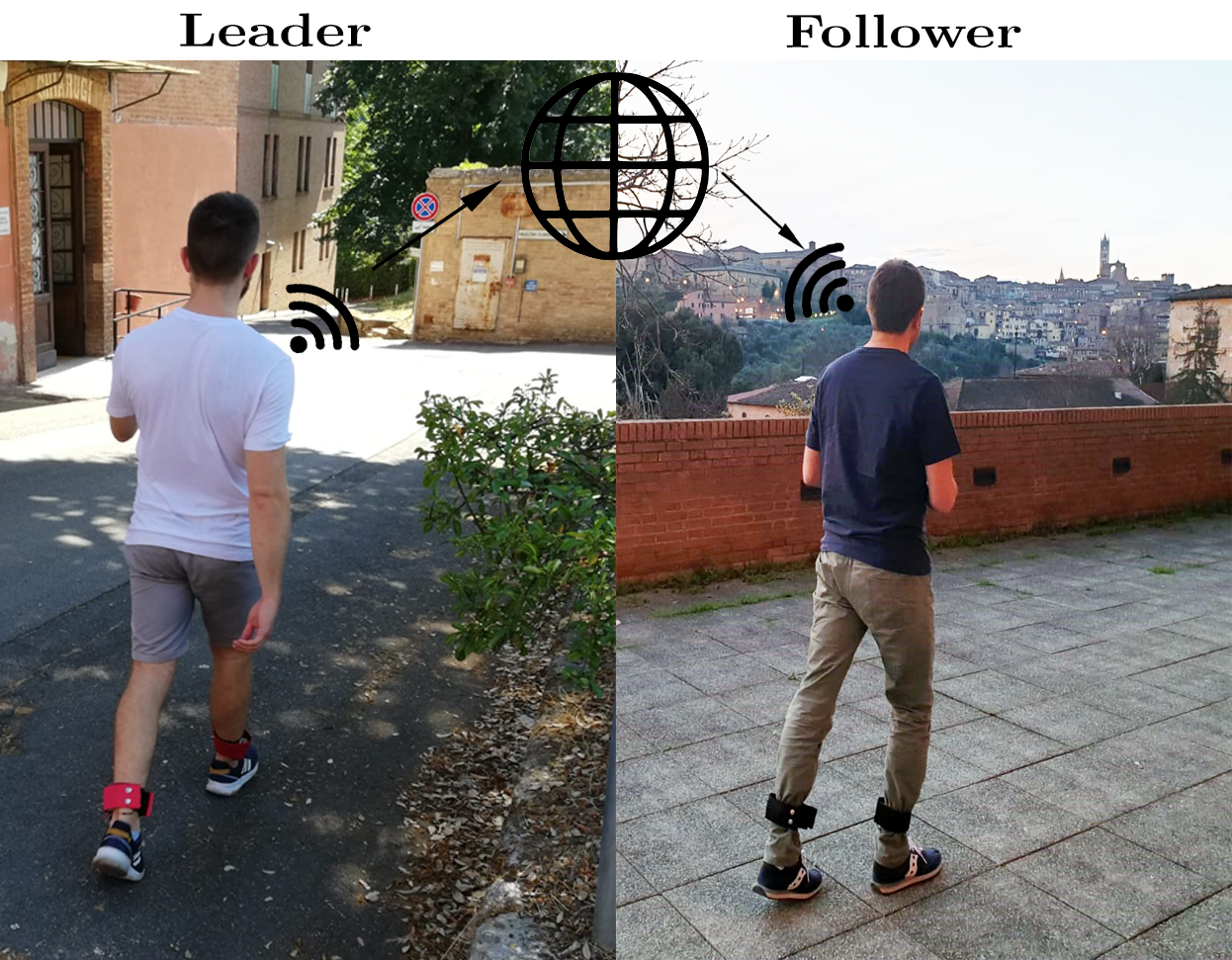} 
        \caption{Human Leader}
        \label{fig:exp2}
    \end{subfigure} %
    \begin{subfigure}{0.32\textwidth}
    \centering
        \includegraphics[width=0.95\textwidth]{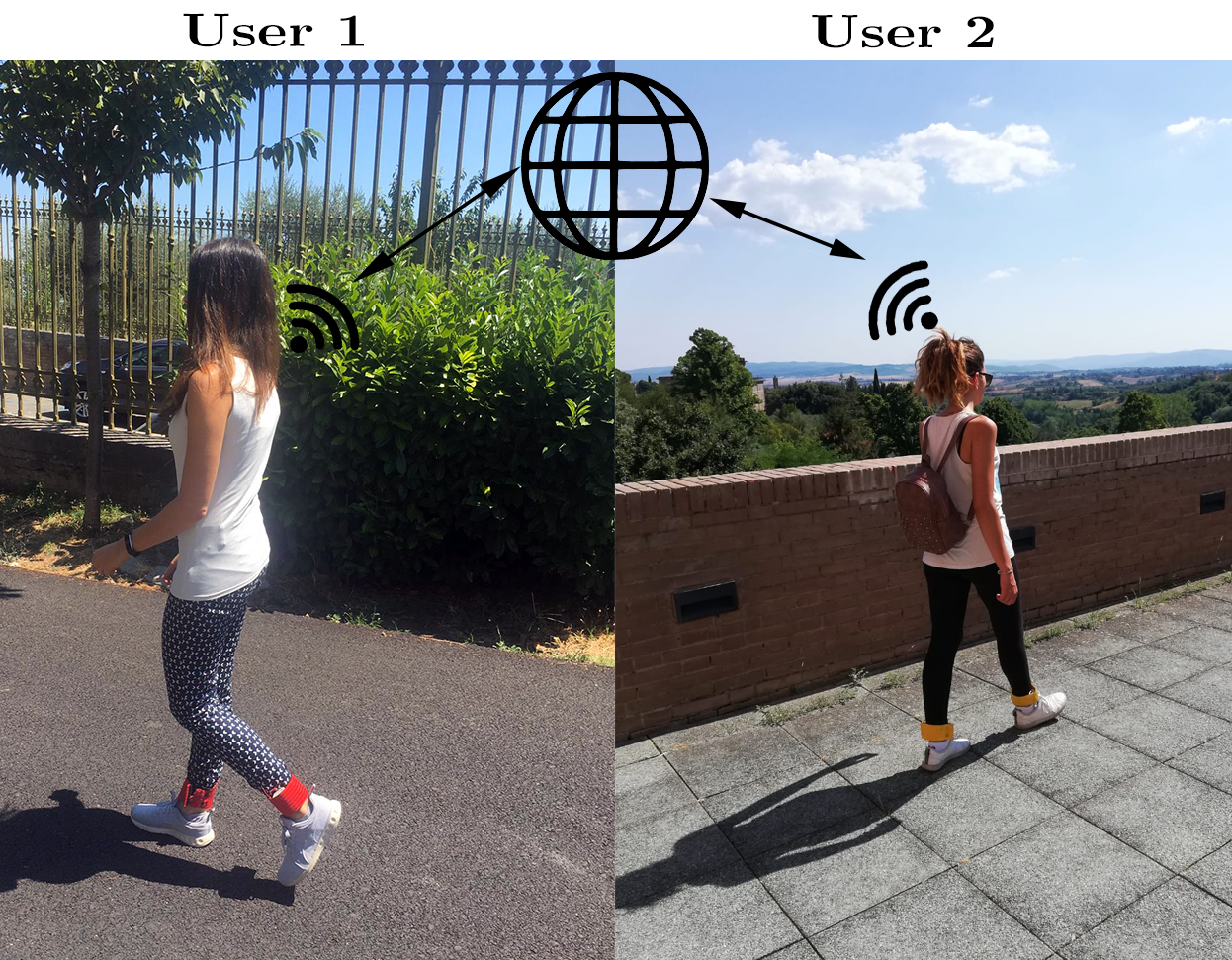} 
        \caption{Peer-to-Peer}
        \label{fig:exp3}
    \end{subfigure}    
    \caption{\TL{In (a) we examined the human capability in entraining to an external rhythm generated by an algorithm. The suggested cadence could be either constant, as in the Artificial Constant Reference experiment (\sect\ref{SEC:VirtualReference}), or time-varying (\sect\ref{SEC:ExpArtificialLeader}). In (b) a representative frame of the \lq\lq Human Leader'' experiment (\sect\ref{SEC:ExpHumanLeader}) is depicted. The follower is asked to synchronize to the gait pace of another human using the proposed system. This experiment differs from the previous one since the human's stride duration is not regular, but features small unpredictable variations. Finally, in the \lq\lq Peer-to-peer'' configuration (c) we tested the bidirectional capability of our system (see \sect\ref{SEC:ExpP2P}). Both participants were sending each other their respective gait cadence (sensed by their anklet devices). 
The direction of the information is graphically represented by the arrows. }}
    \label{fig:manmade}
\end{figure*}

%% file: tablePercentage.tex
\begin{table}[]
\centering
\begin{adjustbox}{center, width=0.9\columnwidth}  
\footnotesize
\begin{tabular}{r|rrr|rrr|c}
\hline \hline
\textbf{Name}    & \multicolumn{3}{c|}{\textbf{Haptics}} & \multicolumn{3}{c|}{\textbf{Haptics + Secondary Task}} & \textbf{Note} \\ \hline
        & 2\%       & 4\%       & 6\%       & 2\%        & 4\%         & 6\%          & \\ \hline
User1   & 66,32     & 99,05     & 100,00    & 47,36      & 84,88       & 92,59        & \\ 
User2   & 89,06     & 100,00    & 100,00    & 88,19      & 100,00      & 100,00       & $\bigstar$\\ 
User3   & 78,90     & 98,00     & 99,80     & 28,85      & 78,19       & 99,60        & \\ 
User4   & 47,35     & 98,83     & 100,00    & 78,66      & 100,00      & 100,00       &  \\ 
User5   & 91,13     & 100,00    & 100,00    & 38,88      & 77,34       & 87,43        & \\ 
User6   & 81,52     & 96,53     & 100,00    & 55,82      & 93,43       & 100,00       & \\ 
User7   & 89,95     & 99,69     & 100,00    & 36,91      & 86,95       & 100,00       & \\ 
User8   & 94,78     & 100,00    & 100,00    & 25,19      & 78,78       & 100,00       & \\ 
User9   & 76,98     & 98,47     & 99,96     & 44,80      & 80,38       & 92,29        & \\ 
User10  & 80,38     & 100,00     & 100,00    & 67,61      & 100,00      & 100,00       & $\bigstar$ \\ 
User11  & 66,58     & 94,41     & 100,00    & 0,00       & 4,20        & 39,82        & \\ 
User12  & 41,50     & 94,30     & 99,52     & 34,45      & 98,54       & 100,00       & \\ 
User13  & 83,28     & 100,00    & 100,00    & 48,96      & 97,32       & 100,00       & \\ 
User14  & 78,77     & 94,68    & 100,00    & 19,07      & 62,93       & 97,33        & \\ 
User15  & 76,01     & 99,50     & 100,00    & 76,01      & 99,52       & 100,00       & \\ 
User16  & 90,76     & 100,00    & 100,00    & 47,63      & 84,67       & 94,27        & \\ 
User17  & 77,10     & 98,56     & 100,00    & 49,64      & 81,36       & 93,68        & \\ 
User18  & 88,35     & 99,60     & 100,00    & 50,88      & 92,36       & 100,00       & \\ 
User19  & 73,78     & 88,66     & 99,09     & 53,58      & 88,04       & 91,48        & \\ 
User20  & 72,27     & 100,00    & 100,00    & 39,28      & 82,75       & 95,68        & \\ \hline



\hline
\textbf{Percentile 25} & \textbf{72.65} & \textbf{96.90}  & \textbf{100.00} & \textbf{35.07} & \textbf{79.18} & \textbf{92.86}  \\
\textbf{Percentile 50} & \textbf{78.84} & \textbf{99.28}  & \textbf{100.00} & \textbf{47.50} & \textbf{85.92} & \textbf{99.80}  \\
\textbf{Percentile 75} & \textbf{88.88} & \textbf{100.00} & \textbf{100.00} & \textbf{55.26} & \textbf{98.24} & \textbf{100.00} \\

\hline \hline
\end{tabular}
\end{adjustbox}
\caption{For each user are reported the percentages of trial duration during which the participant aligned his cadence with the reference rhythm, grouped wrt the considered error bands. Please notice that users tagged with $\bigstar$ are the two high level music players.}
\label{TABLE:synchronization_intervals}
\end{table}

%% file: figures/boxplot.tex
\begin{tikzpicture}[x=1pt,y=1pt]

\definecolor[named]{fillColor}{rgb}{1.00,1.00,1.00}
\path[use as bounding box,fill=fillColor,fill opacity=0.00] (0,0) rectangle (361.35,361.35);
\begin{scope}
\path[clip] (  0.00,  0.00) rectangle (361.35,361.35);
\definecolor[named]{drawColor}{rgb}{1.00,1.00,1.00}
\definecolor[named]{fillColor}{rgb}{1.00,1.00,1.00}

\path[draw=drawColor,line width= 0.6pt,line join=round,line cap=round,fill=fillColor] (  0.00,  0.00) rectangle (361.35,361.35);
\end{scope}
\begin{scope}
\path[clip] ( 42.99, 35.59) rectangle (355.35,302.44);
\definecolor[named]{fillColor}{rgb}{1.00,1.00,1.00}

\path[fill=fillColor] ( 42.99, 35.59) rectangle (355.35,302.44);
\definecolor[named]{drawColor}{rgb}{0.97,0.46,0.43}
\definecolor[named]{fillColor}{rgb}{0.97,0.46,0.43}

\path[draw=drawColor,line width= 0.4pt,line join=round,line cap=round,fill=fillColor] ( 73.22,162.59) circle (  1.96);

\path[draw=drawColor,line width= 0.4pt,line join=round,line cap=round,fill=fillColor] ( 73.22,148.40) circle (  1.96);

\path[draw=drawColor,line width= 0.6pt,line join=round,fill=fillColor] ( 73.22,262.48) -- ( 73.22,277.65);

\path[draw=drawColor,line width= 0.6pt,line join=round,fill=fillColor] ( 73.22,225.79) -- ( 73.22,208.61);
\definecolor[named]{fillColor}{rgb}{1.00,1.00,1.00}

\path[draw=drawColor,line width= 0.6pt,line join=round,line cap=round,fill=fillColor] ( 55.59,262.48) --
	( 55.59,225.79) --
	( 90.85,225.79) --
	( 90.85,262.48) --
	( 55.59,262.48) --
	cycle;
\definecolor[named]{fillColor}{rgb}{0.97,0.46,0.43}

\path[draw=drawColor,line width= 1.1pt,line join=round,fill=fillColor] ( 55.59,238.97) -- ( 90.85,238.97);
\definecolor[named]{drawColor}{rgb}{0.72,0.62,0.00}
\definecolor[named]{fillColor}{rgb}{0.72,0.62,0.00}

\path[draw=drawColor,line width= 0.4pt,line join=round,line cap=round,fill=fillColor] (123.60,262.80) circle (  1.96);

\path[draw=drawColor,line width= 0.6pt,line join=round,fill=fillColor] (123.60,290.31) -- (123.60,290.31);

\path[draw=drawColor,line width= 0.6pt,line join=round,fill=fillColor] (123.60,284.57) -- (123.60,276.49);
\definecolor[named]{fillColor}{rgb}{1.00,1.00,1.00}

\path[draw=drawColor,line width= 0.6pt,line join=round,line cap=round,fill=fillColor] (105.97,290.31) --
	(105.97,284.57) --
	(141.23,284.57) --
	(141.23,290.31) --
	(105.97,290.31) --
	cycle;
\definecolor[named]{fillColor}{rgb}{0.72,0.62,0.00}

\path[draw=drawColor,line width= 1.1pt,line join=round,fill=fillColor] (105.97,288.55) -- (141.23,288.55);
\definecolor[named]{drawColor}{rgb}{0.00,0.73,0.22}
\definecolor[named]{fillColor}{rgb}{0.00,0.73,0.22}

\path[draw=drawColor,line width= 0.4pt,line join=round,line cap=round,fill=fillColor] (173.98,289.83) circle (  1.96);

\path[draw=drawColor,line width= 0.4pt,line join=round,line cap=round,fill=fillColor] (173.98,290.22) circle (  1.96);

\path[draw=drawColor,line width= 0.4pt,line join=round,line cap=round,fill=fillColor] (173.98,289.15) circle (  1.96);

\path[draw=drawColor,line width= 0.4pt,line join=round,line cap=round,fill=fillColor] (173.98,288.11) circle (  1.96);

\path[draw=drawColor,line width= 0.6pt,line join=round,fill=fillColor] (173.98,290.31) -- (173.98,290.31);

\path[draw=drawColor,line width= 0.6pt,line join=round,fill=fillColor] (173.98,290.31) -- (173.98,290.31);
\definecolor[named]{fillColor}{rgb}{1.00,1.00,1.00}

\path[draw=drawColor,line width= 0.6pt,line join=round,line cap=round,fill=fillColor] (156.35,290.31) --
	(156.35,290.31) --
	(191.61,290.31) --
	(191.61,290.31) --
	(156.35,290.31) --
	cycle;
\definecolor[named]{fillColor}{rgb}{0.00,0.73,0.22}

\path[draw=drawColor,line width= 1.1pt,line join=round,fill=fillColor] (156.35,290.31) -- (191.61,290.31);
\definecolor[named]{drawColor}{rgb}{0.00,0.75,0.77}
\definecolor[named]{fillColor}{rgb}{0.00,0.75,0.77}

\path[draw=drawColor,line width= 0.4pt,line join=round,line cap=round,fill=fillColor] (224.36,261.66) circle (  1.96);

\path[draw=drawColor,line width= 0.4pt,line join=round,line cap=round,fill=fillColor] (224.36, 47.72) circle (  1.96);

\path[draw=drawColor,line width= 0.6pt,line join=round,fill=fillColor] (224.36,179.06) -- (224.36,238.54);

\path[draw=drawColor,line width= 0.6pt,line join=round,fill=fillColor] (224.36,135.77) -- (224.36, 93.99);
\definecolor[named]{fillColor}{rgb}{1.00,1.00,1.00}

\path[draw=drawColor,line width= 0.6pt,line join=round,line cap=round,fill=fillColor] (206.73,179.06) --
	(206.73,135.77) --
	(241.99,135.77) --
	(241.99,179.06) --
	(206.73,179.06) --
	cycle;
\definecolor[named]{fillColor}{rgb}{0.00,0.75,0.77}

\path[draw=drawColor,line width= 1.1pt,line join=round,fill=fillColor] (206.73,162.94) -- (241.99,162.94);
\definecolor[named]{drawColor}{rgb}{0.38,0.61,1.00}
\definecolor[named]{fillColor}{rgb}{0.38,0.61,1.00}

\path[draw=drawColor,line width= 0.4pt,line join=round,line cap=round,fill=fillColor] (274.74, 57.91) circle (  1.96);

\path[draw=drawColor,line width= 0.6pt,line join=round,fill=fillColor] (274.74,284.55) -- (274.74,290.31);

\path[draw=drawColor,line width= 0.6pt,line join=round,fill=fillColor] (274.74,241.75) -- (274.74,200.39);
\definecolor[named]{fillColor}{rgb}{1.00,1.00,1.00}

\path[draw=drawColor,line width= 0.6pt,line join=round,line cap=round,fill=fillColor] (257.11,284.55) --
	(257.11,241.75) --
	(292.37,241.75) --
	(292.37,284.55) --
	(257.11,284.55) --
	cycle;
\definecolor[named]{fillColor}{rgb}{0.38,0.61,1.00}

\path[draw=drawColor,line width= 1.1pt,line join=round,fill=fillColor] (257.11,256.14) -- (292.37,256.14);
\definecolor[named]{drawColor}{rgb}{0.96,0.39,0.89}
\definecolor[named]{fillColor}{rgb}{0.96,0.39,0.89}

\path[draw=drawColor,line width= 0.4pt,line join=round,line cap=round,fill=fillColor] (325.12,144.32) circle (  1.96);

\path[draw=drawColor,line width= 0.6pt,line join=round,fill=fillColor] (325.12,290.31) -- (325.12,290.31);

\path[draw=drawColor,line width= 0.6pt,line join=round,fill=fillColor] (325.12,274.32) -- (325.12,259.82);
\definecolor[named]{fillColor}{rgb}{1.00,1.00,1.00}

\path[draw=drawColor,line width= 0.6pt,line join=round,line cap=round,fill=fillColor] (307.49,290.31) --
	(307.49,274.32) --
	(342.75,274.32) --
	(342.75,290.31) --
	(307.49,290.31) --
	cycle;
\definecolor[named]{fillColor}{rgb}{0.96,0.39,0.89}

\path[draw=drawColor,line width= 1.1pt,line join=round,fill=fillColor] (307.49,289.83) -- (342.75,289.83);
\end{scope}
\begin{scope}
\path[clip] (  0.00,  0.00) rectangle (361.35,361.35);
\definecolor[named]{drawColor}{rgb}{0.00,0.00,0.00}

\path[draw=drawColor,line width= 0.6pt,line join=round] ( 42.99, 35.59) --
	( 42.99,302.44);
	\path[draw=drawColor,line width= 0.6pt,line join=round] ( 42.99,302.44) --
	( 355.35, 302.44);
\path[draw=drawColor,line width= 0.6pt,line join=round] ( 355.35, 302.44) --
	( 355.35, 35.59);	
\end{scope}
\begin{scope}
\path[clip] (  0.00,  0.00) rectangle (361.35,361.35);
\definecolor[named]{drawColor}{rgb}{0.00,0.00,0.00}

\node[text=drawColor,anchor=base east,inner sep=0pt, outer sep=0pt, scale=  1.20] at ( 37.59, 43.59) {0\%};

\node[text=drawColor,anchor=base east,inner sep=0pt, outer sep=0pt, scale=  1.20] at ( 37.59,104.24) {25\%};

\node[text=drawColor,anchor=base east,inner sep=0pt, outer sep=0pt, scale=  1.20] at ( 37.59,164.89) {50\%};

\node[text=drawColor,anchor=base east,inner sep=0pt, outer sep=0pt, scale=  1.20] at ( 37.59,225.53) {75\%};

\node[text=drawColor,anchor=base east,inner sep=0pt, outer sep=0pt, scale=  1.20] at ( 37.59,286.18) {100\%};
\node[text=drawColor,inner sep=0pt, outer sep=0pt, scale=  1.50, rotate=90] at ( 6,164.89) {Alignment Percentage};

\end{scope}
\begin{scope}
\path[clip] (  0.00,  0.00) rectangle (361.35,361.35);
\definecolor[named]{drawColor}{rgb}{0.00,0.00,0.00}

\path[draw=drawColor,line width= 0.6pt,line join=round] ( 39.99, 47.72) --
	( 42.99, 47.72);

\path[draw=drawColor,line width= 0.6pt,line join=round] ( 39.99,108.37) --
	( 42.99,108.37);

\path[draw=drawColor,line width= 0.6pt,line join=round] ( 39.99,169.02) --
	( 42.99,169.02);

\path[draw=drawColor,line width= 0.6pt,line join=round] ( 39.99,229.67) --
	( 42.99,229.67);

\path[draw=drawColor,line width= 0.6pt,line join=round] ( 39.99,290.31) --
	( 42.99,290.31);
\end{scope}
\begin{scope}
\path[clip] (  0.00,  0.00) rectangle (361.35,361.35);
\definecolor[named]{drawColor}{rgb}{0.00,0.00,0.00}

\path[draw=drawColor,line width= 0.6pt,line join=round] ( 42.99, 35.59) --
	(355.35, 35.59);
\end{scope}
\begin{scope}
\path[clip] (  0.00,  0.00) rectangle (361.35,361.35);
\definecolor[named]{drawColor}{rgb}{0.00,0.00,0.00}

\path[draw=drawColor,line width= 0.6pt,line join=round] ( 73.22, 32.59) --
	( 73.22, 35.59);

\path[draw=drawColor,line width= 0.6pt,line join=round] (123.60, 32.59) --
	(123.60, 35.59);

\path[draw=drawColor,line width= 0.6pt,line join=round] (173.98, 32.59) --
	(173.98, 35.59);

\path[draw=drawColor,line width= 0.6pt,line join=round] (224.36, 32.59) --
	(224.36, 35.59);

\path[draw=drawColor,line width= 0.6pt,line join=round] (274.74, 32.59) --
	(274.74, 35.59);

\path[draw=drawColor,line width= 0.6pt,line join=round] (325.12, 32.59) --
	(325.12, 35.59);
\end{scope}
\begin{scope}
\path[clip] (  0.00,  0.00) rectangle (361.35,361.35);
\definecolor[named]{drawColor}{rgb}{0.00,0.00,0.00}

\node[text=drawColor,anchor=base,inner sep=0pt, outer sep=0pt, scale=  1.20] at ( 73.22, 21.93) {H$_{2\%}$};

\node[text=drawColor,anchor=base,inner sep=0pt, outer sep=0pt, scale=  1.20] at (123.60, 21.93) {H$_{4\%}$};

\node[text=drawColor,anchor=base,inner sep=0pt, outer sep=0pt, scale=  1.20] at (173.98, 21.93) {H$_{6\%}$};

\node[text=drawColor,anchor=base,inner sep=0pt, outer sep=0pt, scale=  1.20] at (224.36, 21.93) {HS$_{2\%}$};

\node[text=drawColor,anchor=base,inner sep=0pt, outer sep=0pt, scale=  1.20] at (274.74, 21.93) {HS$_{4\%}$};

\node[text=drawColor,anchor=base,inner sep=0pt, outer sep=0pt, scale=  1.20] at (325.12, 21.93) {HS$_{6\%}$};

\node[text=drawColor,anchor=base,inner sep=0pt, outer sep=0pt, scale=  1.50] at (200, 1) {Condition};

\end{scope}

\end{tikzpicture}

%% file: figures/VR1.tex
%
\definecolor{mycolor1}{rgb}{0.00000,0.44700,0.74100}%
\definecolor{mycolor2}{rgb}{0.85000,0.32500,0.09800}%
\definecolor{mycolor3}{rgb}{1.00000,0.00000,1.00000}%
\definecolor{mycolor4}{rgb}{0.92900,0.69400,0.12500}%

\begin{tikzpicture}
\centering
\begin{axis}[%
width=0.95\columnwidth,
height=0.6\columnwidth,,
xtick = {0, 20,40, 60,80, 100},
xticklabels = {50, 70, 90, 110, 130, 150},
xlabel=Time (s),
ylabel=Stride Duration (ms/stride),
xmin=0,
xmax=90,
ymin=850,
ymax=1300,
axis background/.style={fill=white},
legend style={draw=none},
legend columns=2, 
legend style={ font=\scriptsize},
legend cell align={left},
]

%
\addplot [color=mycolor1]
  table[row sep=crcr]{%
0	1030\\
1.16	1030\\
2.231	1030\\
3.35	1030\\
4.429	1030\\
5.54	1030\\
6.65	1030\\
7.77	1030\\
8.872	1030\\
10.01	1030\\
11.111	1030\\
12.21	1030\\
13.329	1030\\
14.46	1030\\
15.55	1030\\
16.678	1030\\
17.77	1030\\
18.869	1030\\
19.95	1030\\
21.032	1030\\
22.14	1030\\
23.27	1030\\
24.378	1030\\
25.481	1030\\
26.571	1030\\
27.683	1030\\
28.812	1030\\
29.93	1030\\
31.039	1030\\
32.17	1030\\
33.32	1030\\
34.43	1030\\
35.552	1030\\
36.678	1030\\
37.819	1030\\
38.951	1030\\
40.056	1030\\
41.066	1030\\
42.039	1030\\
43.093	1030\\
44.195	1030\\
45.101	1030\\
46.127	1030\\
47.168	1030\\
48.198	1030\\
49.232	1030\\
50.258	1030\\
51.261	1030\\
52.22	1030\\
53.3	1030\\
54.312	1030\\
55.39	1030\\
56.409	1030\\
57.45	1030\\
58.47	1030\\
59.491	1030\\
60.499	1030\\
61.492	1030\\
62.559	1030\\
63.672	1030\\
64.744	1030\\
65.614	1030\\
66.702	1030\\
67.729	1030\\
68.791	1030\\
69.784	1030\\
70.854	1030\\
71.885	1030\\
72.924	1030\\
73.952	1030\\
74.954	1030\\
76.005	1030\\
77.034	1030\\
78.044	1030\\
79.043	1030\\
80.167	1030\\
81.228	1030\\
82.264	1030\\
83.244	1030\\
84.271	1030\\
85.324	1030\\
86.332	1030\\
87.392	1030\\
88.364	1030\\
89.436	1030\\
90.454	1030\\
91.483	1030\\
92.514	1030\\
93.572	1030\\
};
\addlegendentry{Reference}

\addplot [color=mycolor2]
  table[row sep=crcr]{%
0	1080\\
1.16	1130\\
2.231	1090\\
3.35	1110\\
4.429	1090\\
5.54	1090\\
6.65	1100\\
7.77	1100\\
8.872	1130\\
10.01	1110\\
11.111	1100\\
12.21	1100\\
13.329	1110\\
14.46	1120\\
15.55	1100\\
16.678	1120\\
17.77	1080\\
18.869	1100\\
19.95	1080\\
21.032	1090\\
22.14	1110\\
23.27	1100\\
24.378	1120\\
25.481	1090\\
26.571	1100\\
27.683	1110\\
28.812	1130\\
29.93	1120\\
31.039	1110\\
32.17	1130\\
33.32	1120\\
34.43	1110\\
35.552	1140\\
36.678	1130\\
37.819	1140\\
38.951	1110\\
40.056	1110\\
41.066	1010\\
42.039	990\\
43.093	1040\\
44.195	1000\\
45.101	1020\\
46.127	990\\
47.168	1040\\
48.198	1030\\
49.232	1040\\
50.258	1010\\
51.261	1010\\
52.22	990\\
53.3	1050\\
54.312	1030\\
55.39	1060\\
56.409	1010\\
57.45	1040\\
58.47	1020\\
59.491	1030\\
60.499	990\\
61.492	1010\\
62.559	1050\\
63.672	1020\\
64.744	1040\\
65.614	1000\\
66.702	1050\\
67.729	1060\\
68.791	1050\\
69.784	1010\\
70.854	1030\\
71.885	1030\\
72.924	1030\\
73.952	1020\\
74.954	1010\\
76.005	1040\\
77.034	1040\\
78.044	990\\
79.043	1010\\
80.167	1020\\
81.228	1060\\
82.264	1020\\
83.244	990\\
84.271	1030\\
85.324	1050\\
86.332	1020\\
87.392	1030\\
88.364	990\\
89.436	1030\\
90.454	1040\\
91.483	1040\\
92.514	1030\\
93.572	1050\\
};
\addlegendentry{User}

\addplot [color=green]
  table[row sep=crcr]{%
37.819	850\\
37.819	1200\\
};
\addlegendentry{Haptics On}

\addplot[area legend, draw=none, fill=blue, fill opacity=0.1]
table[row sep=crcr] {%
x	y\\
37.819	1081.5\\
38.951	1081.5\\
40.056	1081.5\\
41.066	1081.5\\
42.039	1081.5\\
43.093	1081.5\\
44.195	1081.5\\
45.101	1081.5\\
46.127	1081.5\\
47.168	1081.5\\
48.198	1081.5\\
49.232	1081.5\\
50.258	1081.5\\
51.261	1081.5\\
52.22	1081.5\\
53.3	1081.5\\
54.312	1081.5\\
55.39	1081.5\\
56.409	1081.5\\
57.45	1081.5\\
58.47	1081.5\\
59.491	1081.5\\
60.499	1081.5\\
61.492	1081.5\\
62.559	1081.5\\
63.672	1081.5\\
64.744	1081.5\\
65.614	1081.5\\
66.702	1081.5\\
67.729	1081.5\\
68.791	1081.5\\
69.784	1081.5\\
70.854	1081.5\\
71.885	1081.5\\
72.924	1081.5\\
73.952	1081.5\\
74.954	1081.5\\
76.005	1081.5\\
77.034	1081.5\\
78.044	1081.5\\
79.043	1081.5\\
80.167	1081.5\\
81.228	1081.5\\
82.264	1081.5\\
83.244	1081.5\\
84.271	1081.5\\
85.324	1081.5\\
86.332	1081.5\\
87.392	1081.5\\
88.364	1081.5\\
89.436	1081.5\\
90.454	1081.5\\
91.483	1081.5\\
92.514	1081.5\\
93.572	1081.5\\
93.572	978.5\\
92.514	978.5\\
91.483	978.5\\
90.454	978.5\\
89.436	978.5\\
88.364	978.5\\
87.392	978.5\\
86.332	978.5\\
85.324	978.5\\
84.271	978.5\\
83.244	978.5\\
82.264	978.5\\
81.228	978.5\\
80.167	978.5\\
79.043	978.5\\
78.044	978.5\\
77.034	978.5\\
76.005	978.5\\
74.954	978.5\\
73.952	978.5\\
72.924	978.5\\
71.885	978.5\\
70.854	978.5\\
69.784	978.5\\
68.791	978.5\\
67.729	978.5\\
66.702	978.5\\
65.614	978.5\\
64.744	978.5\\
63.672	978.5\\
62.559	978.5\\
61.492	978.5\\
60.499	978.5\\
59.491	978.5\\
58.47	978.5\\
57.45	978.5\\
56.409	978.5\\
55.39	978.5\\
54.312	978.5\\
53.3	978.5\\
52.22	978.5\\
51.261	978.5\\
50.258	978.5\\
49.232	978.5\\
48.198	978.5\\
47.168	978.5\\
46.127	978.5\\
45.101	978.5\\
44.195	978.5\\
43.093	978.5\\
42.039	978.5\\
41.066	978.5\\
40.056	978.5\\
38.951	978.5\\
37.819	978.5\\
}--cycle;
\addlegendentry{$\pm$4\% area}

\end{axis}
\end{tikzpicture}%

%% file: figures/VR2.tex
%
\definecolor{mycolor1}{rgb}{0.00000,0.44700,0.74100}%
\definecolor{mycolor2}{rgb}{0.85000,0.32500,0.09800}%
\definecolor{mycolor3}{rgb}{1.00000,0.00000,1.00000}%
\definecolor{mycolor4}{rgb}{0.92900,0.69400,0.12500}%

\begin{tikzpicture}
\centering
\begin{axis}[%
width=0.95\columnwidth,
height=0.6\columnwidth,,
xtick =       {20 ,40, 60,  80, 100, 120},
xticklabels = {50, 70, 90, 110, 130, 150},
xlabel=Time (s),
ylabel=Stride Duration (ms/stride),
xmin=20,
xmax=110,
ymin=950,
ymax=1300,
axis background/.style={fill=white},
legend style={draw=none},
legend columns=2, 
legend style={ font=\scriptsize},
legend cell align={left},
]

%
\addplot [color=mycolor1]
  table[row sep=crcr]{%
0	1140\\
1.031	1140\\
2.072	1140\\
3.14	    1140\\
4.12	    1140\\
5.162	1140\\
6.222	1140\\
7.242	1140\\
8.272	1140\\
9.332	1140\\
10.382	1140\\
11.413	1140\\
12.442	1140\\
13.53	1140\\
14.572	1140\\
15.632	1140\\
16.686	1140\\
17.705	1140\\
18.742	1140\\
19.775	1140\\
20.802	1140\\
21.852	1140\\
22.872	1140\\
23.881	1140\\
24.92	1140\\
25.942	1140\\
25.972	1140\\
27.002	1140\\
28.041	1140\\
29.052	1140\\
30.062	1140\\
31.092	1140\\
32.122	1140\\
33.166	1140\\
34.172	1140\\
35.212	1140\\
36.212	1140\\
37.252	1140\\
38.3	    1140\\
39.344	1140\\
39.862	1140\\
40.351	1140\\
41.373	1140\\
42.402	1140\\
43.43	1140\\
44.482	1140\\
45.513	1140\\
46.542	1140\\
47.581	1140\\
48.602	1140\\
49.664	1140\\
50.712	1140\\
51.78	1140\\
52.802	1140\\
53.832	1140\\
54.865	1140\\
56.172	1140\\
56.992	1140\\
58.044	1140\\
59.091	1140\\
60.153	1140\\
61.202	1140\\
62.252	1140\\
63.301	1140\\
64.361	1140\\
65.481	1140\\
66.656	1140\\
67.812	1140\\
68.946	1140\\
70.122	1140\\
71.343	1140\\
72.492	1140\\
73.642	1140\\
74.772	1140\\
75.881	1140\\
76.999	1140\\
78.111	1140\\
79.251	1140\\
80.422	1140\\
81.543	1140\\
82.701	1140\\
83.841	1140\\
84.973	1140\\
86.112	1140\\
87.241	1140\\
88.371	1140\\
89.533	1140\\
90.703	1140\\
91.85	1140\\
93.004	1140\\
94.152	1140\\
95.289	1140\\
96.41	1140\\
97.521	1140\\
98.672	1140\\
99.851	1140\\
100.917	1140\\
102.067	1140\\
103.187	1140\\
104.347	1140\\
105.457	1140\\
106.587	1140\\
107.717	1140\\
108.847	1140\\
109.997	1140\\
111.157	1140\\
112.297	1140\\
113.465	1140\\
114.587	1140\\
115.755	1140\\
116.895	1140\\
118.027	1140\\
119.187	1140\\
120.327	1140\\
121.496	1140\\
122.619	1140\\
123.728	1140\\
124.897	1140\\
126.041	1140\\
127.147	1140\\
128.308	1140\\
129.446	1140\\
130.616	1140\\
131.735	1140\\
132.868	1140\\
134.052	1140\\
135.187	1140\\
136.286	1140\\
137.427	1140\\
138.578	1140\\
139.705	1140\\
140.844	1140\\
};
\addlegendentry{Reference}

\addplot [color=mycolor2]
  table[row sep=crcr]{%
0	1070\\
1.031	1040\\
2.072	1040\\
3.14		1030\\
4.12		1035\\
5.162	1040\\
6.222	1035\\
7.242	1020\\
8.272	1030\\
9.332	1040\\
10.382	1050\\
11.413	1020\\
12.442	1040\\
13.53	1060\\
14.572	1050\\
15.632	1050\\
16.686	1030\\
17.705	1040\\
18.742	1040\\
19.775	1030\\
20.802	1030\\
21.852	1040\\
22.872	1020\\
23.881	1030\\
24.92	1000\\
25.942	1000\\
25.972	1060\\
27.002	1030\\
28.041	1030\\
29.052	1020\\
30.062	1010\\
31.092	1030\\
32.122	1020\\
33.166	1030\\
34.172	1030\\
35.212	1030\\
36.212	1010\\
37.252	1040\\
38.3		1040\\
39.344	1030\\
39.862	1030\\
40.351	1010\\
41.373	1020\\
42.402	1040\\
43.43	1020\\
44.482	1040\\
45.513	1030\\
46.542	1020\\
47.581	1030\\
48.602	1040\\
49.664	1050\\
50.712	1040\\
51.78	1040\\
52.802	1030\\
53.832	1030\\
54.865	1040\\
56.172	1040\\
56.992	1070\\
58.044	1060\\
59.091	1050\\
60.153	1040\\
61.202	1050\\
62.252	1050\\
63.301	1070\\
64.361	1040\\
65.481	1100\\
66.656	1150\\
67.812	1180\\
68.946	1130\\
70.122	1170\\
71.343	1160\\
72.492	1140\\
73.642	1150\\
74.772	1120\\
75.881	1100\\
76.999	1120\\
78.111	1120\\
79.251	1140\\
80.422	1150\\
81.543	1130\\
82.701	1130\\
83.841	1140\\
84.973	1120\\
86.112	1150\\
87.241	1120\\
88.371	1130\\
89.533	1160\\
90.703	1160\\
91.85	1130\\
93.004	1150\\
94.152	1150\\
95.289	1130\\
96.41	1130\\
97.521	1120\\
98.672	1140\\
99.851	1140\\
100.917	1110\\
102.067	1130\\
103.187	1120\\
104.347	1130\\
105.457	1120\\
106.587	1130\\
107.717	1140\\
108.847	1110\\
109.997	1150\\
111.157	1160\\
112.297	1150\\
113.465	1150\\
114.587	1140\\
115.755	1140\\
116.895	1150\\
118.027	1140\\
119.187	1170\\
120.327	1140\\
121.496	1140\\
122.619	1120\\
123.728	1110\\
124.897	1150\\
126.041	1130\\
127.147	1130\\
128.308	1150\\
129.446	1130\\
130.616	1170\\
131.735	1120\\
132.868	1140\\
134.052	1140\\
135.187	1140\\
136.286	1130\\
137.427	1110\\
138.578	1140\\
139.705	1120\\
140.844	1130\\
};
\addlegendentry{User}

\addplot [color=green]
  table[row sep=crcr]{%
64.361	850\\
64.361	1220\\
};
\addlegendentry{Haptics On}

\addplot[area legend, draw=none, fill=blue, fill opacity=0.1]
table[row sep=crcr] {%
x	y\\
65.481	1197\\
66.656	1197\\
67.812	1197\\
68.946	1197\\
70.122	1197\\
71.343	1197\\
72.492	1197\\
73.642	1197\\
74.772	1197\\
75.881	1197\\
76.999	1197\\
78.111	1197\\
79.251	1197\\
80.422	1197\\
81.543	1197\\
82.701	1197\\
83.841	1197\\
84.973	1197\\
86.112	1197\\
87.241	1197\\
88.371	1197\\
89.533	1197\\
90.703	1197\\
91.85	1197\\
93.004	1197\\
94.152	1197\\
95.289	1197\\
96.41	1197\\
97.521	1197\\
98.672	1197\\
99.851	1197\\
100.917	1197\\
102.067	1197\\
103.187	1197\\
104.347	1197\\
105.457	1197\\
106.587	1197\\
107.717	1197\\
108.847	1197\\
109.997	1197\\
111.157	1197\\
112.297	1197\\
113.465	1197\\
114.587	1197\\
115.755	1197\\
116.895	1197\\
118.027	1197\\
119.187	1197\\
120.327	1197\\
121.496	1197\\
122.619	1197\\
123.728	1197\\
124.897	1197\\
126.041	1197\\
127.147	1197\\
128.308	1197\\
129.446	1197\\
130.616	1197\\
131.735	1197\\
132.868	1197\\
134.052	1197\\
135.187	1197\\
136.286	1197\\
137.427	1197\\
138.578	1197\\
139.705	1197\\
140.844	1197\\
140.844	1083\\
139.705	1083\\
138.578	1083\\
137.427	1083\\
136.286	1083\\
135.187	1083\\
134.052	1083\\
132.868	1083\\
131.735	1083\\
130.616	1083\\
129.446	1083\\
128.308	1083\\
127.147	1083\\
126.041	1083\\
124.897	1083\\
123.728	1083\\
122.619	1083\\
121.496	1083\\
120.327	1083\\
119.187	1083\\
118.027	1083\\
116.895	1083\\
115.755	1083\\
114.587	1083\\
113.465	1083\\
112.297	1083\\
111.157	1083\\
109.997	1083\\
108.847	1083\\
107.717	1083\\
106.587	1083\\
105.457	1083\\
104.347	1083\\
103.187	1083\\
102.067	1083\\
100.917	1083\\
99.851	1083\\
98.672	1083\\
97.521	1083\\
96.41	1083\\
95.289	1083\\
94.152	1083\\
93.004	1083\\
91.85	1083\\
90.703	1083\\
89.533	1083\\
88.371	1083\\
87.241	1083\\
86.112	1083\\
84.973	1083\\
83.841	1083\\
82.701	1083\\
81.543	1083\\
80.422	1083\\
79.251	1083\\
78.111	1083\\
76.999	1083\\
75.881	1083\\
74.772	1083\\
73.642	1083\\
72.492	1083\\
71.343	1083\\
70.122	1083\\
68.946	1083\\
67.812	1083\\
66.656	1083\\
65.481	1083\\
}--cycle;
\addlegendentry{$\pm$4\% area}

\end{axis}

\end{tikzpicture}%

%% file: figures/boxplot_time_2_review.tex
\begin{tikzpicture}[x=1pt,y=1pt]
\definecolor[named]{fillColor}{rgb}{1.00,1.00,1.00}
\path[use as bounding box,fill=fillColor,fill opacity=0.00] (0,0) rectangle (361.35,361.35);
\begin{scope}
\path[clip] (  0.00,  0.00) rectangle (361.35,361.35);
\definecolor[named]{drawColor}{rgb}{1.00,1.00,1.00}
\definecolor[named]{fillColor}{rgb}{1.00,1.00,1.00}

\path[draw=drawColor,line width= 0.6pt,line join=round,line cap=round,fill=fillColor] (  0.00,  0.00) rectangle (361.35,361.35);
\end{scope}
\begin{scope}
\path[clip] ( 31.00, 35.59) rectangle (355.35,316.90);
\definecolor[named]{fillColor}{rgb}{1.00,1.00,1.00}

\path[fill=fillColor] ( 31.00, 35.59) rectangle (355.35,316.90);
\definecolor[named]{drawColor}{rgb}{0.00,0.69,0.73}
\definecolor[named]{fillColor}{rgb}{0.00,0.69,0.73}

\path[draw=drawColor,line width= 0.4pt,line join=round,line cap=round,fill=fillColor] (119.46,304.11) circle (  1.96);

\path[draw=drawColor,line width= 0.4pt,line join=round,line cap=round,fill=fillColor] (119.46,290.36) circle (  1.96);

\path[draw=drawColor,line width= 0.6pt,line join=round,fill=fillColor] (119.46,163.64) -- (119.46,164.02);

\path[draw=drawColor,line width= 0.6pt,line join=round,fill=fillColor] (119.46,100.60) -- (119.46, 48.38);
\definecolor[named]{fillColor}{rgb}{1.00,1.00,1.00}

\path[draw=drawColor,line width= 0.6pt,line join=round,line cap=round,fill=fillColor] ( 67.85,163.64) --
	( 67.85,100.60) --
	(171.06,100.60) --
	(171.06,163.64) --
	( 67.85,163.64) --
	cycle;
\definecolor[named]{fillColor}{rgb}{0.00,0.69,0.73}

\path[draw=drawColor,line width= 1.1pt,line join=round,fill=fillColor] ( 67.85,134.22) -- (171.06,134.22);
\definecolor[named]{drawColor}{rgb}{0.91,0.72,0.00}
\definecolor[named]{fillColor}{rgb}{0.91,0.72,0.00}

\path[draw=drawColor,line width= 0.6pt,line join=round,fill=fillColor] (266.89,187.45) -- (266.89,233.81);

\path[draw=drawColor,line width= 0.6pt,line join=round,fill=fillColor] (266.89,115.88) -- (266.89, 60.61);
\definecolor[named]{fillColor}{rgb}{1.00,1.00,1.00}

\path[draw=drawColor,line width= 0.6pt,line join=round,line cap=round,fill=fillColor] (215.29,187.45) --
	(215.29,115.88) --
	(318.49,115.88) --
	(318.49,187.45) --
	(215.29,187.45) --
	cycle;
\definecolor[named]{fillColor}{rgb}{0.91,0.72,0.00}

\path[draw=drawColor,line width= 1.1pt,line join=round,fill=fillColor] (215.29,132.94) -- (318.49,132.94);
\end{scope}
\begin{scope}
\path[clip] (  0.00,  0.00) rectangle (361.35,361.35);
\definecolor[named]{drawColor}{rgb}{0.00,0.00,0.00}

\path[draw=drawColor,line width= 0.6pt,line join=round] ( 31.00, 35.59) --
	( 31.00,316.90);
\path[draw=drawColor,line width= 0.6pt,line join=round] ( 31.00, 35.59) --         (355.35, 35.59);
67	
\path[draw=drawColor,line width= 0.6pt,line join=round] ( 31.00, 316.90) -- (355.35, 316.90);
68	
\path[draw=drawColor,line width= 0.6pt,line join=round] (355.35, 316.90) --         ( 355.35, 35.59);    
    
\end{scope}
\begin{scope}
\path[clip] (  0.00,  0.00) rectangle (361.35,361.35);
\definecolor[named]{drawColor}{rgb}{0.00,0.00,0.00}

\node[text=drawColor,anchor=base east,inner sep=0pt, outer sep=0pt, scale=  1.20] at ( 25.60, 76.85) {1};

\node[text=drawColor,anchor=base east,inner sep=0pt, outer sep=0pt, scale=  1.20] at ( 25.60,127.79) {2};

\node[text=drawColor,anchor=base east,inner sep=0pt, outer sep=0pt, scale=  1.20] at ( 25.60,178.74) {3};

\node[text=drawColor,anchor=base east,inner sep=0pt, outer sep=0pt, scale=  1.20] at ( 25.60,229.68) {4};

\node[text=drawColor,anchor=base east,inner sep=0pt, outer sep=0pt, scale=  1.20] at ( 25.60,280.62) {5};
\end{scope}
\begin{scope}
\path[clip] (  0.00,  0.00) rectangle (361.35,361.35);
\definecolor[named]{drawColor}{rgb}{0.00,0.00,0.00}

\path[draw=drawColor,line width= 0.6pt,line join=round] ( 28.00, 80.98) --
	( 31.00, 80.98);

\path[draw=drawColor,line width= 0.6pt,line join=round] ( 28.00,131.93) --
	( 31.00,131.93);

\path[draw=drawColor,line width= 0.6pt,line join=round] ( 28.00,182.87) --
	( 31.00,182.87);

\path[draw=drawColor,line width= 0.6pt,line join=round] ( 28.00,233.81) --
	( 31.00,233.81);

\path[draw=drawColor,line width= 0.6pt,line join=round] ( 28.00,284.75) --
	( 31.00,284.75);
\end{scope}
\begin{scope}
\path[clip] (  0.00,  0.00) rectangle (361.35,361.35);
\definecolor[named]{drawColor}{rgb}{0.00,0.00,0.00}

\path[draw=drawColor,line width= 0.6pt,line join=round] ( 31.00, 35.59) --
	(355.35, 35.59);
\end{scope}
\begin{scope}
\path[clip] (  0.00,  0.00) rectangle (361.35,361.35);
\definecolor[named]{drawColor}{rgb}{0.00,0.00,0.00}

\path[draw=drawColor,line width= 0.6pt,line join=round] (119.46, 32.59) --
	(119.46, 35.59);

\path[draw=drawColor,line width= 0.6pt,line join=round] (266.89, 32.59) --
	(266.89, 35.59);
\end{scope}
\begin{scope}
\path[clip] (  0.00,  0.00) rectangle (361.35,361.35);
\definecolor[named]{drawColor}{rgb}{0.00,0.00,0.00}

\node[text=drawColor,anchor=base,inner sep=0pt, outer sep=0pt, scale=  1.20] at (119.46, 21.93) {$-10\%$};

\node[text=drawColor,anchor=base,inner sep=0pt, outer sep=0pt, scale=  1.20] at (266.89, 21.93) {$+10\%$};
\end{scope}
\begin{scope}
\path[clip] (  0.00,  0.00) rectangle (361.35,361.35);
\definecolor[named]{drawColor}{rgb}{0.00,0.00,0.00}

\node[text=drawColor,anchor=base,inner sep=0pt, outer sep=0pt, scale=  1.50] at (193.17,  0) {Reference Variation from Baseline Stride Duration};
\end{scope}
\begin{scope}
\path[clip] (  0.00,  0.00) rectangle (361.35,361.35);
\definecolor[named]{drawColor}{rgb}{0.00,0.00,0.00}

\node[text=drawColor,rotate= 90.00,anchor=base,inner sep=0pt, outer sep=0pt, scale=  1.50] at ( 10 ,176.25) {Time for Alignment (s)};
\end{scope}
\begin{scope}
\path[clip] (  0.00,  0.00) rectangle (361.35,361.35);
\definecolor[named]{fillColor}{rgb}{1.00,1.00,1.00}

\path[fill=fillColor] (126.38,328.90) rectangle (259.97,355.35);
\end{scope}

\end{tikzpicture}

%% file: figures/boxplot_percent_review.tex
\begin{tikzpicture}[x=1pt,y=1pt]
\definecolor[named]{fillColor}{rgb}{1.00,1.00,1.00}
\begin{scope}
\path[clip] (  0.00,  0.00) rectangle (361.35,361.35);
\definecolor[named]{drawColor}{rgb}{1.00,1.00,1.00}
\definecolor[named]{fillColor}{rgb}{1.00,1.00,1.00}

\path[draw=drawColor,line width= 0.6pt,line join=round,line cap=round,fill=fillColor] (  0.00,  0.00) rectangle (361.35,361.35);
\end{scope}
\begin{scope}
\path[clip] ( 42.99, 35.59) rectangle (355.35,316.90);
\definecolor[named]{fillColor}{rgb}{1.00,1.00,1.00}

\path[fill=fillColor] ( 42.99, 35.59) rectangle (355.35,316.90);
\definecolor[named]{drawColor}{rgb}{0.00,0.69,0.73}
\definecolor[named]{fillColor}{rgb}{0.00,0.69,0.73}

\path[draw=drawColor,line width= 0.6pt,line join=round,fill=fillColor] (128.18,304.11) -- (128.18,304.11);

\path[draw=drawColor,line width= 0.6pt,line join=round,fill=fillColor] (128.18,109.04) -- (128.18, 66.30);
\definecolor[named]{fillColor}{rgb}{1.00,1.00,1.00}

\path[draw=drawColor,line width= 0.6pt,line join=round,line cap=round,fill=fillColor] ( 78.49,304.11) --
	( 78.49,109.04) --
	(177.87,109.04) --
	(177.87,304.11) --
	( 78.49,304.11) --
	cycle;
\definecolor[named]{fillColor}{rgb}{0.00,0.69,0.73}

\path[draw=drawColor,line width= 1.1pt,line join=round,fill=fillColor] ( 78.49,239.69) -- (177.87,239.69);
\definecolor[named]{drawColor}{rgb}{0.91,0.72,0.00}
\definecolor[named]{fillColor}{rgb}{0.91,0.72,0.00}

\path[draw=drawColor,line width= 0.6pt,line join=round,fill=fillColor] (270.16,304.11) -- (270.16,304.11);

\path[draw=drawColor,line width= 0.6pt,line join=round,fill=fillColor] (270.16,148.76) -- (270.16, 48.38);
\definecolor[named]{fillColor}{rgb}{1.00,1.00,1.00}

\path[draw=drawColor,line width= 0.6pt,line join=round,line cap=round,fill=fillColor] (220.47,304.11) --
	(220.47,148.76) --
	(319.85,148.76) --
	(319.85,304.11) --
	(220.47,304.11) --
	cycle;
\definecolor[named]{fillColor}{rgb}{0.91,0.72,0.00}

\path[draw=drawColor,line width= 1.1pt,line join=round,fill=fillColor] (220.47,258.34) -- (319.85,258.34);
\end{scope}
\begin{scope}
\path[clip] (  0.00,  0.00) rectangle (361.35,361.35);
\definecolor[named]{drawColor}{rgb}{0.00,0.00,0.00}

\end{scope}
\begin{scope}
\path[clip] (  0.00,  0.00) rectangle (361.35,361.35);
\definecolor[named]{drawColor}{rgb}{0.00,0.00,0.00}

\node[text=drawColor,anchor=base east,inner sep=0pt, outer sep=0pt, scale=  1.20] at ( 37.59,106.24) {$96\%$};

\node[text=drawColor,anchor=base east,inner sep=0pt, outer sep=0pt, scale=  1.20] at ( 37.59,203.11) {$98\%$};

\node[text=drawColor,anchor=base east,inner sep=0pt, outer sep=0pt, scale=  1.20] at ( 37.59,299.98) {$100\%$};
\end{scope}
\begin{scope}
\path[clip] (  0.00,  0.00) rectangle (361.35,361.35);
\definecolor[named]{drawColor}{rgb}{0.00,0.00,0.00}

\path[draw=drawColor,line width= 0.6pt,line join=round] ( 39.99,110.38) --
	( 42.99,110.38);

\path[draw=drawColor,line width= 0.6pt,line join=round] ( 39.99,207.24) --
	( 42.99,207.24);

\path[draw=drawColor,line width= 0.6pt,line join=round] ( 39.99,304.11) --
	( 42.99,304.11);
\end{scope}
\begin{scope}
\path[clip] (  0.00,  0.00) rectangle (361.35,361.35);
\definecolor[named]{drawColor}{rgb}{0.00,0.00,0.00}

\path[draw=drawColor,line width= 0.6pt,line join=round] ( 42.99, 35.59) -- 	( 42.99,316.90);
\path[draw=drawColor,line width= 0.6pt,line join=round] ( 42.99, 35.59) -- 	(355.35, 35.59);
\path[draw=drawColor,line width= 0.6pt,line join=round] ( 42.99,316.90) -- 	( 355.35, 316.90);
\path[draw=drawColor,line width= 0.6pt,line join=round] ( 355.35, 316.90) -- 	( 355.35, 35.59);	

\end{scope}
\begin{scope}
\path[clip] (  0.00,  0.00) rectangle (361.35,361.35);
\definecolor[named]{drawColor}{rgb}{0.00,0.00,0.00}

\path[draw=drawColor,line width= 0.6pt,line join=round] (128.18, 32.59) --
	(128.18, 35.59);

\path[draw=drawColor,line width= 0.6pt,line join=round] (270.16, 32.59) --
	(270.16, 35.59);
\end{scope}
\begin{scope}
\path[clip] (  0.00,  0.00) rectangle (361.35,361.35);
\definecolor[named]{drawColor}{rgb}{0.00,0.00,0.00}

\node[text=drawColor,anchor=base,inner sep=0pt, outer sep=0pt, scale=  1.20] at (128.18, 21.93) {$-10\%$};

\node[text=drawColor,anchor=base,inner sep=0pt, outer sep=0pt, scale=  1.20] at (270.16, 21.93) {$+10\%$};
\end{scope}
\begin{scope}
\path[clip] (  0.00,  0.00) rectangle (361.35,361.35);
\definecolor[named]{drawColor}{rgb}{0.00,0.00,0.00}

\node[text=drawColor,anchor=base,inner sep=0pt, outer sep=0pt, scale=  1.50] at (199.17,  0) {Reference Variation from Baseline Stride Duration};
\end{scope}
\begin{scope}
\path[clip] (  0.00,  0.00) rectangle (361.35,361.35);
\definecolor[named]{drawColor}{rgb}{0.00,0.00,0.00}

\node[text=drawColor,rotate= 90.00,anchor=base,inner sep=0pt, outer sep=0pt, scale=  1.50] at ( 10,176.25) {Alignment Percentage};
\end{scope}
\begin{scope}
\path[clip] (  0.00,  0.00) rectangle (361.35,361.35);
\definecolor[named]{fillColor}{rgb}{1.00,1.00,1.00}

\path[fill=fillColor] (132.38,328.90) rectangle (265.97,355.35);
\end{scope}
\end{tikzpicture}

%% file: table_constant_rhythm.tex
\TL{
\begin{table}[]
\centering
\begin{adjustbox}{center, width=0.9\columnwidth}  
\begin{tabular}{cccccc}
\hline \hline 
\textbf{User} & \textbf{\begin{tabular}[c]{@{}c@{}}Baseline\\ Stride Dur.\\ (ms/stride)\end{tabular}} & \textbf{\begin{tabular}[c]{@{}c@{}}Time for\\ Alignment\\ (-$10\%$) (s)\end{tabular}} &  \textbf{\begin{tabular}[c]{@{}c@{}}Alignment\\ Percentage\\(-$10\%$) \%\end{tabular}} &  \textbf{\begin{tabular}[c]{@{}c@{}}Time for\\ Alignment\\ (+$10\%$) (s)\end{tabular}} &  \textbf{\begin{tabular}[c]{@{}c@{}}Alignment\\ Percentage\\(+$10\%$) \%\end{tabular}}\\
\hline
U1 & \ 995  & 0.4 & 100.00 & 3.5 & \ 98.11 \\
U2 & 1146 & 1.3 & 100.00 & 1.5 & 100.00 \\
U3 & 1060 & 1.5 & \ 95.23 & 4.0 & \ 94.72 \\
U4 & 1021 & 2.6 & \ 95.09 & 3.1 & 100.00 \\
U5 & 1078 & 2.4 & 100.00 & 3.1 & 100.00 \\
U6 & \ 975  & 1.4 & 100.00 & 2.1 & \ 97.32 \\
U7 & 1107 & 2.6 & \ 97.34 & 1.8 & 100.00 \\
U8 & 1098 & 5.4 & 100.00 & 0.6 & \ 95.21 \\
U9 & 1100 & 1.7 & \ 97.66 & 1.9 & 100.00 \\
U10 & 1208 & 5.1 &\ 96.38 & 1.6 & \ 98.01 \\
\hline \hline
\end{tabular}
\end{adjustbox}
\caption{\TL{Data extracted from experimental data are listed in the table. The baseline stride duration represents the average cadence measured during the haptic-off trials. Time for alignment and Alignment percentages are then reported for fast ($-10\%$) and slow cadence condition ($+10\%$).}}
\label{tab:constant}
\end{table}
}

%% file: table_botFollower.tex
\begin{table}[]
\centering
\begin{adjustbox}{center, width=\columnwidth}  
\begin{tabular}{cccc} 
\hline \hline 
\textbf{User} & \textbf{\begin{tabular}[c]{@{}c@{}}Baseline\\ Stride Duration\\ (ms/stride)\end{tabular}} & \textbf{\begin{tabular}[c]{@{}c@{}}Time for\\ Alignment\\ (s)\end{tabular}} &  \textbf{\begin{tabular}[c]{@{}c@{}}Alignment\\ Percentage\\ \%\end{tabular}} \\
\hline
U1  & 1206 & 4.0 & 100\% \\
U2  & 1336 & 4.7 & 100\% \\
U3  & \ 955  & 3.6 & \ 99\%  \\
U4  & \ 998  & 1.0 & \ 97\%  \\
U5  & \ 905  & 3.8 & \ 99\%  \\
U6  & \ 920  & 0.3 & \ 98\%  \\
U7  & \ 879  & 4.8 & \ 99\%  \\
U8  & 1006 & 2.9 &  \ 98\%  \\
U9  & 1060 & 4.2 & \ 95\%  \\
U10 & \ 986  & 5.6 & 100\% \\
U11 & \ 881  & 4.5 & \ 99\%  \\
U12 & \ 885  & 4.5 & \ 99\%  \\
U13 & \ 980  & 1.5 & \ 99\%  \\
U14 & 1009 & 4.3 & \ 98\%  \\
U15 & 1168 & 1.5 & \ 94\% \\
U16 & 1077 & 5.5 & \ 99\%  \\
U17 & \ 961  & 2.1 & \ 98\%  \\
U18 & 1058 & 2.5 & 100\% \\
U19 & 1297 & 2.7 &\  99\%  \\
U20 & 1251 & 5.4 & \ 99\% \\
\hline \hline
\end{tabular}                                   
\end{adjustbox}
\caption{Artificial Leader. For each user we report data from the experimental validation. The first column details the user's comfortable cadence (\iec  cadence without haptic suggestion). In the second column we report the total time needed to align the actual cadence to the displayed one. The last column describes the percentage of time in which the subject was aligned with the suggested rhythm.}
\label{tab:botFollower}
\end{table}

%% file: figures/botFollower2.tex
%
%
\definecolor{mycolor1}{rgb}{0.00000,0.44700,0.74100}%
\definecolor{mycolor2}{rgb}{0.85000,0.32500,0.09800}%
\begin{tikzpicture}
\centering
\begin{axis}[%
width=\columnwidth,
height=0.8\columnwidth,,
xlabel=Time (s),
ylabel=Stride Duration (ms/stride),
xmin=0,
xmax=600,
ymin=850,
ymax=1300,
axis background/.style={fill=white},
legend style={draw=none},
legend columns=2, 
legend style={ font=\scriptsize},
legend cell align={left},
]
\addplot [color=mycolor1]
  table[row sep=crcr]{%
0	1090\\
1.029	1090\\
2.088	1090\\
3.118	1090\\
4.188	1090\\
5.25	1068\\
6.268	1046\\
6.68	1024\\
7.358	1002\\
8.397	980\\
9.468	980\\
10.538	980\\
11.598	980\\
12.649	980\\
13.679	980\\
14.706	980\\
15.771	980\\
16.818	980\\
17.848	980\\
18.851	980\\
19.876	980\\
20.906	980\\
21.917	980\\
22.938	980\\
23.968	980\\
24.994	980\\
26.038	980\\
27.058	980\\
28.071	980\\
29.118	980\\
30.153	980\\
31.163	980\\
32.199	980\\
33.228	980\\
34.288	980\\
35.346	980\\
36.382	970\\
37.417	960\\
37.681	950\\
38.448	940\\
39.477	930\\
40.548	930\\
41.573	930\\
42.609	930\\
43.649	930\\
44.698	930\\
45.756	930\\
46.808	930\\
47.868	930\\
48.904	930\\
49.917	930\\
50.947	930\\
52.016	930\\
53.068	930\\
54.088	930\\
55.118	930\\
56.148	930\\
57.178	930\\
58.208	930\\
59.229	930\\
60.288	930\\
61.358	930\\
62.356	930\\
63.378	930\\
64.426	930\\
65.428	930\\
66.496	930\\
67.527	930\\
68.557	930\\
69.568	930\\
70.609	928\\
71.659	926\\
71.685	924\\
72.689	922\\
73.717	920\\
74.789	920\\
75.839	920\\
76.862	920\\
77.908	920\\
78.946	920\\
79.949	920\\
80.991	920\\
82.029	920\\
83.09	920\\
84.12	920\\
85.159	920\\
86.158	920\\
87.203	920\\
88.228	920\\
89.238	920\\
90.249	920\\
91.288	920\\
92.358	920\\
93.399	920\\
94.494	920\\
95.538	920\\
96.568	920\\
97.624	920\\
98.668	920\\
99.728	920\\
100.778	920\\
101.838	920\\
102.888	920\\
103.938	930\\
104.989	940\\
105.683	950\\
106.018	960\\
107.098	970\\
108.166	970\\
109.201	970\\
110.286	970\\
111.344	970\\
112.445	970\\
113.468	970\\
114.548	970\\
115.608	970\\
116.676	970\\
117.738	970\\
118.806	970\\
119.898	970\\
120.966	970\\
122.037	970\\
123.088	970\\
124.138	970\\
125.226	970\\
126.272	970\\
127.338	970\\
128.437	970\\
129.516	970\\
130.571	970\\
131.702	970\\
132.758	970\\
133.758	970\\
134.679	970\\
135.586	986\\
136.538	1002\\
136.681	1018\\
137.498	1034\\
138.47	1050\\
139.548	1050\\
140.659	1050\\
141.748	1050\\
142.838	1050\\
143.878	1050\\
144.908	1050\\
145.978	1050\\
147.028	1050\\
148.088	1050\\
149.116	1050\\
150.159	1050\\
151.208	1050\\
152.228	1050\\
153.278	1050\\
154.341	1050\\
155.358	1050\\
156.421	1050\\
157.449	1050\\
158.518	1050\\
159.56	1050\\
160.62	1050\\
161.628	1050\\
162.708	1050\\
163.728	1050\\
164.798	1050\\
165.876	1050\\
166.908	1050\\
167.958	1050\\
169.054	1030\\
170.079	1010\\
170.688	990\\
171.127	970\\
172.188	950\\
173.196	950\\
174.197	950\\
175.065	950\\
175.978	950\\
176.908	950\\
177.828	950\\
178.788	950\\
179.758	950\\
180.688	950\\
181.628	950\\
182.548	950\\
183.478	950\\
184.408	950\\
185.359	950\\
186.325	950\\
187.278	950\\
188.258	950\\
189.208	950\\
190.138	950\\
191.077	950\\
192.036	950\\
192.958	950\\
193.88	950\\
194.829	950\\
195.81	950\\
196.778	950\\
197.749	950\\
198.678	950\\
199.689	950\\
200.629	950\\
201.588	940\\
202.52	930\\
202.683	920\\
203.449	910\\
204.388	900\\
205.262	900\\
206.119	900\\
207.015	900\\
207.908	900\\
208.838	900\\
209.738	900\\
210.619	900\\
211.546	900\\
212.459	900\\
213.396	900\\
214.291	900\\
215.189	900\\
216.088	900\\
217.028	900\\
217.919	900\\
218.819	900\\
219.721	900\\
220.639	900\\
221.538	900\\
222.446	900\\
223.318	900\\
224.168	900\\
225.049	900\\
225.918	900\\
226.788	900\\
227.688	900\\
228.598	900\\
229.54	900\\
230.419	900\\
231.317	900\\
232.218	900\\
233.126	900\\
234.026	928\\
234.919	956\\
235.684	984\\
235.818	1012\\
236.728	1040\\
237.658	1040\\
238.721	1040\\
239.826	1040\\
240.888	1040\\
241.928	1040\\
242.967	1040\\
244.02	1040\\
245.061	1040\\
246.133	1040\\
247.188	1040\\
248.216	1040\\
249.248	1040\\
250.316	1040\\
251.348	1040\\
252.426	1040\\
253.466	1040\\
254.501	1040\\
255.516	1040\\
256.557	1040\\
257.588	1040\\
258.648	1040\\
259.668	1040\\
260.708	1040\\
261.843	1040\\
262.79	1040\\
263.828	1034\\
264.838	1028\\
265.687	1022\\
265.888	1016\\
266.948	1010\\
267.97	1010\\
268.978	1010\\
269.988	1010\\
271	1010\\
272.016	1010\\
272.989	1010\\
274.027	1010\\
274.988	1010\\
275.997	1010\\
276.98	1010\\
277.988	1010\\
279.039	1010\\
280.018	1010\\
281.076	1010\\
282.088	1010\\
283.098	1010\\
284.108	1010\\
285.127	1010\\
286.148	1010\\
287.12	1010\\
288.112	1010\\
289.138	1010\\
290.156	1010\\
291.178	1010\\
292.198	1010\\
293.228	1010\\
294.228	1028\\
295.238	1046\\
295.69	1064\\
296.247	1082\\
297.254	1100\\
298.288	1100\\
299.419	1100\\
300.578	1100\\
301.678	1100\\
302.789	1100\\
303.93	1100\\
305.047	1100\\
306.136	1100\\
307.217	1100\\
308.228	1100\\
309.308	1100\\
310.468	1100\\
311.679	1100\\
312.785	1100\\
313.877	1100\\
314.958	1100\\
316.059	1100\\
317.188	1100\\
318.308	1100\\
319.418	1100\\
320.468	1100\\
321.588	1100\\
322.708	1100\\
323.786	1100\\
324.858	1100\\
325.958	1100\\
327.058	1094\\
328.166	1088\\
328.69	1082\\
329.238	1076\\
330.338	1070\\
331.388	1070\\
332.518	1070\\
333.577	1070\\
334.627	1070\\
335.682	1070\\
336.747	1070\\
337.818	1070\\
338.908	1070\\
339.987	1070\\
341.058	1070\\
342.159	1070\\
343.198	1070\\
344.271	1070\\
345.366	1070\\
346.398	1070\\
347.447	1070\\
348.498	1070\\
349.596	1070\\
350.658	1070\\
351.736	1070\\
352.788	1070\\
353.868	1070\\
354.926	1070\\
356.017	1070\\
357.078	1070\\
358.12	1070\\
359.167	1070\\
360.276	1062\\
361.346	1054\\
361.687	1046\\
362.446	1038\\
363.488	1030\\
364.507	1030\\
365.586	1030\\
366.608	1030\\
367.559	1030\\
368.568	1030\\
369.621	1030\\
370.671	1030\\
371.728	1030\\
372.759	1030\\
373.772	1030\\
374.785	1030\\
375.775	1030\\
376.782	1030\\
377.818	1030\\
378.841	1030\\
379.877	1030\\
380.917	1030\\
381.961	1030\\
382.958	1030\\
384.027	1030\\
385.049	1030\\
386.078	1030\\
387.108	1030\\
388.107	1030\\
389.138	1030\\
390.17	1008\\
391.203	986\\
391.695	964\\
392.23	942\\
393.252	920\\
394.275	920\\
395.08	920\\
395.99	920\\
396.879	920\\
397.787	920\\
398.684	920\\
399.595	920\\
400.514	920\\
401.437	920\\
402.365	920\\
403.269	920\\
404.214	920\\
405.131	920\\
406.067	920\\
406.982	920\\
407.929	920\\
408.854	920\\
409.816	920\\
410.635	920\\
411.606	920\\
412.509	920\\
413.354	920\\
414.283	920\\
415.232	920\\
416.282	920\\
417.125	920\\
418.074	920\\
418.974	920\\
419.916	920\\
420.788	920\\
421.739	920\\
422.649	920\\
423.555	922\\
424.468	924\\
424.69	926\\
425.377	928\\
426.316	930\\
427.227	930\\
428.158	930\\
429.096	930\\
429.987	930\\
430.907	930\\
431.897	930\\
432.828	930\\
433.787	930\\
434.707	930\\
435.647	930\\
436.556	930\\
437.477	930\\
438.388	930\\
439.317	930\\
440.223	930\\
441.178	930\\
442.108	930\\
443.045	930\\
443.967	930\\
444.879	930\\
445.835	930\\
446.787	930\\
447.725	930\\
448.647	930\\
449.587	930\\
450.528	930\\
451.427	930\\
452.365	930\\
453.287	930\\
454.221	930\\
455.116	930\\
456.047	930\\
457.01	938\\
457.98	946\\
458.693	954\\
458.947	962\\
460.008	970\\
461.114	970\\
462.137	970\\
463.165	970\\
464.197	970\\
465.228	970\\
466.26	970\\
467.278	970\\
468.277	970\\
469.341	970\\
470.347	970\\
471.387	970\\
472.427	970\\
473.458	970\\
474.462	970\\
475.526	970\\
476.556	970\\
477.567	970\\
478.597	970\\
479.627	970\\
480.644	970\\
481.676	970\\
482.707	970\\
483.717	970\\
484.768	970\\
485.827	970\\
486.827	970\\
487.807	970\\
488.857	966\\
489.896	962\\
490.695	958\\
490.947	954\\
491.935	950\\
492.926	950\\
494.027	950\\
495.087	950\\
496.125	950\\
497.167	950\\
498.257	950\\
499.327	950\\
500.366	950\\
501.419	950\\
502.468	950\\
503.503	950\\
504.569	950\\
505.63	950\\
506.639	950\\
507.699	950\\
508.698	950\\
509.766	950\\
510.833	950\\
511.912	950\\
512.959	950\\
513.986	950\\
515.046	950\\
516.043	950\\
517.077	950\\
518.09	950\\
519.12	972\\
520.127	994\\
520.695	1016\\
521.13	1038\\
522.175	1060\\
523.193	1060\\
524.239	1060\\
525.264	1060\\
526.28	1060\\
527.313	1060\\
528.299	1060\\
529.3	1060\\
530.32	1060\\
531.313	1060\\
532.31	1060\\
533.31	1060\\
534.321	1060\\
535.319	1060\\
536.328	1060\\
537.359	1060\\
538.352	1060\\
539.341	1060\\
540.316	1060\\
541.32	1060\\
542.3	1060\\
543.304	1060\\
544.31	1060\\
545.301	1060\\
546.318	1060\\
547.311	1060\\
548.32	1060\\
549.332	1052\\
550.331	1044\\
550.7	1036\\
551.329	1028\\
552.35	1020\\
553.374	1020\\
554.35	1020\\
555.352	1020\\
556.351	1020\\
557.36	1020\\
558.381	1020\\
559.391	1020\\
560.399	1020\\
561.391	1020\\
562.437	1020\\
563.48	1020\\
564.462	1020\\
565.468	1020\\
566.423	1020\\
567.442	1020\\
568.459	1020\\
569.45	1020\\
570.431	1020\\
571.42	1020\\
572.445	1020\\
573.452	1020\\
574.467	1020\\
575.439	1020\\
576.441	1020\\
577.432	1020\\
578.44	1020\\
579.471	1008\\
580.479	996\\
580.694	984\\
581.455	972\\
582.45	960\\
583.458	960\\
584.46	960\\
585.442	960\\
586.44	960\\
587.442	960\\
588.439	960\\
589.431	960\\
590.431	960\\
};
\addlegendentry{Artificial Leader}

\addplot [color=mycolor2]
  table[row sep=crcr]{%
0	1040\\
1.029	1033.33333333333\\
2.088	1044\\
3.118	1046\\
4.188	1046\\
5.25	1046\\
6.268	1048\\
6.68	1040\\
7.358	1046\\
8.397	1054\\
9.468	1060\\
10.538	1056\\
11.598	1056\\
12.649	1046\\
13.679	1042\\
14.706	1038\\
15.771	1036\\
16.818	1028\\
17.848	1024\\
18.851	1018\\
19.876	1016\\
20.906	1016\\
21.917	1022\\
22.938	1024\\
23.968	1026\\
24.994	1026\\
26.038	1026\\
27.058	1026\\
28.071	1028\\
29.118	1026\\
30.153	1028\\
31.163	1028\\
32.199	1034\\
33.228	1038\\
34.288	1042\\
35.346	1044\\
36.382	1048\\
37.417	1042\\
37.681	1036\\
38.448	1034\\
39.477	1032\\
40.548	1030\\
41.573	1034\\
42.609	1038\\
43.649	1040\\
44.698	1044\\
45.756	1048\\
46.808	1042\\
47.868	1038\\
48.904	1038\\
49.917	1032\\
50.947	1030\\
52.016	1036\\
53.068	1036\\
54.088	1036\\
55.118	1036\\
56.148	1030\\
57.178	1026\\
58.208	1030\\
59.229	1030\\
60.288	1028\\
61.358	1028\\
62.356	1032\\
63.378	1028\\
64.426	1026\\
65.428	1028\\
66.496	1032\\
67.527	1030\\
68.557	1030\\
69.568	1030\\
70.609	1030\\
71.659	1030\\
71.685	1032\\
72.689	1036\\
73.717	1042\\
74.789	1042\\
75.839	1042\\
76.862	1038\\
77.908	1034\\
78.946	1030\\
79.949	1030\\
80.991	1030\\
82.029	1030\\
83.09	1034\\
84.12	1030\\
85.159	1028\\
86.158	1022\\
87.203	1024\\
88.228	1018\\
89.238	1022\\
90.249	1030\\
91.288	1038\\
92.358	1044\\
93.399	1052\\
94.494	1052\\
95.538	1048\\
96.568	1048\\
97.624	1046\\
98.668	1050\\
99.728	1052\\
100.778	1052\\
101.838	1050\\
102.888	1050\\
103.938	1046\\
104.989	1046\\
105.683	1048\\
106.018	1050\\
107.098	1050\\
108.166	1052\\
109.201	1056\\
110.286	1064\\
111.344	1060\\
112.445	1064\\
113.468	1062\\
114.548	1062\\
115.608	1056\\
116.676	1064\\
117.738	1066\\
118.806	1070\\
119.898	1068\\
120.966	1068\\
122.037	1064\\
123.088	1062\\
124.138	1056\\
125.226	1058\\
126.272	1060\\
127.338	1066\\
128.437	1066\\
129.516	1078\\
130.571	1080\\
131.702	1066\\
132.758	1032\\
133.758	1000\\
134.679	966\\
135.586	936\\
136.538	928\\
136.681	940\\
137.498	976\\
138.47	1012\\
139.548	1044\\
140.659	1068\\
141.748	1080\\
142.838	1070\\
143.878	1058\\
144.908	1052\\
145.978	1046\\
147.028	1044\\
148.088	1044\\
149.116	1040\\
150.159	1036\\
151.208	1034\\
152.228	1038\\
153.278	1036\\
154.341	1036\\
155.358	1036\\
156.421	1040\\
157.449	1040\\
158.518	1044\\
159.56	1044\\
160.62	1048\\
161.628	1044\\
162.708	1048\\
163.728	1048\\
164.798	1050\\
165.876	1048\\
166.908	1052\\
167.958	1050\\
169.054	1050\\
170.079	1050\\
170.688	1050\\
171.127	1040\\
172.188	1014\\
173.196	992\\
174.197	968\\
175.065	944\\
175.978	928\\
176.908	934\\
177.828	936\\
178.788	942\\
179.758	944\\
180.688	940\\
181.628	934\\
182.548	932\\
183.478	930\\
184.408	932\\
185.359	942\\
186.325	952\\
187.278	954\\
188.258	956\\
189.208	954\\
190.138	948\\
191.077	940\\
192.036	934\\
192.958	934\\
193.88	938\\
194.829	948\\
195.81	954\\
196.778	958\\
197.749	964\\
198.678	964\\
199.689	956\\
200.629	948\\
201.588	944\\
202.52	936\\
202.683	930\\
203.449	918\\
204.388	902\\
205.262	892\\
206.119	890\\
207.015	888\\
207.908	888\\
208.838	896\\
209.738	906\\
210.619	906\\
211.546	906\\
212.459	908\\
213.396	910\\
214.291	908\\
215.189	902\\
216.088	902\\
217.028	902\\
217.919	906\\
218.819	906\\
219.721	904\\
220.639	902\\
221.538	898\\
222.446	886\\
223.318	880\\
224.168	872\\
225.049	868\\
225.918	872\\
226.788	884\\
227.688	894\\
228.598	900\\
229.54	898\\
230.419	900\\
231.317	898\\
232.218	892\\
233.126	896\\
234.026	900\\
234.919	898\\
235.684	898\\
235.818	906\\
236.728	940\\
237.658	978\\
238.721	1014\\
239.826	1044\\
240.888	1060\\
241.928	1054\\
242.967	1048\\
244.02	1046\\
245.061	1044\\
246.133	1046\\
247.188	1044\\
248.216	1036\\
249.248	1036\\
250.316	1044\\
251.348	1046\\
252.426	1040\\
253.466	1044\\
254.501	1042\\
255.516	1032\\
256.557	1034\\
257.588	1038\\
258.648	1038\\
259.668	1032\\
260.708	1034\\
261.843	1030\\
262.79	1032\\
263.828	1030\\
264.838	1032\\
265.687	1030\\
265.888	1032\\
266.948	1030\\
267.97	1024\\
268.978	1018\\
269.988	1016\\
271	998\\
272.016	1006\\
272.989	996\\
274.027	1000\\
274.988	992\\
275.997	1002\\
276.98	994\\
277.988	1004\\
279.039	1006\\
280.018	1016\\
281.076	1018\\
282.088	1016\\
283.098	1016\\
284.108	1010\\
285.127	1000\\
286.148	994\\
287.12	996\\
288.112	1000\\
289.138	1004\\
290.156	1012\\
291.178	1018\\
292.198	1016\\
293.228	1012\\
294.228	1008\\
295.238	1006\\
295.69	1004\\
296.247	1008\\
297.254	1034\\
298.288	1068\\
299.419	1084\\
300.578	1106\\
301.678	1126\\
302.789	1120\\
303.93	1104\\
305.047	1098\\
306.136	1082\\
307.217	1070\\
308.228	1084\\
309.308	1108\\
310.468	1114\\
311.679	1124\\
312.785	1128\\
313.877	1114\\
314.958	1096\\
316.059	1102\\
317.188	1106\\
318.308	1104\\
319.418	1104\\
320.468	1100\\
321.588	1094\\
322.708	1088\\
323.786	1088\\
324.858	1090\\
325.958	1088\\
327.058	1088\\
328.166	1088\\
328.69	1090\\
329.238	1084\\
330.338	1082\\
331.388	1078\\
332.518	1076\\
333.577	1068\\
334.627	1066\\
335.682	1064\\
336.747	1066\\
337.818	1070\\
338.908	1074\\
339.987	1076\\
341.058	1070\\
342.159	1072\\
343.198	1070\\
344.271	1064\\
345.366	1058\\
346.398	1062\\
347.447	1060\\
348.498	1062\\
349.596	1066\\
350.658	1066\\
351.736	1066\\
352.788	1062\\
353.868	1062\\
354.926	1062\\
356.017	1060\\
357.078	1060\\
358.12	1064\\
359.167	1066\\
360.276	1072\\
361.346	1084\\
361.687	1078\\
362.446	1066\\
363.488	1062\\
364.507	1048\\
365.586	1022\\
366.608	1016\\
367.559	1020\\
368.568	1014\\
369.621	1024\\
370.671	1034\\
371.728	1036\\
372.759	1028\\
373.772	1020\\
374.785	1006\\
375.775	1006\\
376.782	1008\\
377.818	1016\\
378.841	1026\\
379.877	1030\\
380.917	1030\\
381.961	1034\\
382.958	1032\\
384.027	1026\\
385.049	1026\\
386.078	1022\\
387.108	1018\\
388.107	1020\\
389.138	1022\\
390.17	1024\\
391.203	1026\\
391.695	1026\\
392.23	1014\\
393.252	976\\
394.275	952\\
395.08	930\\
395.99	902\\
396.879	886\\
397.787	898\\
398.684	902\\
399.595	902\\
400.514	908\\
401.437	912\\
402.365	918\\
403.269	918\\
404.214	920\\
405.131	924\\
406.067	924\\
406.982	918\\
407.929	910\\
408.854	912\\
409.816	912\\
410.635	912\\
411.606	908\\
412.509	912\\
413.354	916\\
414.283	920\\
415.232	926\\
416.282	932\\
417.125	934\\
418.074	924\\
418.974	918\\
419.916	910\\
420.788	912\\
421.739	914\\
422.649	920\\
423.555	920\\
424.468	922\\
424.69	920\\
425.377	916\\
426.316	916\\
427.227	918\\
428.158	914\\
429.096	918\\
429.987	932\\
430.907	934\\
431.897	938\\
432.828	944\\
433.787	940\\
434.707	928\\
435.647	924\\
436.556	918\\
437.477	914\\
438.388	914\\
439.317	918\\
440.223	924\\
441.178	926\\
442.108	926\\
443.045	928\\
443.967	928\\
444.879	930\\
445.835	932\\
446.787	936\\
447.725	938\\
448.647	936\\
449.587	928\\
450.528	926\\
451.427	924\\
452.365	920\\
453.287	916\\
454.221	922\\
455.116	924\\
456.047	926\\
457.01	930\\
457.98	944\\
458.693	972\\
458.947	998\\
460.008	1020\\
461.114	1038\\
462.137	1048\\
463.165	1038\\
464.197	1028\\
465.228	1022\\
466.26	1022\\
467.278	1024\\
468.277	1022\\
469.341	1024\\
470.347	1028\\
471.387	1028\\
472.427	1024\\
473.458	1030\\
474.462	1030\\
475.526	1028\\
476.556	1030\\
477.567	1030\\
478.597	1024\\
479.627	1022\\
480.644	1022\\
481.676	1020\\
482.707	1024\\
483.717	1028\\
484.768	1028\\
485.827	1018\\
486.827	1020\\
487.807	1018\\
488.857	1016\\
489.896	1022\\
490.695	1028\\
490.947	1022\\
491.935	1034\\
492.926	1038\\
494.027	1036\\
495.087	1048\\
496.125	1060\\
497.167	1054\\
498.257	1052\\
499.327	1054\\
500.366	1052\\
501.419	1048\\
502.468	1044\\
503.503	1050\\
504.569	1042\\
505.63	1034\\
506.639	1036\\
507.699	1040\\
508.698	1038\\
509.766	1048\\
510.833	1056\\
511.912	1050\\
512.959	1044\\
513.986	1036\\
515.046	1028\\
516.043	1018\\
517.077	1018\\
518.09	1020\\
519.12	1024\\
520.127	1020\\
520.695	1028\\
521.13	1030\\
522.175	1026\\
523.193	1022\\
524.239	1028\\
525.264	1024\\
526.28	1016\\
527.313	1014\\
528.299	1010\\
529.3	1002\\
530.32	1000\\
531.313	1002\\
532.31	1000\\
533.31	998\\
534.321	1000\\
535.319	1002\\
536.328	1004\\
537.359	1000\\
538.352	994\\
539.341	992\\
540.316	986\\
541.32	984\\
542.3	986\\
543.304	996\\
544.31	998\\
545.301	998\\
546.318	998\\
547.311	1000\\
548.32	1000\\
549.332	1002\\
550.331	1004\\
550.7	1006\\
551.329	1008\\
552.35	1002\\
553.374	998\\
554.35	998\\
555.352	998\\
556.351	1000\\
557.36	1004\\
558.381	1008\\
559.391	1010\\
560.399	1012\\
561.391	1016\\
562.437	1014\\
563.48	1012\\
564.462	1002\\
565.468	1000\\
566.423	994\\
567.442	996\\
568.459	992\\
569.45	998\\
570.431	998\\
571.42	992\\
572.445	992\\
573.452	994\\
574.467	994\\
575.439	992\\
576.441	998\\
577.432	1002\\
578.44	1002\\
579.471	1002\\
580.479	998\\
580.694	998\\
581.455	994\\
582.45	994\\
583.458	994\\
584.46	998\\
585.442	992\\
586.44	990\\
587.442	992\\
588.439	992\\
589.431	993.333333333333\\
590.431	990\\
};
\addlegendentry{User}

\addplot[area legend, draw=none, fill=blue, fill opacity=0.1]
table[row sep=crcr] {%
x	y\\
132		1018.5\\
133.758	1018.5\\
134.679	1018.5\\
135.586	1035.3\\
136.538	1052.1\\
136.681	1068.9\\
137.498	1085.7\\
138.47	1102.5\\
139.548	1102.5\\
140.659	1102.5\\
141.748	1102.5\\
142.838	1102.5\\
143.878	1102.5\\
144.908	1102.5\\
145.978	1102.5\\
147.028	1102.5\\
148.088	1102.5\\
149.116	1102.5\\
150.159	1102.5\\
151.208	1102.5\\
152.228	1102.5\\
153.278	1102.5\\
154.341	1102.5\\
155.358	1102.5\\
156.421	1102.5\\
157.449	1102.5\\
158.518	1102.5\\
159.56	1102.5\\
160.62	1102.5\\
161.628	1102.5\\
162.708	1102.5\\
163.728	1102.5\\
164.798	1102.5\\
165.876	1102.5\\
166.908	1102.5\\
167.958	1102.5\\
169.054	1081.5\\
170.079	1060.5\\
170.688	1039.5\\
171.127	1018.5\\
172.188	997.5\\
173.196	997.5\\
174.197	997.5\\
175.065	997.5\\
175.978	997.5\\
176.908	997.5\\
177.828	997.5\\
178.788	997.5\\
179.758	997.5\\
180.688	997.5\\
181.628	997.5\\
182.548	997.5\\
183.478	997.5\\
184.408	997.5\\
185.359	997.5\\
186.325	997.5\\
187.278	997.5\\
188.258	997.5\\
189.208	997.5\\
190.138	997.5\\
191.077	997.5\\
192.036	997.5\\
192.958	997.5\\
193.88	997.5\\
194.829	997.5\\
195.81	997.5\\
196.778	997.5\\
197.749	997.5\\
198.678	997.5\\
199.689	997.5\\
200.629	997.5\\
201.588	987\\
202.52	976.5\\
202.683	966\\
203.449	955.5\\
204.388	945\\
205.262	945\\
206.119	945\\
207.015	945\\
207.908	945\\
208.838	945\\
209.738	945\\
210.619	945\\
211.546	945\\
212.459	945\\
213.396	945\\
214.291	945\\
215.189	945\\
216.088	945\\
217.028	945\\
217.919	945\\
218.819	945\\
219.721	945\\
220.639	945\\
221.538	945\\
222.446	945\\
223.318	945\\
224.168	945\\
225.049	945\\
225.918	945\\
226.788	945\\
227.688	945\\
228.598	945\\
229.54	945\\
230.419	945\\
231.317	945\\
232.218	945\\
233.126	945\\
234.026	974.4\\
234.919	1003.8\\
235.684	1033.2\\
235.818	1062.6\\
236.728	1092\\
237.658	1092\\
238.721	1092\\
239.826	1092\\
240.888	1092\\
241.928	1092\\
242.967	1092\\
244.02	1092\\
245.061	1092\\
246.133	1092\\
247.188	1092\\
248.216	1092\\
249.248	1092\\
250.316	1092\\
251.348	1092\\
252.426	1092\\
253.466	1092\\
254.501	1092\\
255.516	1092\\
256.557	1092\\
257.588	1092\\
258.648	1092\\
259.668	1092\\
260.708	1092\\
261.843	1092\\
262.79	1092\\
263.828	1085.7\\
264.838	1079.4\\
265.687	1073.1\\
265.888	1066.8\\
266.948	1060.5\\
267.97	1060.5\\
268.978	1060.5\\
269.988	1060.5\\
271	1060.5\\
272.016	1060.5\\
272.989	1060.5\\
274.027	1060.5\\
274.988	1060.5\\
275.997	1060.5\\
276.98	1060.5\\
277.988	1060.5\\
279.039	1060.5\\
280.018	1060.5\\
281.076	1060.5\\
282.088	1060.5\\
283.098	1060.5\\
284.108	1060.5\\
285.127	1060.5\\
286.148	1060.5\\
287.12	1060.5\\
288.112	1060.5\\
289.138	1060.5\\
290.156	1060.5\\
291.178	1060.5\\
292.198	1060.5\\
293.228	1060.5\\
294.228	1079.4\\
295.238	1098.3\\
295.69	1117.2\\
296.247	1136.1\\
297.254	1155\\
298.288	1155\\
299.419	1155\\
300.578	1155\\
301.678	1155\\
302.789	1155\\
303.93	1155\\
305.047	1155\\
306.136	1155\\
307.217	1155\\
308.228	1155\\
309.308	1155\\
310.468	1155\\
311.679	1155\\
312.785	1155\\
313.877	1155\\
314.958	1155\\
316.059	1155\\
317.188	1155\\
318.308	1155\\
319.418	1155\\
320.468	1155\\
321.588	1155\\
322.708	1155\\
323.786	1155\\
324.858	1155\\
325.958	1155\\
327.058	1148.7\\
328.166	1142.4\\
328.69	1136.1\\
329.238	1129.8\\
330.338	1123.5\\
331.388	1123.5\\
332.518	1123.5\\
333.577	1123.5\\
334.627	1123.5\\
335.682	1123.5\\
336.747	1123.5\\
337.818	1123.5\\
338.908	1123.5\\
339.987	1123.5\\
341.058	1123.5\\
342.159	1123.5\\
343.198	1123.5\\
344.271	1123.5\\
345.366	1123.5\\
346.398	1123.5\\
347.447	1123.5\\
348.498	1123.5\\
349.596	1123.5\\
350.658	1123.5\\
351.736	1123.5\\
352.788	1123.5\\
353.868	1123.5\\
354.926	1123.5\\
356.017	1123.5\\
357.078	1123.5\\
358.12	1123.5\\
359.167	1123.5\\
360.276	1115.1\\
361.346	1106.7\\
361.687	1098.3\\
362.446	1089.9\\
363.488	1081.5\\
364.507	1081.5\\
365.586	1081.5\\
366.608	1081.5\\
367.559	1081.5\\
368.568	1081.5\\
369.621	1081.5\\
370.671	1081.5\\
371.728	1081.5\\
372.759	1081.5\\
373.772	1081.5\\
374.785	1081.5\\
375.775	1081.5\\
376.782	1081.5\\
377.818	1081.5\\
378.841	1081.5\\
379.877	1081.5\\
380.917	1081.5\\
381.961	1081.5\\
382.958	1081.5\\
384.027	1081.5\\
385.049	1081.5\\
386.078	1081.5\\
387.108	1081.5\\
388.107	1081.5\\
389.138	1081.5\\
390.17	1058.4\\
391.203	1035.3\\
391.695	1012.2\\
392.23	989.1\\
393.252	966\\
394.275	966\\
395.08	966\\
395.99	966\\
396.879	966\\
397.787	966\\
398.684	966\\
399.595	966\\
400.514	966\\
401.437	966\\
402.365	966\\
403.269	966\\
404.214	966\\
405.131	966\\
406.067	966\\
406.982	966\\
407.929	966\\
408.854	966\\
409.816	966\\
410.635	966\\
411.606	966\\
412.509	966\\
413.354	966\\
414.283	966\\
415.232	966\\
416.282	966\\
417.125	966\\
418.074	966\\
418.974	966\\
419.916	966\\
420.788	966\\
421.739	966\\
422.649	966\\
423.555	968.1\\
424.468	970.2\\
424.69	972.3\\
425.377	974.4\\
426.316	976.5\\
427.227	976.5\\
428.158	976.5\\
429.096	976.5\\
429.987	976.5\\
430.907	976.5\\
431.897	976.5\\
432.828	976.5\\
433.787	976.5\\
434.707	976.5\\
435.647	976.5\\
436.556	976.5\\
437.477	976.5\\
438.388	976.5\\
439.317	976.5\\
440.223	976.5\\
441.178	976.5\\
442.108	976.5\\
443.045	976.5\\
443.967	976.5\\
444.879	976.5\\
445.835	976.5\\
446.787	976.5\\
447.725	976.5\\
448.647	976.5\\
449.587	976.5\\
450.528	976.5\\
451.427	976.5\\
452.365	976.5\\
453.287	976.5\\
454.221	976.5\\
455.116	976.5\\
456.047	976.5\\
456.047	883.5\\
455.116	883.5\\
454.221	883.5\\
453.287	883.5\\
452.365	883.5\\
451.427	883.5\\
450.528	883.5\\
449.587	883.5\\
448.647	883.5\\
447.725	883.5\\
446.787	883.5\\
445.835	883.5\\
444.879	883.5\\
443.967	883.5\\
443.045	883.5\\
442.108	883.5\\
441.178	883.5\\
440.223	883.5\\
439.317	883.5\\
438.388	883.5\\
437.477	883.5\\
436.556	883.5\\
435.647	883.5\\
434.707	883.5\\
433.787	883.5\\
432.828	883.5\\
431.897	883.5\\
430.907	883.5\\
429.987	883.5\\
429.096	883.5\\
428.158	883.5\\
427.227	883.5\\
426.316	883.5\\
425.377	881.6\\
424.69	879.7\\
424.468	877.8\\
423.555	875.9\\
422.649	874\\
421.739	874\\
420.788	874\\
419.916	874\\
418.974	874\\
418.074	874\\
417.125	874\\
416.282	874\\
415.232	874\\
414.283	874\\
413.354	874\\
412.509	874\\
411.606	874\\
410.635	874\\
409.816	874\\
408.854	874\\
407.929	874\\
406.982	874\\
406.067	874\\
405.131	874\\
404.214	874\\
403.269	874\\
402.365	874\\
401.437	874\\
400.514	874\\
399.595	874\\
398.684	874\\
397.787	874\\
396.879	874\\
395.99	874\\
395.08	874\\
394.275	874\\
393.252	874\\
392.23	894.9\\
391.695	915.8\\
391.203	936.7\\
390.17	957.6\\
389.138	978.5\\
388.107	978.5\\
387.108	978.5\\
386.078	978.5\\
385.049	978.5\\
384.027	978.5\\
382.958	978.5\\
381.961	978.5\\
380.917	978.5\\
379.877	978.5\\
378.841	978.5\\
377.818	978.5\\
376.782	978.5\\
375.775	978.5\\
374.785	978.5\\
373.772	978.5\\
372.759	978.5\\
371.728	978.5\\
370.671	978.5\\
369.621	978.5\\
368.568	978.5\\
367.559	978.5\\
366.608	978.5\\
365.586	978.5\\
364.507	978.5\\
363.488	978.5\\
362.446	986.1\\
361.687	993.7\\
361.346	1001.3\\
360.276	1008.9\\
359.167	1016.5\\
358.12	1016.5\\
357.078	1016.5\\
356.017	1016.5\\
354.926	1016.5\\
353.868	1016.5\\
352.788	1016.5\\
351.736	1016.5\\
350.658	1016.5\\
349.596	1016.5\\
348.498	1016.5\\
347.447	1016.5\\
346.398	1016.5\\
345.366	1016.5\\
344.271	1016.5\\
343.198	1016.5\\
342.159	1016.5\\
341.058	1016.5\\
339.987	1016.5\\
338.908	1016.5\\
337.818	1016.5\\
336.747	1016.5\\
335.682	1016.5\\
334.627	1016.5\\
333.577	1016.5\\
332.518	1016.5\\
331.388	1016.5\\
330.338	1016.5\\
329.238	1022.2\\
328.69	1027.9\\
328.166	1033.6\\
327.058	1039.3\\
325.958	1045\\
324.858	1045\\
323.786	1045\\
322.708	1045\\
321.588	1045\\
320.468	1045\\
319.418	1045\\
318.308	1045\\
317.188	1045\\
316.059	1045\\
314.958	1045\\
313.877	1045\\
312.785	1045\\
311.679	1045\\
310.468	1045\\
309.308	1045\\
308.228	1045\\
307.217	1045\\
306.136	1045\\
305.047	1045\\
303.93	1045\\
302.789	1045\\
301.678	1045\\
300.578	1045\\
299.419	1045\\
298.288	1045\\
297.254	1045\\
296.247	1027.9\\
295.69	1010.8\\
295.238	993.7\\
294.228	976.6\\
293.228	959.5\\
292.198	959.5\\
291.178	959.5\\
290.156	959.5\\
289.138	959.5\\
288.112	959.5\\
287.12	959.5\\
286.148	959.5\\
285.127	959.5\\
284.108	959.5\\
283.098	959.5\\
282.088	959.5\\
281.076	959.5\\
280.018	959.5\\
279.039	959.5\\
277.988	959.5\\
276.98	959.5\\
275.997	959.5\\
274.988	959.5\\
274.027	959.5\\
272.989	959.5\\
272.016	959.5\\
271	959.5\\
269.988	959.5\\
268.978	959.5\\
267.97	959.5\\
266.948	959.5\\
265.888	965.2\\
265.687	970.9\\
264.838	976.6\\
263.828	982.3\\
262.79	988\\
261.843	988\\
260.708	988\\
259.668	988\\
258.648	988\\
257.588	988\\
256.557	988\\
255.516	988\\
254.501	988\\
253.466	988\\
252.426	988\\
251.348	988\\
250.316	988\\
249.248	988\\
248.216	988\\
247.188	988\\
246.133	988\\
245.061	988\\
244.02	988\\
242.967	988\\
241.928	988\\
240.888	988\\
239.826	988\\
238.721	988\\
237.658	988\\
236.728	988\\
235.818	961.4\\
235.684	934.8\\
234.919	908.2\\
234.026	881.6\\
233.126	855\\
232.218	855\\
231.317	855\\
230.419	855\\
229.54	855\\
228.598	855\\
227.688	855\\
226.788	855\\
225.918	855\\
225.049	855\\
224.168	855\\
223.318	855\\
222.446	855\\
221.538	855\\
220.639	855\\
219.721	855\\
218.819	855\\
217.919	855\\
217.028	855\\
216.088	855\\
215.189	855\\
214.291	855\\
213.396	855\\
212.459	855\\
211.546	855\\
210.619	855\\
209.738	855\\
208.838	855\\
207.908	855\\
207.015	855\\
206.119	855\\
205.262	855\\
204.388	855\\
203.449	864.5\\
202.683	874\\
202.52	883.5\\
201.588	893\\
200.629	902.5\\
199.689	902.5\\
198.678	902.5\\
197.749	902.5\\
196.778	902.5\\
195.81	902.5\\
194.829	902.5\\
193.88	902.5\\
192.958	902.5\\
192.036	902.5\\
191.077	902.5\\
190.138	902.5\\
189.208	902.5\\
188.258	902.5\\
187.278	902.5\\
186.325	902.5\\
185.359	902.5\\
184.408	902.5\\
183.478	902.5\\
182.548	902.5\\
181.628	902.5\\
180.688	902.5\\
179.758	902.5\\
178.788	902.5\\
177.828	902.5\\
176.908	902.5\\
175.978	902.5\\
175.065	902.5\\
174.197	902.5\\
173.196	902.5\\
172.188	902.5\\
171.127	921.5\\
170.688	940.5\\
170.079	959.5\\
169.054	978.5\\
167.958	997.5\\
166.908	997.5\\
165.876	997.5\\
164.798	997.5\\
163.728	997.5\\
162.708	997.5\\
161.628	997.5\\
160.62	997.5\\
159.56	997.5\\
158.518	997.5\\
157.449	997.5\\
156.421	997.5\\
155.358	997.5\\
154.341	997.5\\
153.278	997.5\\
152.228	997.5\\
151.208	997.5\\
150.159	997.5\\
149.116	997.5\\
148.088	997.5\\
147.028	997.5\\
145.978	997.5\\
144.908	997.5\\
143.878	997.5\\
142.838	997.5\\
141.748	997.5\\
140.659	997.5\\
139.548	997.5\\
138.47	997.5\\
137.498	982.3\\
136.681	967.1\\
136.538	951.9\\
135.586	936.7\\
134.679	921.5\\
133.7 	921.5\\
132		921.5\\
}--cycle;
\addlegendentry{$\pm$4\% area}

\addplot [color=green]
  table[row sep=crcr]{%
132	850\\
132	1200\\
};
\addlegendentry{Haptics On/Off}

\addplot [color=green]
  table[row sep=crcr]{%
454	850\\
454	1200\\
};

\end{axis}

\end{tikzpicture}%

%% file: table_leaderFollower.tex
\begin{table*}[t]
\centering
\begin{adjustbox}{center, width=2\columnwidth}  
\begin{tabular}{cccccccc}
\hline \hline
\textbf{Leader} & \textbf{Follower} & \begin{tabular}[c]{@{}c@{}}\textbf{Leader B}\\ \textbf{Stride Duration}\\ \textbf{(ms/stride)}\end{tabular} & \begin{tabular}[c]{@{}c@{}}\textbf{Follower B}\\ \textbf{Stride Duration}\\ (\textbf{ms/stride)}\end{tabular} & \begin{tabular}[c]{@{}c@{}}\textbf{Mean H}\\ \textbf{Stride Duration}\\ \textbf{(ms/stride})\end{tabular} & \begin{tabular}[c]{@{}c@{}}\textbf{STD H}\\ \textbf{Stride Duration}\\ \textbf{(ms/stride)}\end{tabular} & \begin{tabular}[c]{@{}c@{}}\textbf{Time for}\\ \textbf{Alignment}\\ \textbf{(s)}\end{tabular} & \begin{tabular}[c]{@{}c@{}}\textbf{Alignment}\\ \textbf{Percentage}\\ \textbf{\%}\end{tabular} \\
\hline
U1	&	U2	&	1023	&	1141	&	1022	&	19	&	2.0	&	97\%	\\
U3	&	U4	&	1145	&	1035	&	1151	&	21	&	4.4	&	99\%	\\
U5	&	U6	&	1025	&	1136	&	1037	&	17	&	3.5	&	94\%	\\
U7	&	U8	&	1014	&	1081	&	1017	&	16	&	3.2	&	98\%	\\
U9	&	U10	&	1050	&	1082	&	1042	&	18	&	3.9	&	96\%	\\
U11	&	U12	&	1113	&	1106	&	1110	&	17	&	3.2	&	99\%	\\
U13	&	U14	& 	1020	&	1070	&	1024	&	16	&	2.5	&	95\%	\\
U15	&	U16	&	1086	&	1030	&	1078	&	20	&	3.9	&	98\%	\\
U17	&	U18	&	1055	&	1101	&	1059	&	23	&	3.0	&	94\%	\\
U19	&	U20	&	1018	&	1081	&	1030	&	20	&	3.5	&	98\%	\\
\hline
\multicolumn{2}{c}{\textbf{AVERAGE}} & & & & \textbf{18.7}& \textbf{3.31}  & \textbf{96.8\%} \\ \hline \hline
\end{tabular}
\end{adjustbox}
\caption{Human Leader. The table details data regarding the experiment with the human leader, where \textbf{B} and \textbf{H} stand for baseline and haptic-on condition, respectively. Mean cadences estimated in the first part of the trial (without haptics) are reported for both the users, and compared to the average walking rhythm during the phase with the haptic stimulation (Mean H). It is important to notice that the standard deviation of the mean cadence during the phase with haptic stimulation is comparable with the variability registered in the comfortable cadence walking.}
\label{tab:LeaderFollower}
\end{table*}

%% file: table_masterMaster.tex
\begin{table*}[]
\centering
\begin{adjustbox}{center, width=1.85\columnwidth}  
\begin{tabular}{cccccccccccc}
\hline \hline	
\multicolumn{1}{l}{\textbf{User1}} & \multicolumn{1}{l}{\textbf{User2}} & \multicolumn{1}{c}{\textbf{\begin{tabular}[c]{@{}c@{}}User1 B \\  Stride Duration \\ (ms/stride)\end{tabular}}} & \multicolumn{1}{c}{\textbf{\begin{tabular}[c]{@{}c@{}}User2 B \\ Stride Duration\\  (ms/stride)\end{tabular}}} & \multicolumn{1}{c}{\textbf{\begin{tabular}[c]{@{}c@{}}Mean H \\ Stride Duration\\ (ms/stride)\end{tabular}}} & \multicolumn{1}{c}{\textbf{\begin{tabular}[c]{@{}c@{}}STD H\\ Stride Duration\\ (ms/stride)\end{tabular}}} & \multicolumn{1}{c}{\textbf{\begin{tabular}[c]{@{}c@{}}Time for \\ Alignment \\(s)\end{tabular}}} & \multicolumn{1}{c}{\textbf{\begin{tabular}[c]{@{}c@{}} Alignment \\ percentage\\ \%\end{tabular}}} & \multicolumn{1}{c}{\textbf{\begin{tabular}[c]{@{}c@{}}User1 \\ Variation \\ \% \end{tabular}}} & \multicolumn{1}{c}{\textbf{\begin{tabular}[c]{@{}c@{}}User2 \\ Variation\\ \% \end{tabular}}} \\
\hline 
U1  & U2  & 1082 & 1158 & 1136 & 18 & 3.6 & 95\% & 5\% & 2\% \\
U3  & U4  & 1060 & 1002 & 1023 & 25 & 5.2 & 96\% & 4\% & 2\% \\
U5  & U6  & 1139 & 1136 & 1190 & 32 & 4.2 & 96\% & 4\% & 5\% \\
U7  & U8  & 1004 & 1070 & 1083 & 25 & 2.8 & 91\% & 7\% & 1\% \\
U9  & U10 & 1039 & \ 982  & 1046 & 31 & 1.1 & 92\% & 1\% & 6\% \\
U11 & U12 & 1067 & 1047 & 1090 & 35 & 5.6 & 93\% & 2\% & 4\% \\
U13 & U14 & 996  & 1167 & 1087 & 33 & 11.6 & 99\% & 8\% & 7\% \\
U15 & U16 & 1102 & 1119 & 1145 & 33 & 7.3 & 86\% & 4\% & 2\% \\
U17 & U18 & 1032 & 1103 & 1067 & 22 & 4.9 & 99\% & 3\% & 3\% \\
U19 & U20 & 1082 & 1197 & 1116 & 24 & 2.9 & 98\% & 3\% & 7\% \\
\hline
\multicolumn{2}{c}{\textbf{AVERAGE}	}	&	&	&	& \textbf{28}	& \textbf{4.9}	& \textbf{94.5\%}	& \textbf{4.1\%}	& \textbf{3.9\%}	\\ \hline \hline 
\end{tabular}
\end{adjustbox}
\caption{Peer-to-peer. The table details data regarding the final experiment, where \textbf{B} and \textbf{H} stand for baseline and haptic-on condition, respectively. Mean cadences estimated in the first part of the trial (without haptics) are reported for both the users, and compared to the average walking rhythm during the phase with the haptic stimulation (Mean H). Also in this experiment is important to notice that the standard deviation of the mean cadence during the phase with haptic stimulation is comparable with the variability registered in the comfortable cadence walking.}
\label{tab:peerToPeer}
\end{table*}

%% file: figures/peerToPeer.tex
%
\definecolor{mycolor1}{rgb}{0.00000,0.44700,0.74100}%
\definecolor{mycolor2}{rgb}{0.85000,0.32500,0.09800}%
\begin{tikzpicture}

\begin{axis}[%
width=0.95\columnwidth,
height=0.8\columnwidth,,
xlabel=Time (s),
ylabel=Stride Duration (ms/stride),
xmin=0,
xmax=400,
ymin=700,
ymax=1300,
axis background/.style={fill=white},
legend style={draw=none, align=left, font=\scriptsize}, 
legend columns=2, 
legend cell align={left},
]

\addplot [color=mycolor1]
  table[row sep=crcr]{%
0 		1023.\\
1.118	1023.3333333333333\\
1.826	1023\\
2.38	1023\\
2.683	1023\\
3.548	1023.666666666667\\
3.574	1023.33333333333\\
4.526	1050\\
4.728	1076.66666666667\\
5.398	1080\\
5.869	1076.66666666667\\
7.013	1073.33333333333\\
7.039	1070\\
7.426	1056.66666666667\\
8.209	1056.66666666667\\
8.249	1056.66666666667\\
9.134	1053.33333333333\\
9.328	1050\\
10.092	1066.66666666667\\
10.488	1073.33333333333\\
11.034	1073.33333333333\\
11.639	1076.66666666667\\
12.085	1076.66666666667\\
12.799	1066.66666666667\\
12.895	1056.66666666667\\
13.833	1056.66666666667\\
13.93	1056.66666666667\\
14.745	1050\\
15.058	1043.33333333333\\
15.728	1046.66666666667\\
16.2	1046.66666666667\\
16.78	1046.66666666667\\
17.369	1056.66666666667\\
18.549	1066.66666666667\\
19.699	1066.66666666667\\
20.851	1066.66666666667\\
21.989	1066.66666666667\\
23.17	1063.33333333333\\
24.299	1053.33333333333\\
24.536	1056.66666666667\\
24.536	1050\\
24.536	1046.66666666667\\
24.54	1043.33333333333\\
24.54	1043.33333333333\\
24.54	1033.33333333333\\
24.541	1036.66666666667\\
24.693	1036.66666666667\\
25.076	1043.33333333333\\
25.45	1053.33333333333\\
25.998	1056.66666666667\\
26.597	1056.66666666667\\
26.914	1056.66666666667\\
27.719	1056.66666666667\\
27.914	1056.66666666667\\
28.849	1056.66666666667\\
28.889	1056.66666666667\\
29.722	1053.33333333333\\
30.008	1046.66666666667\\
30.696	1043.33333333333\\
31.168	1043.33333333333\\
31.637	1043.33333333333\\
32.348	1046.66666666667\\
32.562	1053.33333333333\\
33.48	1063.33333333333\\
33.514	1073.33333333333\\
34.417	1076.66666666667\\
34.658	1080\\
35.332	1083.33333333333\\
35.796	1080\\
36.28	1076.66666666667\\
36.968	1080\\
37.179	1070\\
38.132	1060\\
38.238	1050\\
39.133	1043.33333333333\\
39.306	1036.66666666667\\
40.122	1050\\
40.428	1060\\
41.214	1063.33333333333\\
41.639	1063.33333333333\\
41.994	1070\\
42.769	1066.66666666667\\
42.864	1063.33333333333\\
43.773	1070\\
43.938	1076.66666666667\\
44.704	1073.33333333333\\
45.099	1073.33333333333\\
45.637	1070\\
46.28	1063.33333333333\\
46.583	1060\\
47.449	1063.33333333333\\
47.492	1066.66666666667\\
48.412	1066.66666666667\\
48.609	1066.66666666667\\
49.382	1063.33333333333\\
49.769	1056.66666666667\\
50.302	1053.33333333333\\
50.926	1056.66666666667\\
51.778	1060\\
52.069	1063.33333333333\\
52.309	1066.66666666667\\
53.229	1066.66666666667\\
54.051	1066.66666666667\\
54.379	1063.33333333333\\
55.076	1060\\
55.256	1053.33333333333\\
55.53	1046.66666666667\\
55.965	1040\\
56.659	1036.66666666667\\
56.925	1036.66666666667\\
57.799	1043.33333333333\\
57.835	1050\\
58.835	1056.66666666667\\
58.894	1063.33333333333\\
59.736	1063.33333333333\\
60.048	1060\\
60.686	1056.66666666667\\
61.18	1053.33333333333\\
61.625	1050\\
62.322	1053.33333333333\\
62.543	1056.66666666667\\
63.433	1060\\
63.499	1063.33333333333\\
64.363	1066.66666666667\\
64.647	1066.66666666667\\
65.343	1066.66666666667\\
65.819	1066.66666666667\\
66.274	1063.33333333333\\
66.96	1063.33333333333\\
67.223	1063.33333333333\\
68.133	1063.33333333333\\
68.139	1063.33333333333\\
69.033	1063.33333333333\\
69.286	1060\\
69.953	1063.33333333333\\
70.44	1066.66666666667\\
70.874	1070\\
71.6	1070\\
71.783	1070\\
72.723	1060\\
72.823	1050\\
73.683	1040\\
73.955	1036.66666666667\\
75.15	1036.66666666667\\
75.152	1040\\
75.949	1046.66666666667\\
76.289	1053.33333333333\\
76.843	1060\\
77.449	1063.33333333333\\
77.57	1066.66666666667\\
78.502	1066.66666666667\\
78.598	1066.66666666667\\
79.514	1063.33333333333\\
79.762	1060\\
80.374	1056.66666666667\\
80.398	1056.66666666667\\
80.919	1056.66666666667\\
81.461	1060\\
82.077	1060\\
82.393	1060\\
83.17	1053.33333333333\\
83.223	1046.66666666667\\
84.186	1036.66666666667\\
84.379	1030\\
85.118	1023.33333333333\\
85.598	1020\\
86.727	1016.66666666667\\
87.194	1023.33333333333\\
87.898	1030\\
87.994	1036.66666666667\\
88.965	1043.33333333333\\
89.041	1053.33333333333\\
89.29	1056.66666666667\\
89.857	1060\\
90.219	1063.33333333333\\
90.813	1063.33333333333\\
91.409	1060\\
92.509	1056.66666666667\\
92.61	1053.33333333333\\
93.054	1050\\
93.749	1053.33333333333\\
93.902	1056.66666666667\\
94.57	1060\\
94.91	1063.33333333333\\
95.474	1066.66666666667\\
96.077	1066.66666666667\\
96.53	1063.33333333333\\
97.21	1056.66666666667\\
97.439	1050\\
98.316	1046.66666666667\\
98.379	1043.33333333333\\
99.262	1043.33333333333\\
99.499	1050\\
99.549	1056.66666666667\\
100.242	1060\\
100.628	1060\\
101.779	1060\\
102.206	1056.66666666667\\
102.399	1053.33333333333\\
102.899	1046.66666666667\\
102.918	1043.33333333333\\
103.872	1050\\
104.086	1056.66666666667\\
104.752	1066.66666666667\\
105.296	1080\\
105.683	1080\\
106.46	1073.33333333333\\
106.638	1066.66666666667\\
107.598	1053.33333333333\\
107.617	1040\\
108.554	1040\\
108.749	1036.66666666667\\
109.502	1036.66666666667\\
109.858	1040\\
110.438	1036.66666666667\\
111.029	1033.33333333333\\
111.412	1033.33333333333\\
112.169	1030\\
112.38	1026.66666666667\\
113.318	1033.33333333333\\
113.369	1040\\
114.273	1043.33333333333\\
114.517	1046.66666666667\\
115.228	1046.66666666667\\
115.618	1043.33333333333\\
116.167	1040\\
116.78	1036.66666666667\\
117.133	1033.33333333333\\
117.969	1033.33333333333\\
118.086	1033.33333333333\\
119.041	1036.66666666667\\
119.109	1040\\
119.999	1043.33333333333\\
120.258	1046.66666666667\\
120.929	1053.33333333333\\
121.419	1056.66666666667\\
121.858	1053.33333333333\\
122.58	1050\\
122.808	1050\\
123.68	1050\\
123.748	1050\\
124.698	1050\\
124.859	1050\\
125.689	1050\\
126.01	1046.66666666667\\
126.599	1043.33333333333\\
127.169	1046.66666666667\\
127.538	1053.33333333333\\
128.388	1056.66666666667\\
128.671	1056.66666666667\\
129.487	1050\\
129.622	1043.33333333333\\
130.535	1036.66666666667\\
130.619	1030\\
131.786	1026.66666666667\\
131.82	1030\\
132.7	1036.66666666667\\
132.938	1040\\
133.824	1046.66666666667\\
133.959	1053.33333333333\\
134.429	1050\\
135.227	1043.33333333333\\
135.33	1000\\
136.189	953.333333333333\\
136.895	900\\
137.219	856.666666666667\\
137.713	850.666666666667\\
138.37	849.333333333333\\
138.992	843.333333333333\\
139.517	880\\
139.977	926.666666666667\\
140.619	970\\
141.146	1006.66666666667\\
141.708	1010\\
141.908	990\\
142.709	960\\
143.25	923.333333333333\\
143.76	893.333333333333\\
144.256	866.666666666667\\
144.839	863.333333333333\\
145.358	870\\
145.9	883.333333333333\\
146.423	903.333333333333\\
146.937	920\\
147.465	936.666666666667\\
148.04	940\\
149.12	943.333333333333\\
149.47	936.666666666667\\
150.189	930\\
150.751	930\\
151.228	930\\
151.254	930\\
151.923	930\\
152.299	930\\
153.076	926.666666666667\\
153.398	926.666666666667\\
153.905	926.666666666667\\
154.47	930\\
154.99	933.333333333333\\
155.509	933.333333333333\\
156.547	926.666666666667\\
156.645	920\\
157.137	913.333333333333\\
157.565	906.666666666667\\
158.158	903.333333333333\\
158.639	906.666666666667\\
159.25	910\\
159.68	913.333333333333\\
160.329	920\\
160.756	923.333333333333\\
161.394	926.666666666667\\
161.858	930\\
162.474	940\\
162.94	946.666666666667\\
163.455	950\\
164.06	953.333333333333\\
164.651	950\\
165.139	940\\
165.731	930\\
166.16	923.333333333333\\
167.258	916.666666666667\\
168.305	916.666666666667\\
168.339	913.333333333333\\
169.39	910\\
169.423	910\\
169.794	910\\
170.183	913.333333333333\\
170.408	920\\
171.042	926.666666666667\\
171.458	930\\
172.064	933.333333333333\\
172.539	933.333333333333\\
173.482	933.333333333333\\
173.638	933.333333333333\\
174.247	933.333333333333\\
174.707	933.333333333333\\
175.242	930\\
175.779	926.666666666667\\
176.353	923.333333333333\\
176.828	920\\
177.919	913.333333333333\\
178.989	910\\
179.822	913.333333333333\\
180.066	916.666666666667\\
180.562	920\\
180.978	926.666666666667\\
180.978	933.333333333333\\
181.072	933.333333333333\\
181.642	933.333333333333\\
182.169	933.333333333333\\
182.833	930\\
183.218	926.666666666667\\
183.84	926.666666666667\\
184.319	926.666666666667\\
184.863	923.333333333333\\
185.35	923.333333333333\\
185.963	923.333333333333\\
186.419	920\\
187.233	920\\
187.509	923.333333333333\\
188.167	926.666666666667\\
188.528	930\\
189.159	930\\
189.619	926.666666666667\\
190.257	923.333333333333\\
190.686	920\\
191.749	916.666666666667\\
192.141	916.666666666667\\
192.771	916.666666666667\\
192.799	916.666666666667\\
193.473	916.666666666667\\
193.879	916.666666666667\\
194.533	916.666666666667\\
194.969	916.666666666667\\
195.615	916.666666666667\\
196.046	916.666666666667\\
196.727	916.666666666667\\
197.157	916.666666666667\\
198.187	916.666666666667\\
198.209	916.666666666667\\
198.905	920\\
199.258	923.333333333333\\
200.279	930\\
200.339	936.666666666667\\
201.21	943.333333333333\\
201.378	943.333333333333\\
202.48	943.333333333333\\
202.497	940\\
203.54	936.666666666667\\
204.371	930\\
204.618	926.666666666667\\
204.966	923.333333333333\\
205.351	920\\
205.709	916.666666666667\\
206.431	916.666666666667\\
206.78	916.666666666667\\
207.414	916.666666666667\\
207.858	916.666666666667\\
208.567	913.333333333333\\
208.938	910\\
209.684	906.666666666667\\
209.998	903.333333333333\\
210.711	903.333333333333\\
211.072	906.666666666667\\
211.802	916.666666666667\\
212.18	926.666666666667\\
212.837	930\\
213.258	930\\
213.962	930\\
214.329	923.333333333333\\
215.022	916.666666666667\\
215.409	916.666666666667\\
216.486	920\\
216.556	923.333333333333\\
217.129	926.666666666667\\
217.569	930\\
218.196	933.333333333333\\
218.67	933.333333333333\\
219.278	930\\
219.74	926.666666666667\\
220.335	923.333333333333\\
220.819	916.666666666667\\
221.87	910\\
222.311	913.333333333333\\
222.909	916.666666666667\\
223.228	920\\
223.615	923.333333333333\\
224.019	926.666666666667\\
224.649	926.666666666667\\
225.088	926.666666666667\\
225.732	920\\
226.149	916.666666666667\\
226.859	913.333333333333\\
227.22	906.666666666667\\
227.919	900\\
228.268	900\\
229.019	900\\
229.346	900\\
230.072	903.333333333333\\
230.389	906.666666666667\\
231.261	913.333333333333\\
231.478	920\\
232.214	926.666666666667\\
232.548	926.666666666667\\
233.709	926.666666666667\\
234.078	923.333333333333\\
234.705	916.666666666667\\
234.809	910\\
235.501	910\\
235.89	910\\
236.577	913.333333333333\\
236.989	920\\
237.623	920\\
238.07	916.666666666667\\
238.688	913.333333333333\\
239.159	906.666666666667\\
239.833	900\\
240.229	900\\
240.901	900\\
241.309	900\\
242.027	903.333333333333\\
242.379	906.666666666667\\
243.095	910\\
243.43	913.333333333333\\
244.156	916.666666666667\\
244.559	913.333333333333\\
245.637	910\\
245.962	913.333333333333\\
246.739	920\\
246.776	926.666666666667\\
247.443	930\\
247.81	933.333333333333\\
248.448	933.333333333333\\
248.901	930\\
249.573	923.333333333333\\
250	923.333333333333\\
250.616	926.666666666667\\
251.069	926.666666666667\\
251.665	926.666666666667\\
252.099	930\\
252.701	923.333333333333\\
253.189	913.333333333333\\
253.833	903.333333333333\\
254.277	893.333333333333\\
255.289	886.666666666667\\
255.388	890\\
256.225	893.333333333333\\
256.43	896.666666666667\\
257.509	903.333333333333\\
257.766	906.666666666667\\
258.57	913.333333333333\\
258.609	920\\
259.349	923.333333333333\\
259.716	923.333333333333\\
260.394	920\\
260.82	913.333333333333\\
261.494	906.666666666667\\
261.928	903.333333333333\\
263.03	900\\
263.079	903.333333333333\\
264.087	896.666666666667\\
264.325	890\\
264.663	886.666666666667\\
265.19	883.333333333333\\
265.799	880\\
266.276	886.666666666667\\
266.909	896.666666666667\\
267.418	903.333333333333\\
267.979	913.333333333333\\
268.509	913.333333333333\\
269.079	913.333333333333\\
270.129	910\\
271.209	900\\
271.227	886.666666666667\\
272.367	880\\
272.518	873.333333333333\\
273.51	873.333333333333\\
273.789	880\\
274.595	896.666666666667\\
275.14	913.333333333333\\
275.647	930\\
276.41	940\\
276.715	953.333333333333\\
277.708	960\\
277.743	966.666666666667\\
278.733	980\\
278.917	993.333333333333\\
279.703	1000\\
280.179	1003.33333333333\\
280.693	1000\\
281.44	990\\
281.703	983.333333333333\\
282.68	980\\
282.713	976.666666666667\\
283.723	983.333333333333\\
283.919	990\\
284.713	993.333333333333\\
285.187	996.666666666667\\
285.837	996.666666666667\\
286.388	993.333333333333\\
286.882	993.333333333333\\
287.619	1000\\
288.128	1006.66666666667\\
288.778	1013.33333333333\\
288.799	1020\\
289.766	1023.33333333333\\
290	1013.33333333333\\
291.23	1003.33333333333\\
291.661	996.666666666667\\
291.846	990\\
292.46	990\\
292.641	996.666666666667\\
293.623	1003.33333333333\\
293.648	1010\\
294.629	1020\\
294.89	1023.33333333333\\
295.569	1020\\
296.099	1016.66666666667\\
296.539	1026.66666666667\\
297.3	1033.33333333333\\
297.469	1040\\
297.749	1050\\
298.459	1050\\
298.461	1036.66666666667\\
299.409	1023.33333333333\\
299.64	1010\\
300.429	993.333333333333\\
300.849	986.666666666667\\
301.449	983.333333333333\\
302.079	980\\
302.449	980\\
303.278	983.333333333333\\
303.459	986.666666666667\\
304.419	990\\
304.51	993.333333333333\\
305.439	996.666666666667\\
305.739	1003.33333333333\\
306.379	1010\\
306.91	1013.33333333333\\
307.359	1013.33333333333\\
308.15	1010\\
308.351	1003.33333333333\\
309.348	1003.33333333333\\
309.351	1003.33333333333\\
310.331	1003.33333333333\\
310.48	1003.33333333333\\
311.321	1006.66666666667\\
311.709	1003.33333333333\\
312.899	1000\\
313.634	1000\\
314.126	1000\\
315.359	1000\\
315.367	1010\\
315.796	1020\\
316.321	1023.33333333333\\
316.598	1026.66666666667\\
316.798	1030\\
317.41	1020\\
317.86	1010\\
318.394	1010\\
319.099	1010\\
319.378	1010\\
320.3	1013.33333333333\\
320.304	1016.66666666667\\
321.336	1020\\
321.557	1023.33333333333\\
322.267	1023.33333333333\\
322.757	1023.33333333333\\
323.246	1016.66666666667\\
323.99	1010\\
324.168	1003.33333333333\\
325.206	993.333333333333\\
325.22	983.333333333333\\
326.219	973.333333333333\\
326.44	960\\
327.27	956.666666666667\\
327.72	960\\
328.35	960\\
328.949	966.666666666667\\
329.415	990\\
330.2	986.666666666667\\
330.273	983.333333333333\\
331.295	983.333333333333\\
331.439	990\\
332.738	976.666666666667\\
333.038	980\\
333.517	966.666666666667\\
333.948	953.333333333333\\
334.469	933.333333333333\\
335.24	916.666666666667\\
336.404	906.666666666667\\
336.479	920\\
336.793	933.333333333333\\
337.581	953.333333333333\\
337.62	973.333333333333\\
338.574	983.333333333333\\
338.859	986.666666666667\\
339.59	990\\
340.028	986.666666666667\\
340.607	986.666666666667\\
341.218	990\\
341.952	986.666666666667\\
342.46	990\\
342.625	993.333333333333\\
343.607	996.666666666667\\
343.649	1000\\
344.573	1010\\
344.869	1013.33333333333\\
345.532	1020\\
346.06	1023.33333333333\\
346.521	1020\\
347.259	1020\\
348.038	1020\\
348.455	1016.66666666667\\
348.516	1013.33333333333\\
349.426	1020\\
349.64	1023.33333333333\\
350.369	1030\\
350.851	1036.66666666667\\
351.348	1036.66666666667\\
352.019	1033.33333333333\\
352.541	1030\\
353.22	1023.33333333333\\
353.266	1016.66666666667\\
354.435	1016.66666666667\\
354.49	1016.66666666667\\
355.24	1020\\
355.62	1023.33333333333\\
356.211	1030\\
356.809	1036.66666666667\\
357.336	1043.33333333333\\
358.046	1043.33333333333\\
359.22	1043.33333333333\\
359.612	1040\\
360.418	1036.66666666667\\
360.421	1030\\
360.891	1026.66666666667\\
361.392	1023.33333333333\\
361.631	1020\\
362.095	1013.33333333333\\
362.878	1010\\
363.156	1013.33333333333\\
364.095	1016.66666666667\\
364.214	1023.33333333333\\
364.547	1033.33333333333\\
364.728	1043.33333333333\\
365.338	1046.66666666667\\
365.834	1046.66666666667\\
366.59	1043.33333333333\\
366.611	1040\\
367.571	1033.33333333333\\
367.779	1026.66666666667\\
368.541	1026.66666666667\\
369.02	1026.66666666667\\
369.512	1030\\
370.228	1040\\
370.431	1050\\
371.371	1050\\
371.479	1050\\
372.331	1050\\
372.668	1046.66666666667\\
373.261	1040\\
373.879	1040\\
374.553	1043.33333333333\\
375.099	1043.33333333333\\
375.389	1043.33333333333\\
376.296	1043.33333333333\\
376.328	1043.33333333333\\
377.224	1043.33333333333\\
377.489	1043.33333333333\\
378.254	1043.33333333333\\
378.67	1046.66666666667\\
379.204	1046.66666666667\\
379.938	1040\\
380.087	1033.33333333333\\
381.056	1030\\
381.15	1026.66666666667\\
382.032	1020\\
382.362	1016.66666666667\\
383.006	1016.66666666667\\
383.581	1013.33333333333\\
384.186	1003.33333333333\\
384.75	1000\\
385.589	980\\
385.939	956.666666666667\\
386.526	943.333333333333\\
387.139	946.666666666667\\
387.174	950\\
388.171	970\\
388.339	990\\
389.136	1006.66666666667\\
389.52	1013.33333333333\\
390.101	1016.66666666667\\
390.748	1020\\
391.627	1023.33333333333\\
391.957	1020\\
392.014	1016.66666666667\\
393.035	1016.66666666667\\
393.186	1016.66666666667\\
394.03	1013.33333333333\\
394.368	1010\\
395.013	1006.66666666667\\
395.621	1013.33333333333\\
396.841	1016.66666666667\\
397.281	1023.33333333333\\
397.915	1026.66666666667\\
398.061	1033.33333333333\\
398.177	1030\\
398.837	1030\\
399.25	1030\\
399.786	1033.33333333333\\
400.44	1033.33333333333\\
400.987	1030\\
401.621	1033.33333333333\\
401.731	1036.66666666667\\
402.823	1033.33333333333\\
402.83	1030\\
403.746	1026.66666666667\\
403.998	1016.66666666667\\
404.712	1003.33333333333\\
405.179	996.666666666667\\
405.765	986.666666666667\\
406.419	980\\
406.932	983.333333333333\\
407.549	990\\
407.654	993.333333333333\\
408.73	1000\\
409.527	1010\\
409.928	1016.66666666667\\
410.087	1023.33333333333\\
410.563	1026.66666666667\\
411.129	1030\\
411.593	1030\\
412.34	1026.66666666667\\
412.485	1023.33333333333\\
413.446	1030\\
413.539	1036.66666666667\\
414.451	1036.66666666667\\
414.752	1033.33333333333\\
415.426	1026.66666666667\\
415.976	1016.66666666667\\
416.443	1010\\
417.19	1013.33333333333\\
417.345	1016.66666666667\\
418.258	1023.33333333333\\
418.409	1030\\
419.262	1033.33333333333\\
419.659	1033.33333333333\\
420.233	1026.66666666667\\
420.9	1020\\
421.456	1016.66666666667\\
422.089	1013.33333333333\\
422.133	1010\\
423.142	1016.66666666667\\
423.329	1023.33333333333\\
424.085	1020\\
424.549	1016.66666666667\\
425.06	1013.33333333333\\
425.82	1006.66666666667\\
426.5	1000\\
427.076	1003.33333333333\\
427.077	1006.66666666667\\
428.057	1010\\
428.318	1013.33333333333\\
429.038	1013.33333333333\\
429.549	1010\\
430.029	1006.66666666667\\
430.81	1000\\
431.012	1000\\
};
\addlegendentry{User1}

\addplot [color=mycolor2]
  table[row sep=crcr]{%
0	870\\
1.118	850.777777777778\\
1.826	850\\
2.38	850\\
2.683	836.666666666667\\
3.548	823.333333333333\\
3.574	836.666666666667\\
4.526	850\\
4.728	850\\
5.398	850\\
5.869	850\\
7.013	850\\
7.039	853.333333333333\\
7.426	856.666666666667\\
8.209	860\\
8.249	863.333333333333\\
9.134	866.666666666667\\
9.328	866.666666666667\\
10.092	866.666666666667\\
10.488	866.666666666667\\
11.034	866.666666666667\\
11.639	866.666666666667\\
12.085	866.666666666667\\
12.799	866.666666666667\\
12.895	870\\
13.833	873.333333333333\\
13.93	876.666666666667\\
14.745	880\\
15.058	880\\
15.728	876.666666666667\\
16.2	873.333333333333\\
16.78	866.666666666667\\
17.369	860\\
18.549	860\\
19.699	860\\
20.851	860\\
21.989	863.333333333333\\
23.17	866.666666666667\\
24.299	866.666666666667\\
24.536	866.666666666667\\
24.536	866.666666666667\\
24.536	866.666666666667\\
24.54	866.666666666667\\
24.54	866.666666666667\\
24.54	866.666666666667\\
24.541	866.666666666667\\
24.693	866.666666666667\\
25.076	866.666666666667\\
25.45	866.666666666667\\
25.998	866.666666666667\\
26.597	866.666666666667\\
26.914	866.666666666667\\
27.719	866.666666666667\\
27.914	866.666666666667\\
28.849	866.666666666667\\
28.889	870\\
29.722	873.333333333333\\
30.008	870\\
30.696	866.666666666667\\
31.168	866.666666666667\\
31.637	863.333333333333\\
32.348	856.666666666667\\
32.562	856.666666666667\\
33.48	856.666666666667\\
33.514	856.666666666667\\
34.417	856.666666666667\\
34.658	856.666666666667\\
35.332	856.666666666667\\
35.796	856.666666666667\\
36.28	853.333333333333\\
36.968	853.333333333333\\
37.179	856.666666666667\\
38.132	860\\
38.238	863.333333333333\\
39.133	866.666666666667\\
39.306	863.333333333333\\
40.122	860\\
40.428	856.666666666667\\
41.214	853.333333333333\\
41.639	853.333333333333\\
41.994	856.666666666667\\
42.769	860\\
42.864	860\\
43.773	860\\
43.938	856.666666666667\\
44.704	853.333333333333\\
45.099	850\\
45.637	850\\
46.28	850\\
46.583	850\\
47.449	850\\
47.492	853.333333333333\\
48.412	856.666666666667\\
48.609	856.666666666667\\
49.382	856.666666666667\\
49.769	860\\
50.302	860\\
50.926	860\\
51.778	863.333333333333\\
52.069	866.666666666667\\
52.309	866.666666666667\\
53.229	866.666666666667\\
54.051	866.666666666667\\
54.379	866.666666666667\\
55.076	866.666666666667\\
55.256	866.666666666667\\
55.53	870\\
55.965	873.333333333333\\
56.659	876.666666666667\\
56.925	880\\
57.799	883.333333333333\\
57.835	886.666666666667\\
58.835	890\\
58.894	890\\
59.736	890\\
60.048	886.666666666667\\
60.686	880\\
61.18	873.333333333333\\
61.625	870\\
62.322	866.666666666667\\
62.543	863.333333333333\\
63.433	860\\
63.499	856.666666666667\\
64.363	853.333333333333\\
64.647	850\\
65.343	850\\
65.819	853.333333333333\\
66.274	856.666666666667\\
66.96	860\\
67.223	860\\
68.133	860\\
68.139	860\\
69.033	860\\
69.286	860\\
69.953	863.333333333333\\
70.44	863.333333333333\\
70.874	860\\
71.6	856.666666666667\\
71.783	853.333333333333\\
72.723	850\\
72.823	850\\
73.683	846.666666666667\\
73.955	843.333333333333\\
75.15	840\\
75.152	843.333333333333\\
75.949	846.666666666667\\
76.289	850\\
76.843	853.333333333333\\
77.449	856.666666666667\\
77.57	860\\
78.502	863.333333333333\\
78.598	863.333333333333\\
79.514	863.333333333333\\
79.762	863.333333333333\\
80.374	856.666666666667\\
80.398	850\\
80.919	853.333333333333\\
81.461	856.666666666667\\
82.077	860\\
82.393	860\\
83.17	860\\
83.223	856.666666666667\\
84.186	853.333333333333\\
84.379	846.666666666667\\
85.118	850\\
85.598	853.333333333333\\
86.727	856.666666666667\\
87.194	860\\
87.898	866.666666666667\\
87.994	863.333333333333\\
88.965	860\\
89.041	856.666666666667\\
89.29	853.333333333333\\
89.857	850\\
90.219	846.666666666667\\
90.813	843.333333333333\\
91.409	843.333333333333\\
92.509	843.333333333333\\
92.61	843.333333333333\\
93.054	846.666666666667\\
93.749	850\\
93.902	850\\
94.57	850\\
94.91	853.333333333333\\
95.474	856.666666666667\\
96.077	863.333333333333\\
96.53	870\\
97.21	876.666666666667\\
97.439	873.333333333333\\
98.316	870\\
98.379	863.333333333333\\
99.262	863.333333333333\\
99.499	863.333333333333\\
99.549	866.666666666667\\
100.242	873.333333333333\\
100.628	880\\
101.779	880\\
102.206	880\\
102.399	876.666666666667\\
102.899	870\\
102.918	863.333333333333\\
103.872	856.666666666667\\
104.086	850\\
104.752	850\\
105.296	850\\
105.683	850\\
106.46	850\\
106.638	853.333333333333\\
107.598	856.666666666667\\
107.617	860\\
108.554	863.333333333333\\
108.749	870\\
109.502	873.333333333333\\
109.858	873.333333333333\\
110.438	873.333333333333\\
111.029	873.333333333333\\
111.412	870\\
112.169	866.666666666667\\
112.38	863.333333333333\\
113.318	860\\
113.369	860\\
114.273	860\\
114.517	860\\
115.228	863.333333333333\\
115.618	866.666666666667\\
116.167	866.666666666667\\
116.78	860\\
117.133	853.333333333333\\
117.969	846.666666666667\\
118.086	850\\
119.041	853.333333333333\\
119.109	860\\
119.999	866.666666666667\\
120.258	873.333333333333\\
120.929	870\\
121.419	866.666666666667\\
121.858	866.666666666667\\
122.58	870\\
122.808	873.333333333333\\
123.68	876.666666666667\\
123.748	873.333333333333\\
124.698	870\\
124.859	866.666666666667\\
125.689	863.333333333333\\
126.01	860\\
126.599	863.333333333333\\
127.169	863.333333333333\\
127.538	860\\
128.388	860\\
128.671	860\\
129.487	860\\
129.622	863.333333333333\\
130.535	866.666666666667\\
130.619	863.333333333333\\
131.786	860\\
131.82	866.666666666667\\
132.7	873.333333333333\\
132.938	893.333333333333\\
133.824	916.666666666667\\
133.959	903.333333333333\\
134.429	880\\
135.227	906.666666666667\\
135.33	920\\
136.189	920\\
136.895	956.666666666667\\
137.219	976.666666666667\\
137.713	946.666666666667\\
138.37	913.333333333333\\
138.992	893.333333333333\\
139.517	880\\
139.977	883.333333333333\\
140.619	890\\
141.146	900\\
141.708	923.333333333333\\
141.908	940\\
142.709	953.333333333333\\
143.25	963.333333333333\\
143.76	963.333333333333\\
144.256	950\\
144.839	940\\
145.358	933.333333333333\\
145.9	933.333333333333\\
146.423	943.333333333333\\
146.937	943.333333333333\\
147.465	943.333333333333\\
148.04	943.333333333333\\
149.12	936.666666666667\\
149.47	930\\
150.189	936.666666666667\\
150.751	940\\
151.228	943.333333333333\\
151.254	940\\
151.923	936.666666666667\\
152.299	926.666666666667\\
153.076	916.666666666667\\
153.398	916.666666666667\\
153.905	923.333333333333\\
154.47	930\\
154.99	943.333333333333\\
155.509	956.666666666667\\
156.547	960\\
156.645	966.666666666667\\
157.137	973.333333333333\\
157.565	963.333333333333\\
158.158	953.333333333333\\
158.639	950\\
159.25	943.333333333333\\
159.68	930\\
160.329	930\\
160.756	930\\
161.394	923.333333333333\\
161.858	916.666666666667\\
162.474	916.666666666667\\
162.94	913.333333333333\\
163.455	910\\
164.06	906.666666666667\\
164.651	903.333333333333\\
165.139	920\\
165.731	923.333333333333\\
166.16	926.666666666667\\
167.258	930\\
168.305	940\\
168.339	930\\
169.39	936.666666666667\\
169.423	943.333333333333\\
169.794	953.333333333333\\
170.183	956.666666666667\\
170.408	960\\
171.042	963.333333333333\\
171.458	953.333333333333\\
172.064	940\\
172.539	930\\
173.482	920\\
173.638	913.333333333333\\
174.247	920\\
174.707	923.333333333333\\
175.242	923.333333333333\\
175.779	930\\
176.353	930\\
176.828	930\\
177.919	933.333333333333\\
178.989	936.666666666667\\
179.822	933.333333333333\\
180.066	933.333333333333\\
180.562	933.333333333333\\
180.978	940\\
180.978	946.666666666667\\
181.072	943.333333333333\\
181.642	940\\
182.169	940\\
182.833	933.333333333333\\
183.218	926.666666666667\\
183.84	930\\
184.319	936.666666666667\\
184.863	940\\
185.35	940\\
185.963	940\\
186.419	936.666666666667\\
187.233	930\\
187.509	933.333333333333\\
188.167	940\\
188.528	936.666666666667\\
189.159	936.666666666667\\
189.619	940\\
190.257	930\\
190.686	920\\
191.749	920\\
192.141	926.666666666667\\
192.771	930\\
192.799	930\\
193.473	930\\
193.879	933.333333333333\\
194.533	930\\
194.969	926.666666666667\\
195.615	930\\
196.046	930\\
196.727	926.666666666667\\
197.157	923.333333333333\\
198.187	920\\
198.209	920\\
198.905	923.333333333333\\
199.258	926.666666666667\\
200.279	930\\
200.339	940\\
201.21	936.666666666667\\
201.378	933.333333333333\\
202.48	930\\
202.497	926.666666666667\\
203.54	916.666666666667\\
204.371	916.666666666667\\
204.618	916.666666666667\\
204.966	916.666666666667\\
205.351	916.666666666667\\
205.709	920\\
206.431	923.333333333333\\
206.78	926.666666666667\\
207.414	930\\
207.858	930\\
208.567	926.666666666667\\
208.938	926.666666666667\\
209.684	926.666666666667\\
209.998	923.333333333333\\
210.711	923.333333333333\\
211.072	923.333333333333\\
211.802	920\\
212.18	916.666666666667\\
212.837	916.666666666667\\
213.258	916.666666666667\\
213.962	916.666666666667\\
214.329	920\\
215.022	923.333333333333\\
215.409	926.666666666667\\
216.486	930\\
216.556	930\\
217.129	926.666666666667\\
217.569	926.666666666667\\
218.196	926.666666666667\\
218.67	923.333333333333\\
219.278	923.333333333333\\
219.74	923.333333333333\\
220.335	926.666666666667\\
220.819	930\\
221.87	936.666666666667\\
222.311	943.333333333333\\
222.909	950\\
223.228	940\\
223.615	930\\
224.019	926.666666666667\\
224.649	923.333333333333\\
225.088	920\\
225.732	926.666666666667\\
226.149	933.333333333333\\
226.859	933.333333333333\\
227.22	933.333333333333\\
227.919	933.333333333333\\
228.268	933.333333333333\\
229.019	933.333333333333\\
229.346	933.333333333333\\
230.072	933.333333333333\\
230.389	933.333333333333\\
231.261	933.333333333333\\
231.478	930\\
232.214	920\\
232.548	910\\
233.709	900\\
234.078	893.333333333333\\
234.705	890\\
234.809	896.666666666667\\
235.501	903.333333333333\\
235.89	906.666666666667\\
236.577	906.666666666667\\
236.989	910\\
237.623	910\\
238.07	910\\
238.688	913.333333333333\\
239.159	920\\
239.833	923.333333333333\\
240.229	926.666666666667\\
240.901	930\\
241.309	933.333333333333\\
242.027	933.333333333333\\
242.379	933.333333333333\\
243.095	933.333333333333\\
243.43	930\\
244.156	926.666666666667\\
244.559	923.333333333333\\
245.637	916.666666666667\\
245.962	910\\
246.739	906.666666666667\\
246.776	906.666666666667\\
247.443	906.666666666667\\
247.81	910\\
248.448	913.333333333333\\
248.901	913.333333333333\\
249.573	910\\
250	913.333333333333\\
250.616	916.666666666667\\
251.069	923.333333333333\\
251.665	933.333333333333\\
252.099	940\\
252.701	940\\
253.189	936.666666666667\\
253.833	930\\
254.277	923.333333333333\\
255.289	920\\
255.388	920\\
256.225	920\\
256.43	920\\
257.509	920\\
257.766	920\\
258.57	916.666666666667\\
258.609	913.333333333333\\
259.349	910\\
259.716	906.666666666667\\
260.394	903.333333333333\\
260.82	900\\
261.494	903.333333333333\\
261.928	906.666666666667\\
263.03	913.333333333333\\
263.079	920\\
264.087	926.666666666667\\
264.325	923.333333333333\\
264.663	920\\
265.19	916.666666666667\\
265.799	913.333333333333\\
266.276	900\\
266.909	893.333333333333\\
267.418	893.333333333333\\
267.979	890\\
268.509	886.666666666667\\
269.079	786.666666666667\\
270.129	776.666666666667\\
271.209	766\\
271.227	756\\
272.367	750\\
272.518	740\\
273.51	750\\
273.789	753.333333333333\\
274.595	750\\
275.14	756.666666666667\\
275.647	760\\
276.41	760\\
276.715	766.666666666667\\
277.708	773.333333333333\\
277.743	783.333333333333\\
278.733	793.333333333333\\
278.917	796.666666666667\\
279.703	800\\
280.179	810\\
280.693	806.666666666667\\
281.44	796.666666666667\\
281.703	796.666666666667\\
282.68	796.666666666667\\
282.713	793.333333333333\\
283.723	790\\
283.919	790\\
284.713	790\\
285.187	796.666666666667\\
285.837	800\\
286.388	803.333333333333\\
286.882	810\\
287.619	816.666666666667\\
288.128	820\\
288.778	823.333333333333\\
288.799	826.666666666667\\
289.766	823.333333333333\\
290	820\\
291.23	813.333333333333\\
291.661	810\\
291.846	806.666666666667\\
292.46	810\\
292.641	816.666666666667\\
293.623	823.333333333333\\
293.648	820\\
294.629	816.666666666667\\
294.89	816.666666666667\\
295.569	813.333333333333\\
296.099	813.333333333333\\
296.539	820\\
297.3	826.666666666667\\
297.469	833.333333333333\\
297.749	840\\
298.459	843.333333333333\\
298.461	846.666666666667\\
299.409	850\\
299.64	846.666666666667\\
300.429	843.333333333333\\
300.849	833.333333333333\\
301.449	823.333333333333\\
302.079	816.666666666667\\
302.449	813.333333333333\\
303.278	810\\
303.459	813.333333333333\\
304.419	816.666666666667\\
304.51	816.666666666667\\
305.439	816.666666666667\\
305.739	820\\
306.379	823.333333333333\\
306.91	823.333333333333\\
307.359	823.333333333333\\
308.15	826.666666666667\\
308.351	826.666666666667\\
309.348	826.666666666667\\
309.351	840\\
310.331	853.333333333333\\
310.48	846.666666666667\\
311.321	850\\
311.709	853.333333333333\\
312.899	836.666666666667\\
313.634	823.333333333333\\
314.126	826.666666666667\\
315.359	820\\
315.367	813.333333333333\\
315.796	810\\
316.321	803.333333333333\\
316.598	796.666666666667\\
316.798	793.333333333333\\
317.41	790\\
317.86	793.333333333333\\
318.394	796.666666666667\\
319.099	803.333333333333\\
319.378	806.666666666667\\
320.3	810\\
320.304	810\\
321.336	810\\
321.557	810\\
322.267	810\\
322.757	806.666666666667\\
323.246	806.666666666667\\
323.99	806.666666666667\\
324.168	806.666666666667\\
325.206	806.666666666667\\
325.22	806.666666666667\\
326.219	806.666666666667\\
326.44	803.333333333333\\
327.27	796.666666666667\\
327.72	793.333333333333\\
328.35	793.333333333333\\
328.949	793.333333333333\\
329.415	796.666666666667\\
330.2	800\\
330.273	800\\
331.295	796.666666666667\\
331.439	793.333333333333\\
332.738	790\\
333.038	790\\
333.517	790\\
333.948	790\\
334.469	790\\
335.24	793.333333333333\\
336.404	793.333333333333\\
336.479	793.333333333333\\
336.793	806.666666666667\\
337.581	820\\
337.62	823.333333333333\\
338.574	826.666666666667\\
338.859	833.333333333333\\
339.59	830\\
340.028	830\\
340.607	836.666666666667\\
341.218	830\\
341.952	820\\
342.46	810\\
342.625	810\\
343.607	810\\
343.649	816.666666666667\\
344.573	823.333333333333\\
344.869	833.333333333333\\
345.532	830\\
346.06	823.333333333333\\
346.521	823.333333333333\\
347.259	823.333333333333\\
348.038	820\\
348.455	816.666666666667\\
348.516	823.333333333333\\
349.426	830\\
349.64	830\\
350.369	830\\
350.851	836.666666666667\\
351.348	836.666666666667\\
352.019	833.333333333333\\
352.541	836.666666666667\\
353.22	836.666666666667\\
353.266	830\\
354.435	823.333333333333\\
354.49	823.333333333333\\
355.24	823.333333333333\\
355.62	826.666666666667\\
356.211	830\\
356.809	830\\
357.336	833.333333333333\\
358.046	836.666666666667\\
359.22	840\\
359.612	836.666666666667\\
360.418	836.666666666667\\
360.421	830\\
360.891	823.333333333333\\
361.392	816.666666666667\\
361.631	813.333333333333\\
362.095	810\\
362.878	813.333333333333\\
363.156	816.666666666667\\
364.095	820\\
364.214	826.666666666667\\
364.547	826.666666666667\\
364.728	820\\
365.338	810\\
365.834	800\\
366.59	790\\
366.611	793.333333333333\\
367.571	796.666666666667\\
367.779	800\\
368.541	803.333333333333\\
369.02	810\\
369.512	810\\
370.228	810\\
370.431	813.333333333333\\
371.371	816.666666666667\\
371.479	816.666666666667\\
372.331	816.666666666667\\
372.668	820\\
373.261	823.333333333333\\
373.879	823.333333333333\\
374.553	823.333333333333\\
375.099	823.333333333333\\
375.389	820\\
376.296	816.666666666667\\
376.328	820\\
377.224	823.333333333333\\
377.489	830\\
378.254	836.666666666667\\
378.67	830\\
379.204	820\\
379.938	810\\
380.087	803.333333333333\\
381.056	796.666666666667\\
381.15	806.666666666667\\
382.032	816.666666666667\\
382.362	826.666666666667\\
383.006	830\\
383.581	833.333333333333\\
384.186	833.333333333333\\
384.75	833.333333333333\\
385.589	833.333333333333\\
385.939	836.666666666667\\
386.526	840\\
387.139	843.333333333333\\
387.174	843.333333333333\\
388.171	843.333333333333\\
388.339	840\\
389.136	836.666666666667\\
389.52	833.333333333333\\
390.101	833.333333333333\\
390.748	826.666666666667\\
391.627	820\\
391.957	813.333333333333\\
392.014	813.333333333333\\
393.035	813.333333333333\\
393.186	816.666666666667\\
394.03	820\\
394.368	820\\
395.013	813.333333333333\\
395.621	806.666666666667\\
396.841	803.333333333333\\
397.281	806.666666666667\\
397.915	813.333333333333\\
398.061	820\\
398.177	826.666666666667\\
398.837	833.333333333333\\
399.25	833.333333333333\\
399.786	833.333333333333\\
400.44	833.333333333333\\
400.987	833.333333333333\\
401.621	833.333333333333\\
401.731	836.666666666667\\
402.823	840\\
402.83	840\\
403.746	840\\
403.998	840\\
404.712	836.666666666667\\
405.179	830\\
405.765	826.666666666667\\
406.419	833.333333333333\\
406.932	840\\
407.549	843.333333333333\\
407.654	850\\
408.73	850\\
409.527	840\\
409.928	830\\
410.087	826.666666666667\\
410.563	823.333333333333\\
411.129	826.666666666667\\
411.593	830\\
412.34	833.333333333333\\
412.485	833.333333333333\\
413.446	833.333333333333\\
413.539	826.666666666667\\
414.451	820\\
414.752	816.666666666667\\
415.426	813.333333333333\\
415.976	810\\
416.443	813.333333333333\\
417.19	816.666666666667\\
417.345	816.666666666667\\
418.258	816.666666666667\\
418.409	810\\
419.262	803.333333333333\\
419.659	803.333333333333\\
420.233	803.333333333333\\
420.9	806.666666666667\\
421.456	816.666666666667\\
422.089	826.666666666667\\
422.133	823.333333333333\\
423.142	820\\
423.329	816.666666666667\\
424.085	813.333333333333\\
424.549	803.333333333333\\
425.06	800\\
425.82	796.666666666667\\
426.5	793.333333333333\\
427.076	790\\
427.077	793.333333333333\\
428.057	796.666666666667\\
428.318	800\\
429.038	800\\
429.549	796.666666666667\\
430.029	793.333333333333\\
430.81	788.888888888889\\
431.012	783.333333333333\\
};
\addlegendentry{User2}

\addplot[area legend, draw=none, fill=blue, fill opacity=0.1]
table[row sep=crcr] {%
x	y\\
132.7	1002.75\\
132.938	1015\\
133.824	1030.75\\
133.959	1027.25\\
134.429	1013.25\\
135.227	1023.75\\
135.33	1008\\
136.189	983.5\\
136.895	974.75\\
137.219	962.5\\
137.713	925.75\\
138.37	906.5\\
138.992	911.75\\
139.517	924\\
139.977	950.25\\
140.619	976.5\\
141.146	1001\\
141.708	1015\\
141.908	1013.25\\
142.709	1004.5\\
143.25	990.5\\
143.76	974.75\\
144.256	953.75\\
144.839	946.75\\
145.358	946.75\\
145.9	953.75\\
146.423	969.5\\
146.937	978.25\\
147.465	987\\
148.04	988.75\\
149.12	987\\
149.47	980\\
150.189	980\\
150.751	981.75\\
151.228	983.5\\
151.254	981.75\\
151.923	980\\
152.299	974.75\\
153.076	967.75\\
153.398	967.75\\
153.905	971.25\\
154.47	976.5\\
154.99	985.25\\
155.509	992.25\\
156.547	990.5\\
156.645	990.5\\
157.137	990.5\\
157.565	981.75\\
158.158	974.75\\
158.639	974.75\\
159.25	973\\
159.68	967.75\\
160.329	971.25\\
160.756	973\\
161.394	971.25\\
161.858	969.5\\
162.474	974.75\\
162.94	976.5\\
163.455	976.5\\
164.06	976.5\\
164.651	973\\
165.139	976.5\\
165.731	973\\
166.16	971.25\\
167.258	969.5\\
168.305	974.75\\
168.339	967.75\\
169.39	969.5\\
169.423	973\\
169.794	978.25\\
170.183	981.75\\
170.408	987\\
171.042	992.25\\
171.458	988.75\\
172.064	983.5\\
172.539	978.25\\
173.482	973\\
173.638	969.5\\
174.247	973\\
174.707	974.75\\
175.242	973\\
175.779	974.75\\
176.353	973\\
176.828	971.25\\
177.919	969.5\\
178.989	969.5\\
179.822	969.5\\
180.066	971.25\\
180.562	973\\
180.978	980\\
180.978	987\\
181.072	985.25\\
181.642	983.5\\
182.169	983.5\\
182.833	978.25\\
183.218	973\\
183.84	974.75\\
184.319	978.25\\
184.863	978.25\\
185.35	978.25\\
185.963	978.25\\
186.419	974.75\\
187.233	971.25\\
187.509	974.75\\
188.167	980\\
188.528	980\\
189.159	980\\
189.619	980\\
190.257	973\\
190.686	966\\
191.749	964.25\\
192.141	967.75\\
192.771	969.5\\
192.799	969.5\\
193.473	969.5\\
193.879	971.25\\
194.533	969.5\\
194.969	967.75\\
195.615	969.5\\
196.046	969.5\\
196.727	967.75\\
197.157	966\\
198.187	964.25\\
198.209	964.25\\
198.905	967.75\\
199.258	971.25\\
200.279	976.5\\
200.339	985.25\\
201.21	987\\
201.378	985.25\\
202.48	983.5\\
202.497	980\\
203.54	973\\
204.371	969.5\\
204.618	967.75\\
204.966	966\\
205.351	964.25\\
205.709	964.25\\
206.431	966\\
206.78	967.75\\
207.414	969.5\\
207.858	969.5\\
208.567	966\\
208.938	964.25\\
209.684	962.5\\
209.998	959\\
210.711	959\\
211.072	960.75\\
211.802	964.25\\
212.18	967.75\\
212.837	969.5\\
213.258	969.5\\
213.962	969.5\\
214.329	967.75\\
215.022	966\\
215.409	967.75\\
216.486	971.25\\
216.556	973\\
217.129	973\\
217.569	974.75\\
218.196	976.5\\
218.67	974.75\\
219.278	973\\
219.74	971.25\\
220.335	971.25\\
220.819	969.5\\
221.87	969.5\\
222.311	974.75\\
222.909	980\\
223.228	976.5\\
223.615	973\\
224.019	973\\
224.649	971.25\\
225.088	969.5\\
225.732	969.5\\
226.149	971.25\\
226.859	969.5\\
227.22	966\\
227.919	962.5\\
228.268	962.5\\
229.019	962.5\\
229.346	962.5\\
230.072	964.25\\
230.389	966\\
231.261	969.5\\
231.478	971.25\\
232.214	969.5\\
232.548	964.25\\
233.709	959\\
234.078	953.75\\
234.705	948.5\\
234.809	948.5\\
235.501	952\\
235.89	953.75\\
236.577	955.5\\
236.989	960.75\\
237.623	960.75\\
238.07	959\\
238.688	959\\
239.159	959\\
239.833	957.25\\
240.229	959\\
240.901	960.75\\
241.309	962.5\\
242.027	964.25\\
242.379	966\\
243.095	967.75\\
243.43	967.75\\
244.156	967.75\\
244.559	964.25\\
245.637	959\\
245.962	957.25\\
246.739	959\\
246.776	962.5\\
247.443	964.25\\
247.81	967.75\\
248.448	969.5\\
248.901	967.75\\
249.573	962.5\\
250	964.25\\
250.616	967.75\\
251.069	971.25\\
251.665	976.5\\
252.099	981.75\\
252.701	978.25\\
253.189	971.25\\
253.833	962.5\\
254.277	953.75\\
255.289	948.5\\
255.388	950.25\\
256.225	952\\
256.43	953.75\\
257.509	957.25\\
257.766	959\\
258.57	960.75\\
258.609	962.5\\
259.349	962.5\\
259.716	960.75\\
260.394	957.25\\
260.82	952\\
261.494	950.25\\
261.928	950.25\\
263.03	952\\
263.079	957.25\\
264.087	957.25\\
264.325	952\\
264.663	948.5\\
265.19	945\\
265.799	941.5\\
266.276	938\\
266.909	939.75\\
266.909	850.25\\
266.276	848.666666666667\\
265.799	851.833333333333\\
265.19	855\\
264.663	858.166666666667\\
264.325	861.333333333333\\
264.087	866.083333333333\\
263.079	866.083333333333\\
263.03	861.333333333333\\
261.928	859.75\\
261.494	859.75\\
260.82	861.333333333333\\
260.394	866.083333333333\\
259.716	869.25\\
259.349	870.833333333333\\
258.609	870.833333333333\\
258.57	869.25\\
257.766	867.666666666667\\
257.509	866.083333333333\\
256.43	862.916666666667\\
256.225	861.333333333333\\
255.388	859.75\\
255.289	858.166666666667\\
254.277	862.916666666667\\
253.833	870.833333333333\\
253.189	878.75\\
252.701	885.083333333333\\
252.099	888.25\\
251.665	883.5\\
251.069	878.75\\
250.616	875.583333333333\\
250	872.416666666667\\
249.573	870.833333333333\\
248.901	875.583333333333\\
248.448	877.166666666667\\
247.81	875.583333333333\\
247.443	872.416666666667\\
246.776	870.833333333333\\
246.739	867.666666666667\\
245.962	866.083333333333\\
245.637	867.666666666667\\
244.559	872.416666666667\\
244.156	875.583333333333\\
243.43	875.583333333333\\
243.095	875.583333333333\\
242.379	874\\
242.027	872.416666666667\\
241.309	870.833333333333\\
240.901	869.25\\
240.229	867.666666666667\\
239.833	866.083333333333\\
239.159	867.666666666667\\
238.688	867.666666666667\\
238.07	867.666666666667\\
237.623	869.25\\
236.989	869.25\\
236.577	864.5\\
235.89	862.916666666667\\
235.501	861.333333333333\\
234.809	858.166666666667\\
234.705	858.166666666667\\
234.078	862.916666666667\\
233.709	867.666666666667\\
232.548	872.416666666667\\
232.214	877.166666666667\\
231.478	878.75\\
231.261	877.166666666667\\
230.389	874\\
230.072	872.416666666667\\
229.346	870.833333333333\\
229.019	870.833333333333\\
228.268	870.833333333333\\
227.919	870.833333333333\\
227.22	874\\
226.859	877.166666666667\\
226.149	878.75\\
225.732	877.166666666667\\
225.088	877.166666666667\\
224.649	878.75\\
224.019	880.333333333333\\
223.615	880.333333333333\\
223.228	883.5\\
222.909	886.666666666667\\
222.311	881.916666666667\\
221.87	877.166666666667\\
220.819	877.166666666667\\
220.335	878.75\\
219.74	878.75\\
219.278	880.333333333333\\
218.67	881.916666666667\\
218.196	883.5\\
217.569	881.916666666667\\
217.129	880.333333333333\\
216.556	880.333333333333\\
216.486	878.75\\
215.409	875.583333333333\\
215.022	874\\
214.329	875.583333333333\\
213.962	877.166666666667\\
213.258	877.166666666667\\
212.837	877.166666666667\\
212.18	875.583333333333\\
211.802	872.416666666667\\
211.072	869.25\\
210.711	867.666666666667\\
209.998	867.666666666667\\
209.684	870.833333333333\\
208.938	872.416666666667\\
208.567	874\\
207.858	877.166666666667\\
207.414	877.166666666667\\
206.78	875.583333333333\\
206.431	874\\
205.709	872.416666666667\\
205.351	872.416666666667\\
204.966	874\\
204.618	875.583333333333\\
204.371	877.166666666667\\
203.54	880.333333333333\\
202.497	886.666666666667\\
202.48	889.833333333333\\
201.378	891.416666666667\\
201.21	893\\
200.339	891.416666666667\\
200.279	883.5\\
199.258	878.75\\
198.905	875.583333333333\\
198.209	872.416666666667\\
198.187	872.416666666667\\
197.157	874\\
196.727	875.583333333333\\
196.046	877.166666666667\\
195.615	877.166666666667\\
194.969	875.583333333333\\
194.533	877.166666666667\\
193.879	878.75\\
193.473	877.166666666667\\
192.799	877.166666666667\\
192.771	877.166666666667\\
192.141	875.583333333333\\
191.749	872.416666666667\\
190.686	874\\
190.257	880.333333333333\\
189.619	886.666666666667\\
189.159	886.666666666667\\
188.528	886.666666666667\\
188.167	886.666666666667\\
187.509	881.916666666667\\
187.233	878.75\\
186.419	881.916666666667\\
185.963	885.083333333333\\
185.35	885.083333333333\\
184.863	885.083333333333\\
184.319	885.083333333333\\
183.84	881.916666666667\\
183.218	880.333333333333\\
182.833	885.083333333333\\
182.169	889.833333333333\\
181.642	889.833333333333\\
181.072	891.416666666667\\
180.978	893\\
180.978	886.666666666667\\
180.562	880.333333333333\\
180.066	878.75\\
179.822	877.166666666667\\
178.989	877.166666666667\\
177.919	877.166666666667\\
176.828	878.75\\
176.353	880.333333333333\\
175.779	881.916666666667\\
175.242	880.333333333333\\
174.707	881.916666666667\\
174.247	880.333333333333\\
173.638	877.166666666667\\
173.482	880.333333333333\\
172.539	885.083333333333\\
172.064	889.833333333333\\
171.458	894.583333333333\\
171.042	897.75\\
170.408	893\\
170.183	888.25\\
169.794	885.083333333333\\
169.423	880.333333333333\\
169.39	877.166666666667\\
168.339	875.583333333333\\
168.305	881.916666666667\\
167.258	877.166666666667\\
166.16	878.75\\
165.731	880.333333333333\\
165.139	883.5\\
164.651	880.333333333333\\
164.06	883.5\\
163.455	883.5\\
162.94	883.5\\
162.474	881.916666666667\\
161.858	877.166666666667\\
161.394	878.75\\
160.756	880.333333333333\\
160.329	878.75\\
159.68	875.583333333333\\
159.25	880.333333333333\\
158.639	881.916666666667\\
158.158	881.916666666667\\
157.565	888.25\\
157.137	896.166666666667\\
156.645	896.166666666667\\
156.547	896.166666666667\\
155.509	897.75\\
154.99	891.416666666667\\
154.47	883.5\\
153.905	878.75\\
153.398	875.583333333333\\
153.076	875.583333333333\\
152.299	881.916666666667\\
151.923	886.666666666667\\
151.254	888.25\\
151.228	889.833333333333\\
150.751	888.25\\
150.189	886.666666666667\\
149.47	886.666666666667\\
149.12	893\\
148.04	894.583333333333\\
147.465	893\\
146.937	885.083333333333\\
146.423	877.166666666667\\
145.9	862.916666666667\\
145.358	856.583333333333\\
144.839	856.583333333333\\
144.256	862.916666666667\\
143.76	881.916666666667\\
143.25	896.166666666667\\
142.709	908.833333333333\\
141.908	916.75\\
141.708	918.333333333333\\
141.146	905.666666666667\\
140.619	883.5\\
139.977	859.75\\
139.517	836\\
138.992	824.916666666667\\
138.37	820.166666666667\\
137.713	837.583333333333\\
137.219	870.833333333333\\
136.895	881.916666666667\\
136.189	889.833333333333\\
135.33	912\\
135.227	926.25\\
134.429	916.75\\
133.959	929.416666666667\\
133.824	932.583333333333\\
132.938	918.333333333333\\
132.7	907.25\\
}--cycle;
\addlegendentry{$\pm$4\% area}

\addplot [color=green]
  table[row sep=crcr]{%
132.7	700\\
132.7	1200\\
};
\addlegendentry{Haptics On/Off}

\addplot [color=green]
  table[row sep=crcr]{%
266.909	700\\
266.909	1200\\
};

\end{axis}
\end{tikzpicture}%

%% file: appendixNike.tex
\section*{Appendix: interfacing and using Nike+iPod Kit \label{Appendix:1}}
In this section	 we briefly detail the interfacing procedure for using the Nike+iPod sensor for customized application. The kit (approx. 29\$ (USD)) contains two modules: a tiny sensor to be placed in the shoe and a receiver to be used with iPod. When the user walks or runs, the piezo-sensor estimates and wirelessly transmits information about the user's gait to the receiver.
Following the result presented in \cite{Saponas2006},  we modified an iPod female connector by soldering wires from the serial pins on the iPod connector to our adapter, adjusted the voltage accordingly, and powered with \unit{3.3}{V}. We then plugged a
 Nike+iPod receiver into our female connector replacing the Ipod with a PC running an \textit{ad-hoc} developed software. This caused the receiver to start sending packets over the serial connection to our computer, allowing us to monitor the measured cadence.
Acosta \eal in \cite{acosta2011} and Kane \eal in \cite{Kane2010} validated the accuracy of the \textit{Nike+ Wireless Sport Kit} to estimate pace (min/km), and distance (km) during treadmill walking and running. 
Results showed that the \textit{Nike+} device overestimated the speed of level walking at \unit[3.3]{$km / h$} about 20\%, underestimated the speed of level walking at \unit[6.6]{$km / h$} by 12\%, but correctly estimated the speed of level walking at \unit[4.9]{$km / h$}, and level running at all speeds ($p<0.05$). Similar results were found for distance estimation.
Starting from the preliminary results presented in \cite{Saponas2006} we developed a device for receiving and decoding messages from th \textit{Nike+ }sensor.
We designed and build an \textit{ad-hoc} PCB for connecting the  receiver with an Arduino based micro-controller.
We can split the developed code in two main parts. The former acquires information from the sensor and sends the computed cadence to a remote server using internet, the latter receives the information about the partner rhythm and activates the motors correspondingly.
Two serial communications were created in order to communicate at the same time with the sensor and the smartphone.
The communication between the pedometer and our system starts sending a header packet of 8 byte. This packet puts the sensor in active mode and the stream of data is enabled. We observed that the sensor streams a packet of 34 bytes every seconds. We collected and analyzed  several packets from multiple sensors, noticing some common bytes.
A representative packet is the following:
\texttt{FF 55 1E 09 0D 0D 01 24 F2 1D 30 A3 A1 97 E3 86 C1 F3 39 DC C6 12 5C CE FB 3C 83 0D EE 4C 1F FB F8 38}.
We discovered that \texttt{FF 55} is the packet header, and the payload starts with  \texttt{1E 09 0D 0D 01} for all the sensors and all packets. The packet continues with 27 bytes. The first 26 bytes carries all the information estimated by the pedometer, such us walking steps, running steps, sensor ID, lifetime walking and running miles, etc.
The last byte is used as a check-sum to validate or discard received packets. We tested all the possible combination of packet bytes and checksum type and we found that the last byte is a \textit{8bit 2s Complement} checksum.
The 26 bytes payload are decoded using a library based on the work done by Grinberg \cite{dmitryNike}.
All the sensors use the same radio frequency, and a packet per second is sent regardless the presence of a request or ack from the receivers, thus to use multiple Nike+ we process packet only if the descrambled serial number matches the one associated to the user. One per second Arduino receives the total amount of walked (or run) steps. We exploit this incremental measures to compute the cadence \iec the number of steps per minute. A moving average with time window of 5 seconds is used to have good compromise between  response time and smoothness. As soon as a change in the cadence occurs, the smart-phone (or smart-watch) is notified.
